\documentclass[12pt]{article}
\usepackage{amsmath,amsthm,amsfonts,amssymb}

\parskip=\medskipamount
\newtheorem{definition}{Definition}[section]
\newtheorem{defn}{Definition}[section]

\newtheorem{lem}[definition]{Lemma}

\newtheorem{thm}[definition]{Theorem}

\newtheorem{prop}[definition]{Proposition}

\newtheorem{cor}[definition]{Corollary}
\newtheorem{remark}[definition]{Remark}

 1
\font\ddpp=msbm10  scaled \magstep 1

\def\QED{\hskip0.1em\hfill\null\ \null\nobreak\hfill
\kern3pt\lower1.8pt\vbox{\hrule\hbox   {\vrule\kern1pt\vbox{\kern1.7pt
\hbox{$\scriptstyle   QED$}\kern0.2pt}\kern1pt\vrule}\hrule}}

\def\hfl#1#2{\smash{\mathop{\hbox to 12 mm{\rightarrowfill}}
\limits^{\scriptstyle#1}_{\scriptstyle#2}}}

 1

\begin{document}

\title{{\Large \textbf{SYMMETRIES IN CLASSICAL FIELD THEORY}}}
\author{Manuel DE LE\'ON \thanks{%
mdeleon@imaff.cfmac.csic.es}, \and David MART\'IN DE DIEGO\thanks{%
d.martin@imaff.cfmac.csic.es} \and Aitor SANTAMAR\'IA--MERINO\thanks{%
aitors@imaff.cfmac.csic.es} \bigskip \\
Departamento de Matem\'aticas\\
Instituto de Matem\'aticas y F{\'\i}sica Fundamental\\
Consejo Superior de Investigaciones Cient{\'\i}ficas\\
Serrano 123, 28006 Madrid, SPAIN }
\date{April 3, 2004}
\maketitle

\begin{abstract}
The multisymplectic description of Classical Field Theories is revisited,
including its relation with the presymplectic formalism on the space of
Cauchy data. Both descriptions allow us to give a complete scheme of
classification of infinitesimal symmetries, and to obtain the corresponding
conservation laws.
\end{abstract}


\bigskip

\section{Introduction}

The multisymplectic description of Classical Field Theories goes
back to the end of the sixties, when it was developed by the
Polish school leadered by W. Tulczyjew (see
\cite{BSF,Kijow,KiSz,KT,Sni}), and also independently by P.L
Garc{\'\i}a and A. P\'erez-Rend\'on \cite{PL,sala1,sala2}, and H.
Goldschmidt and S. Sternberg \cite{GS}. From that time, this topic
has continuously deserved a lot of attention mainly after the
paper \cite{CCI}, and more recently in
\cite{FPR,sean1,sean2,PR1,PR2}. A serious attempts to get a full
development of the theory has been done in the monographs
\cite{gimmsy1,gimmsy2} (see also \cite{LR0} for higher order
theories). In addition, multisymplectic setting is proving to be
useful for numerical purposes \cite{MPS}.

The final goal is to obtain a geometric description similar to the
symplectic one for Lagrangian and Hamiltonian mechanics.
Therefore, the first idea was to introduce a generalization of the
symplectic form. The canonical symplectic structure on the
cotangent bundle of a configuration manifold is now replaced by
multisymplectic forms canonically defined on the bundles of
exterior forms on the bundle configuration $\pi: Y \longrightarrow
X$ of the theory in consideration. These geometric structures can
be abstracted to arbitrary manifolds; its study constitutes a new
subject of interest for geometers
\cite{CIL,CIL2,gent2002,Mar1,Mar2} which could give new insights
as it happened with symplectic geometry in the sixties.

On the other hand, if we start with a Lagrangian density, we can
construct first a Lagrangian form from a volume form fixed on the
space-time manifold $X$, and then, using the bundle structure of
the 1-jet prolongation $\pi_{XZ} : Z \longrightarrow X$ of $Y$, we
construct a multisymplectic form on $Z$ (provided that the
Lagrangian is regular).

In this geometric context, one can present the field equations in
two alternative ways: in terms of multivectors (see
\cite{EMR0,EMR1,EMR2,EMR3,EMR4,EMR5,EMR6,EMR7, FPR}), or in terms
of Ehresmann connections \cite{Vilanova,LMMS,LMMS2,gent2002}.

Let us remark that there are alternative approaches using the
so-called polysymplectic structures (see
\cite{sarda0,sarda1,igor,sard-1,sard,SaZa}) or even $n$-symplectic
structures (see \cite{norris} for a recent survey). Here, we shall
present the field equations in terms of Ehresmann connections;
indeed, note that in Lagrangian or Hamiltonian mechanics one looks
for curves, or, in an infinitesimal version, tangent vectors; now,
we look for sections of the corresponding bundles, which
infinitesimally correspond to the horizontal subspaces of
Ehresmann connections. In fact, the Euler-Lagrange equations (more
generally, teh De Donder equations) and Hamilton equations can be
described in a form which is very similar to the corresponding
ones in Mechanics. Both formalisms (Lagrangian and Hamiltonian)
are related via the Legendre transformation. The case of singular
theories is also considered, and a constraint algorithm is
obtained.

Accordingly with these different descriptions, we have different
notions of infinitesimal symmetries (see \cite{OLVER} for a
description based in the calculus of variations). The aim of the
present paper is to classify the different kind of infinitesimal
symmetries and to study their relationship with conservation laws
in the geometric context of multisymplectic geometry and Ehresmann
connections.

In addition, choosing a Cauchy surface, we also develop the
corresponding infinite dimensional setting in the space of Cauchy
data. Both descriptions are related by means of integration along
the Cauchy surfaces, allowing to relate the above symmetries with
the ones of the presymplectic infinite dimensional system.

Let us remark that we consider boundary conditions along the paper.

The paper is structured as follows. Section 2 describe the Lagrangian
setting for the Classical Field Theories of first order using the tools of
jet manifolds, in both regular and singular cases. Multisymplectic forms and
brackets are introduced at the end of the section in order to be used later.
Section 3 is devoted to give a Hamiltonian description for Classical Field
Theories, including the Legendre transformation and the equivalence theorem.
The singular case is also discussed. Section 4 deals with the theory of
Cauchy surfaces for the Classical Field Theory, where the tools that will be
required later are introduced. In particular, the integration method, as a
way to connect the finite dimensional setting and the theory of Cauchy
Surfaces, is discussed in depth. The singular case and the Poisson brackets
are also considered. Section 5 describes thoroughly the different
infinitesimal symmetries for the Lagrangian and Hamiltonian settings, using
the tools that have been described in previous sections. In Section 6, we
discuss the Momentum Map in the finite and infinite dimensional settings.
The paper finishes with section 7, in which we illustrate the concepts
discussed with the examples of the Bosonic string, following the Polyakov
approach, and the Klein-Gordon field.

Along this paper, we shall use the following notations.
$\mathfrak{X}(M)$ will denote the Lie algebra of vector fields on
a manifold $M$, and $\pounds_X$ will be the Lie derivative with
respect to a vector field $X$. The differential of a
differentiable mapping $F: M \longrightarrow N$ will be
indistinctly denoted by $F_*$, $dF$ or $TF$. By $C^\infty(M)$ we
denote the algebra of smooth functions on a manifold $M$.

\section{Lagrangian formalism}

\subsection{The setting for classical field theories}

Consider a fibration $\pi = \pi_{XY} :Y\longrightarrow X$, where
$Y$ is an $(n+1+m)$-dimensional manifold and $X$ is an orientable
$(n+1)$-dimensional manifold. We shall also fix a volume form on
$X$, that will be denoted by $\eta$. We can choose fibered
coordinates $(x^\mu,y^i)$ in $Y$, so that $\pi (x^{\mu}, y^i) =
(x^{\mu})$, and assume that the volume form is $\eta = d^{n+1}x =
dx^0\wedge\ldots\wedge dx^n $. Here, $0 \leq \mu, \nu, ... \leq n$
and $1 \leq i, j, ... \leq m$.

\begin{remark}
\textrm{Time dependent mechanics can be considered as an example of
classical field theory, where $X$ is chosen to be the real line $\hbox{\ddpp
R}$, representing time, and the fibre over $t$ represents the configuration
space at time $t$.}
\end{remark}

We shall also use the following notation:
\begin{equation*}
d^nx_\mu := \iota_{\partial/\partial x^\mu}d^{n+1}x,\quad d^{n-1}x_{\mu\nu}
:= \iota_{\partial/\partial x^\mu}\iota_{\partial/\partial x^\nu}d^{n+1}x,
\quad\ldots
\end{equation*}

The first order jet prolongation $J^{1}\pi$ is the manifold of
classes $j^{1}_x\phi$ of sections $\phi$ of $\pi$ around a point
$x$ of $X$ which have the same Taylor expansion up to order one.
$J^1\pi$ can be viewed as the generalisation of the phase space of
the velocities for classical mechanics. Therefore, $J^{1}\pi$,
which we shall denote by $Z$, is an $(n+1+m+(n+1)m)$-dimensional
manifold. We also define the canonical projections $\pi_{XZ}: Z
\longrightarrow X$ by $\pi_{XZ}(j^{1}_x\phi)=x$, and $\pi_{YZ}: Z
\longrightarrow Y$ by $\pi_{YZ}(j^{1}_x\phi)=\phi(x)$ (see Figure
1). We shall also use the same notation $\eta$ for the pullback of
the chosen volume form $\eta$ on $X$ to $Z$ along the projection.
If we have adapted coordinates $(x^{\mu}, y^i)$ in $Y$, then we
can define induced coordinates in $Z$, given by
$(x^{\mu},y^{i},z^i_{\mu})$, such that
\begin{align*}
x^{\mu}(j^{1}_x\phi) &= x^{\mu}(x) \\
y^i(j^{1}_x\phi) &= y^i(\phi(x)) = \phi^i(x) \\
z^i_{\mu}(j^{1}_x\phi) &= \frac{\partial \phi^i}{\partial x^{\mu}}\Big| _{x}
\end{align*}

\unitlength=1mm \special{em:linewidth 0.4pt} \linethickness{0.4pt}
\begin{picture}(115.00,85.00)(-20,0)
\put(40.00,80.00){\makebox(0,0)[cb]{$Z=J^1\pi_{XY}$}}
\put(00.00,50.00){\makebox(0,0)[cc]{$Y$}}
\put(20.00,20.00){\makebox(0,0)[ct]{$X$}}
\put(36.00,77.67){\vector(-4,-3){32.33}}
\put(2.00,47.33){\vector(2,-3){16.33}}
\put(39.33,76.67){\vector(-1,-3){17.67}}
\put(17.67,68.00){\makebox(0,0)[rb]{$\pi_{YZ}$}}
\put(28.67,51.33){\makebox(0,0)[rb]{$\pi_{XZ}$}}
\put(9.33,38.67){\makebox(0,0)[lb]{$\pi_{XY}$}}
\bezier{168}(16.67,21.67)(-6.00,31.33)(-1.00,47.67)
\put(-1.05,48.67){\vector(0,1){0.00}}
\bezier{292}(22.33,20.33)(54.00,66.33)(42.00,78.00)
\put(42.00,78.00){\vector(-1,1){0.33}}
\put(-1.00,29.67){\makebox(0,0)[rt]{$\phi$}}
\put(43.67,52.33){\makebox(0,0)[lt]{$j^1\phi$}}
\put(110.00,80.00){\makebox(0,0)[cb]{$(x^{\mu}, y^i, z^i_{\mu})$}}
\put(70.00,50.00){\makebox(0,0)[cc]{$(x^{\mu}, y^i)$}}
\put(90.00,20.00){\makebox(0,0)[ct]{$(x^{\mu})$}}
\put(106.00,77.67){\vector(-4,-3){32.33}}
\put(72.00,47.33){\vector(2,-3){16.33}}
\put(109.33,76.67){\vector(-1,-3){17.67}}
\put(60,10){\makebox(0,0)[cc]{ $\displaystyle{\left\{
\begin{array}{l}
\dim X=n+1\\
\dim Y=n+1+m\\
\dim Z=n+1+m+(n+1)m\\
\end{array}
\right. }$ }} \put(60.00,-2){\makebox(0,0)[ct]{Figure 1}}
\end{picture}

\

As usual, one can define the concept of verticality, by defining the
following subbundles:
\begin{align*}
\mathcal{V}_y\pi &:= (T_y\pi)^{-1}(0_x) \\
\mathcal{V}_z\pi_{XZ} &:= (T_z\pi_{XZ})^{-1}(0_x)
\end{align*}
We can consider the more general case in which $X$ is a manifold
with boundary $\partial X$, and we also have boundaries for
manifolds $Y$ and $Z$, given by $\partial Y = \pi^{-1}(\partial
X)$ and $\partial Z = \pi^{-1}_{XZ}(\partial X)$, respectively. A
boundary condition is encoded in a subbundle $B$ of $\partial Z
\longrightarrow \partial X$, and restricting ourselves to sections
$\phi : X \longrightarrow Y$ such that $j^1\phi (\partial
X)\subseteq B$ (see \cite{BSF}).

There are several other alternative (and equivalent) definitions
of the first order jet bundle, such as considering the affine
bundle over $Y$ whose fibre over $y\in \pi^{-1}(x)$ consists of
linear sections of $T\pi_{XY}$, modelled over the vector bundle on
$Y$ whose fibre over $y\in \pi^{-1}(x)$ is the space of linear
maps of $T_xX$ to $\mathcal{V}_y\pi$; in other words, $Z$ is an
affine bundle over $Y$ modelled on the vector bundle $\pi{T^*X}
\otimes_{Y} \mathcal{V}\pi$ (see \cite{gimmsy1,Sa1,Saunders}).

The first order jet bundle is equipped with a geometric object $S_\eta$,
which depends on our choice of the volume form, called vertical endomorphism
(see \cite{Crampin} or \cite{Saunders}). What follows is an alternative way
to define it. First of all, we construct the isomorphism (vertical lift)
\begin{equation*}
v : \pi^*{T^*X}\otimes_{Y} \mathcal{V}\pi \longrightarrow
\mathcal{V}\pi_{YZ}
\end{equation*}
as follows: given $\left. f\in (\pi^*{T^*X}\otimes_{Y} \mathcal{V}
\pi)\right|_{j^1\phi}$ consider the curve $\gamma_{f}: \mathbb{R}
\longrightarrow \pi_{YZ}^{-1}(\pi_{YZ} (j_{x}^{1} \phi))$ given by
\begin{equation*}
\gamma_{f}(t) = j_{x}^{1} \phi + t~ f \; ,
\end{equation*}
for all $t \in \mathbb{R}$. Now define
\begin{equation*}
f^v = \frac{d}{dt}\gamma_f(t)|_{t=0}
\end{equation*}

If $(x^{\mu}, y^{i})$ are fibered coordinates on $Y$ and $f =
\displaystyle f^{i}_{\mu} dx^{\mu}|_{x} \otimes \left.
\frac{\partial}{\partial y^{i}} \right|_{\phi(x)}$ then
\begin{equation*}
f^{v} = f^{i}_{\mu} \left. \frac{\partial}{\partial z^i_{\mu}}
\right|_{j^1_x\phi} \; .
\end{equation*}

Let $x$ be a point of $X$ and $\phi \in \Gamma_{x}(\pi)$, where
$\Gamma_{x}(\pi)$ denotes the set of all local sections around the
point $x$. If $V_{0}, \dots, V_n$ are $n+1$ tangent vectors to
$J^{1}\pi$ at the point $j^{1}_{x} \phi \in Z$, then we have that
$T_{j^{1}_{x} \phi}\pi_{YZ}(V_{i}) - T_x\phi \circ T_{j^{1}_{x}
\phi}\pi_{XZ}(V_{i}) \in (\mathcal{V} \pi)_{\phi(x)}$ (this is the
vertical differential of a vector field on $Z)$. From the volume
form $\eta$, we also construct a family of 1-forms $\eta_{i}$ as
follows:
\begin{equation*}
\eta_{i}(x) = (-1)^{n+1-i} i_{T_{j^{1}_{x} \phi}\pi_{XZ}(V_{0})} ~ \cdots ~
\widehat{ i_{T_{j^{1}_{x} \phi}\pi_{XZ}(V_{i})} } ~ \cdots ~ i_{T_{j^{1}_{x}
\phi}\pi_{XZ}(V_{n})} ~ \eta(x) \; ,
\end{equation*}
where the hat over a term means that it is omitted.

Next, we define the \textbf{vertical endomorphism} $S_{\eta}$ as follows:
\begin{equation*}
(S_{\eta})_{j^{1}_{x}\phi}(V_{0}, \dots , V_{n}) = \sum_{i=0}^{n}~
\left(\eta_{i}(x) \otimes (T_{^{j^{1}_{x} \phi}}\pi_{YZ}(V_{i}) - T_x\phi
\circ T_{j^{1}_{x} \phi}\pi_{XZ}(V_{i}))\right)^{v}
\end{equation*}

Whenever we pick a different volume form $F \eta$, then $(F\eta)_i
= F\eta_i$, whence we also get $S_{F\eta}=FS_\eta$, where $F : X
\longrightarrow \hbox{\ddpp R}$ is nowhere-vanishing smooth
function on $X$.

The vertical endomorphism can be also written in local induced coordinates
as follows
\begin{equation*}
S_\eta = (dy^i - z^i_\mu dx^\mu) \wedge d^nx_\nu \otimes
\frac{\partial }{\partial z^i_\nu}
\end{equation*}

Higher order jet bundles can be defined in a similar manner. The second
order jet bundle, for example, is an $(n+1+m+(n+1)m+\left(
\begin{array}{c}
n+2 \\
2
\end{array}
\right)m$)-dimensional manifold, which has induced coordinates
$(x^\mu,y^i,z^i_\mu, z^i_{\mu\nu})$, where
\begin{equation*}
z^i_{\mu\nu}(j^{1}_{p}\phi) = \frac{\partial^{2} \phi^i}{\partial x^{\mu}
\partial x^{\nu}}\Big| _{p}
\end{equation*}

These bundles allow us to define the \textbf{total derivative} associated to
the partial derivative vector fields, which are locally expressed as
\begin{equation*}
\frac{d}{dx^{\mu}}= \frac{\partial}{\partial x^{\mu}} + z^i_{\mu}
\frac{\partial}{\partial y^i} + z^i_{\mu\nu}
\frac{\partial}{\partial z^i_{\nu}}+\ldots
\end{equation*}

\subsection{Jet prolongation of vector fields}

\begin{defn}
A 1-form $\theta\in \Lambda^1(Z)$ is said to be a \textbf{contact 1-form}
whenever
\begin{equation*}
(j^1\phi)^*\theta=0
\end{equation*}
for every section $\phi$ of $\pi$.
\end{defn}

If $(x^{\mu}, y^i, z^i_{\mu})$ is a system of local coordinates on $Z$, then
the contact forms are locally spanned by the 1-forms
\begin{equation*}
\theta^i= dy^i-z^i_{\mu}\, dx^{\mu}
\end{equation*}

We shall denote by $\mathcal{C}$ the algebraic ideal of the contact forms,
and by $\mathcal{I}(\mathcal{C})$ the differential ideal generated by the
contact forms, in other words, the ideal of the exterior algebra generated
by the contact forms and their differentials.

The distribution determined by the annihilation of the contact forms on $Z$
is called the \textbf{Cartan distribution} and it plays a fundamental role,
since it is the geometrical structure which distinguishes the holonomic
sections (sections which are prolongations of sections of $\pi_{XY})$ from
arbitrary sections of $\pi_{XZ}$ (see \cite%
{Chern,Ka,Krupka,krupkamusi,Kr,Vinogradov} for more details).

\begin{lem}
For any vector field $X$ in $Z$, the following two conditions are equivalent:

(i) For every $Y$ in the Cartan distribution $\pounds_X Y$ lies in the
Cartan distribution; in other words, $X$ preserves the Cartan distribution.

(ii) $X$ preserves $\mathcal{C}$, in other words, for every
$\theta\in \mathcal{C}$, $\pounds_X \theta \in\mathcal{C}$.

If any of the preceding two hold, then $X$ preserves
$\mathcal{I}(\mathcal{C} )$, in other words, for every
$\alpha\in\mathcal{I}(\mathcal{C})$, $\pounds_X
\alpha\in\mathcal{I}(\mathcal{C})$.
\end{lem}

\begin{defn}
Given a vector field $\xi_{Y} \in \mathfrak{X}(Y)$, then its \textbf{1-jet
prolongation} is defined as the unique vector field $\xi_Y^{(1)} \in
\mathfrak{X}(Z)$ projectable onto $\xi_{Y}$ by $\pi_{YZ}$, and which
preserves the Cartan distribution (in other words, $\pounds_{\xi_Y^{(1)}}%
\theta\in\mathcal{C}$ for every contact form $\theta)$.
\end{defn}

If $\xi_{Y}$ is locally expressed as
\begin{equation*}
\xi_{Y} = \xi^\mu_{Y}\frac{\partial }{\partial x^\mu} + \xi^i_Y \frac{%
\partial }{\partial y^i}
\end{equation*}
then the 1-jet prolongation of $\xi_{Y}$ must have the following form
\begin{equation}  \label{eq:vfliftcoord}
\xi_Y^{(1)} = \xi^\mu_Y\frac{\partial }{\partial x^\mu} + \xi^i_Y
\frac{\partial }{\partial y^i} + \left(
\frac{d\xi^i_Y}{dx^\mu}-z^i_\nu\frac{ d\xi^\nu_{Y}}{dx^\mu}
\right) \frac{\partial }{\partial z^i_\mu}
\end{equation}

Assume that the local expression of $\xi_Y^{(1)}$ is
\begin{equation}  \label{ddd}
\xi^{(1)}_{Y} = \xi^\mu_{Y}\frac{\partial }{\partial x^\mu} +
\xi^i_Y \frac{\partial }{\partial y^i} + \xi^i_{\mu Y}
\frac{\partial }{\partial z^i_\mu}
\end{equation}
In order to see that (\ref{ddd}) has the form
(\ref{eq:vfliftcoord}), pick $i\in\{1, 2, \ldots, m\}$, and impose
the second condition
$\pounds_{\xi_Y^{(1)}}\theta^i\in\mathcal{C}$. We have
\begin{align*}
\pounds_{\xi_Y^{(1)}}\theta^i &= \frac{\partial \xi^i_Y}{\partial
x^\mu} dx^\mu+\frac{\partial \xi^i_Y}{\partial y^j}dy^j -
\xi^i_{\mu Y}dx^\mu - z^i_\mu(\frac{\partial \xi^\mu_Y}{\partial
x^\nu}dx^\nu+\frac{\partial
\xi^\mu_Y}{\partial y^j}dy^j) \\
&= (\frac{\partial \xi^i_Y}{\partial y^j}-z^i_\mu\frac{\partial
\xi^\mu_Y}{\partial y^j})dy^j - (-\frac{\partial \xi^i_Y}{\partial
x^\nu}+\xi^i_{\nu Y}+z^i_\mu\frac{\partial \xi^\mu_Y}{\partial
x^\nu})dx^\nu
\end{align*}
Therefore
\begin{equation*}
-\frac{\partial \xi^i_Y}{\partial x^\nu}+\xi^i_{\nu
Y}+z^i_\mu\frac{\partial \xi^\mu_Y}{\partial x^\nu} =
z^j_\nu(\frac{\partial \xi^i_Y}{\partial y^j}
-z^i_\mu\frac{\partial \xi^\mu_Y}{\partial y^j})
\end{equation*}
and we get
\begin{equation*}
\xi^i_{\mu Y} = \frac{d\xi^i_Y}{dx^\mu}-z^i_\nu\frac{d\xi^\nu_{Y}}{dx^\mu}.
\end{equation*}

Vertical lifting is a Lie algebra homomorphism, as we can see in

\begin{prop}
\label{prop:liftLieBracket} For every $\xi,\zeta\in\mathfrak{X}(Y)$,
\begin{equation*}
[\xi,\zeta]^{(1)} = [\xi^{(1)},\zeta^{(1)}]
\end{equation*}
\end{prop}

\textit{Proof}. $[\xi^{(1)},\zeta^{(1)}]$ obviously projects onto
$[\xi,\zeta]$, and if $\alpha$ is a contact form, then
\begin{equation*}
\pounds_{[\xi^{(1)},\zeta^{(1)}]}\alpha =
\pounds_{\xi^{(1)}}\pounds_{\zeta^{(1)}}\alpha -
\pounds_{\zeta^{(1)}}\pounds_{\xi^{(1)}}\alpha
\end{equation*}
which is obviously an element of $\mathcal{C}$. \hfill $\ \ \ \vrule height
1.5ex width.8ex depth.3ex \medskip$

If $\xi_{Y}$ is projectable onto a vector field $\xi_{X}\in
\mathfrak{X}(X)$, there is a natural alternative way of defining
its 1-jet prolongation, which will be used afterwards. If
$\xi_{Y}$ projects onto $\xi_{X}$, having flows $\Phi^Y_t$ and
$\Phi^X_t$ respectively, then $\Phi^Z_t: Z \longrightarrow Z$
defined by $\Phi^Z_t (j^1_x\phi) =
j^1_{\Phi^X_t(x)}(\Phi^Y_t\circ\phi\circ(\Phi^X_t)^{-1})$ is the
flow of the 1-jet prolongation of $\xi_Y$ (see \cite{Saunders} for
further details).

\begin{lem}
\label{lem:PreEL} For every $\pi_{XY}$-projectable vector field
$\xi_Y \in \mathfrak{X}(Y)$ and for any form $\alpha\in \bigwedge
Z$, and any section $\phi: X \longrightarrow Y$ of $\pi$, we have
\begin{equation*}
\frac{d}{dt}\Big|_{t=0}
(j^1(\Phi^Y_t\circ\phi\circ(\Phi^X_t)^{-1}))^{*}\alpha =
(j^1\phi)^{*}\pounds_{\xi_Y^{(1)}}(\alpha)
\end{equation*}
where $\Phi^Y_t$ and $\Phi^X_t$ are the flows induced by $\xi_Y$ and its
projection onto $X$, respectively.
\end{lem}

The proof of this lemma follows in a similar way to the one of Lemma 4.4.5
in \cite{Saunders}.

\subsection{Lagrangian form. Poincar\'e-Cartan forms}

For first order field theories, the dynamical evolution of a
Lagrangian system is described by a \textbf{Lagrangian form}
$\mathcal{L}$ defined on $Z $, which is a semibasic $(n+1)$-form
in $Z$ respect to the $\pi_{XZ}$ projector (in other words, it is
annihilated when applied to at least one $\pi_{XZ}$-vertical
vector). This allows us to define the \textbf{Lagrangian} function
as the unique function $L$ such that $\mathcal{L}=L\eta$.\medskip

Let us introduce the following local notation, that we shall often use.

\begin{defn}
\label{defpp} We denote by
\begin{equation*}
\hat{p}_i^\mu := \frac{\partial L}{\partial z^i_\mu}
\end{equation*}
and by
\begin{equation*}
\hat{p} := L - z^i_\mu \hat{p}_i^\mu
\end{equation*}
\end{defn}

\begin{defn}
For a given Lagrangian form $\mathcal{L}$ and a volume form $\eta$ we define
the \textbf{Poincar\'e-Cartan $(n+1)$-form} as
\begin{equation}  \label{eq:mform}
\Theta_L := \mathcal{L} + (S_{\eta})^{*}(dL)
\end{equation}
\end{defn}

In induced coordinates, it has the following expression
\begin{align*}
\Theta_L &= \Big(L-z^i_{\mu}\frac{\partial L}{\partial
z^i_{\mu}}\Big)
d^{n+1}x +\frac{\partial L}{\partial z^i_\mu}dy^i\wedge d^nx_\mu \\
&= (\hat{p}dx^\mu+\hat{p}^\mu_idy^i)\wedge d^{n}x_\mu \\
&= \mathcal{L} + \hat{p}^\mu_i\theta^i\wedge d^{n}x_\mu
\end{align*}

From this form, we can also define its differential

\begin{defn}
The \textbf{Poincar\'e-Cartan $(n+2)$-form} is defined as
\begin{equation*}
\Omega_L := -d\Theta_L.
\end{equation*}
\end{defn}

In induced coordinates is expressed as follows
\begin{align*}
\Omega_L &= -(dy^i - z^i_{\mu}dx^{\mu})\wedge\Big(\frac{\partial L}{\partial
y^i}d^{n+1}x-d\Big(\frac{\partial L}{\partial z^i_{\mu}}\Big)\wedge
d^{n}x_\mu\Big) \\
&= ( d\hat{p}\wedge dx^\mu + d\hat{p}^\mu_i\wedge dy^i)\wedge d^{n}x_\mu \\
&= -\theta^i\wedge\Big(\frac{\partial L}{\partial
y^i}d^{n+1}x-d\hat{p} ^\mu_i\wedge d^{n}x_\mu\Big)
\end{align*}

\begin{remark}
\textrm{A different choice for the volume form $\eta$ does not
produce changes in the Poincar\'e-Cartan forms. In fact, if we
replace $\eta$ with a new volume form $\bar{\eta}=F\eta$, where
$F$ is a non-vanishing function, we would have $\mathcal{L} = L
\eta = \bar{L} \bar{\eta}$, with $\bar{L}=L/F$ and using the
preceding computations we finally get $\Theta_L = \Theta_{{
\bar{L}}}$. Thus, we could use the notation $\Theta_{\mathcal{L}}$
and $\Omega_{\mathcal{L}}$ (see \cite{EMR0}). }
\end{remark}

At this point, we have to introduce an extra hypothesis on the boundary
condition $B \subseteq \partial Z$, that represents boundary conditions on
the solutions, which is the existence of an $n$-form $\Pi$ on $B$ such that
\begin{equation*}
i_B^*\Theta_L = d\Pi
\end{equation*}
where $i_B: B \longrightarrow Z$ is the inclusion map (see \cite{BSF}).

We can deduce the following properties

\begin{prop}
\label{lemma:verticalOnOmega} The following holds:\newline (i)
$(j^1\phi)^*\pounds_{\xi_Y^{(1)}}(\mathcal{L})=(j^1\phi)^*\pounds_{\xi_Y^{(1)}}(\Theta_L)$
\newline
(ii) For any $z\in Z$ and every two $\pi_{XZ}$-vertical tangent
vectors $v, w\in \mathcal{V}_z\pi_{YZ}$,
\begin{equation*}
\iota_v\iota_w(\Theta_L)_z=0
\end{equation*}
(iii) For any $z\in Z$ and every three $\pi_{XZ}$-vertical tangent
vectors $u, v, w\in \mathcal{V}_z\pi_{YZ}$,
\begin{equation*}
\iota_u\iota_v\iota_w(\Omega_L)_z=0
\end{equation*}
\end{prop}

The following proposition will be useful later.

\begin{prop}
\label{prop:tangentToSection} If $\sigma$ is a section of $\pi_{XZ}$ and $\xi
$ is a vector field in $Z$ tangent to $\sigma$, then
\begin{equation*}
\sigma^*(\iota_\xi\Omega_L)=0
\end{equation*}
\end{prop}

\textit{Proof}. $\xi=T\sigma (\lambda)$ along $\sigma$ for certain
$\lambda\in\mathfrak{X}(X)$. Thus,
\begin{equation*}
\sigma^*(\iota_\xi\Omega_L)=\sigma^*(\iota_{T\sigma (\lambda)}\Omega_L)
=\iota_\lambda (\sigma^*\Omega_L)=0
\end{equation*}
as $\sigma^*\Omega_L=0$.\hfill$\ \ \ \vrule height 1.5ex width.8ex depth.3ex
\medskip$

\subsection{Calculus of variations. Euler-Lagrange equations}

The previously introduced geometric objets will take part in the geometric
description of the dynamics of field theories, more precisely in the
Euler-Lagrange equations, that are traditionally obtained from a variational
problem.

The dynamics of the system is given by sections $\phi$ of $\pi_{XY}$ which
verify the boundary condition $(j^1\phi)(\partial X)\subseteq B$ and that
extremise the \textbf{action integral}
\begin{equation*}
S(\phi)=\int_{(j^1\phi)(C)}\mathcal{L}
\end{equation*}
where $C$ is a compact ($n+1$)-dimensional submanifold of $X$.

Variations of such sections are introduced by small perturbations
of certain section $\phi$ along the trajectories of a vertical or,
in general, a projectable vector field $\xi_Y$; in other words, if
$\Phi^Y_t$ is the flow of $\xi_Y$ and $\Phi_X$ the flow of its
projection, defines the \textbf{variations} of $\phi$ as the
sections $\phi_t := \Phi^Y_t\circ\phi\circ\Phi^X_{-t}$.

\begin{defn}
A section $\phi \in \Gamma (\pi)$ is an \textbf{extremal} of $S$ if
\begin{equation*}
\frac{d}{dt}\Big|_{t=0}\int_{(j^1\phi_t)(C)}\mathcal{L} =
\frac{d}{dt}\Big|_{t=0}\int_{C}(j^{1}\phi_{t})^{*} \mathcal{L} =0
\end{equation*}
for any compact $(n+1)$-dimensional submanifold $C$ of $X$, and for every
projectable vector field $\xi_{Y} \in \mathfrak{X}(Y)$
\end{defn}

Lemma \ref{lem:PreEL} allows us to rewrite to extremality condition as
\begin{equation}  \label{eq:variac1}
\int_{C}(j^{1}\phi)^{*} \pounds_{\xi_Y^{(1)}}(\mathcal{L}) =0
\end{equation}

\begin{thm}
If $\phi$ is an extremal of $L$, then for every $(n+1)$-dimensional compact
submanifold $C$ of $X$, such that $\phi(C)$ lies in a single coordinate
domain $(x^\mu,y^i)$, and for every projectable vector field $\xi_Y$ on $Y$
we have
\begin{align*}
0 &= \int_C (j^2\phi)^*\left[\frac{\partial L}{\partial
y^i}-\frac{d}{dx^\mu}
\frac{\partial L}{\partial z^i_\mu}\right](\xi_Y^i-z^i_\nu\xi_Y^\nu)\eta \\
&+ \int_{\partial C} (j^1\phi)^*(\iota_{\xi_Y^{(1)}}\Theta_L)
\end{align*}
Whenever $\phi$ is an extremal for the variational problem with fixed value
at the boundary of $C$, then $\phi$ satisfies the \textbf{Euler-Lagrange
equations}
\begin{equation*}
(j^{2}\phi)^{*} \left(\frac{\partial L}{\partial
y^i}-\frac{d}{dx^\mu}\frac{\partial L}{\partial z^i_\mu}\right)=0,
\qquad 1\leq i \leq m
\end{equation*}
\end{thm}

\textit{Proof}. A computation on the previous formula gives us
\begin{align*}
\int_{C}(j^{1}\phi)^* \pounds_{\xi_Y^{(1)}}(\mathcal{L}) &=
\int_C(j^{1}\phi)^* \xi_Y^{(1)}(L)\eta+\int_C(j^{1}\phi)^*L
(\pounds_{\xi_Y^{(1)}}(\eta)) \\
&= \int_C(j^{1}\phi)^* \xi_Y^\mu\frac{\partial L}{\partial x^\mu}
\eta+\int_C(j^{1}\phi)^*\xi_Y^i\frac{\partial L}{\partial y^i}\eta \\
&+\int_C(j^1\phi)^*\left[\frac{d}{dx^\mu}\xi_Y^i-z^i_\nu\frac{d}{dx^\mu}
\xi_Y^\nu\right]\frac{\partial L}{\partial z^i_\mu}\eta +
\int_C(j^{1}\phi)^*L (\pounds_{\xi_Y^{(1)}}(\eta)) \\
&= \int_C(j^{1}\phi)^* \xi_Y^\mu\frac{\partial L}{\partial x^\mu}
\eta+\int_C(j^{1}\phi)^*\xi_Y^i\frac{\partial L}{\partial y^i}\eta \\
&+\int_C(j^2\phi)^*\frac{d}{dx^\mu}\left[\xi_Y^i-z^i_\nu\xi_Y^\nu\right]
\frac{\partial L}{\partial
z^i_\mu}\eta+\int_C(j^2\phi)^*\xi_Y^\nu\frac{dz^i_\nu} {dx^\mu}\frac{\partial L}{\partial z^i_\mu}\eta \\
&+ \int_C(j^{1}\phi)^* L\frac{d\xi^\mu_Y}{dx^\mu}\eta \\
&= \int_C(j^{1}\phi)^* \xi_Y^\mu\frac{\partial L}{\partial x^\mu}
\eta+\int_C(j^{1}\phi)^*\xi_Y^i\frac{\partial L}{\partial y^i}\eta \\
&+\int_C(j^2\phi)^*\frac{d}{dx^\mu}\left[\xi_Y^i-z^i_\nu\xi_Y^\nu\right]
\frac{\partial L}{\partial z^i_\mu}\eta+\int_C(j^2\phi)^*\xi_Y^\nu
\frac{dz^i_\nu}{dx^\mu}\frac{\partial L}{\partial z^i_\mu}\eta \\
&+ \int_{\partial C}(j^1\phi)^*L\xi_Y^\mu d^{n}x_\mu
-\int_C(j^{1}\phi)^*\xi_Y^\mu\frac{\partial L}{\partial x^\mu}\eta
-\int_C(j^{1}\phi)^*z^i_\mu\frac{\partial L}{\partial y^i}\xi_Y^\mu\eta \\
&-\int_C(j^2\phi)^*\xi_Y^\mu\frac{dz^i_\nu}{dx^\mu}\frac{\partial
L}{\partial z^i_\nu}\eta \\
&=\int_C(j^{1}\phi)^*\frac{\partial L}{\partial
y^i}(\xi_Y^i-z^i_\mu\xi^\mu_Y)\eta+
\int_C(j^2\phi)^*\frac{d}{dx^\mu}\left[\xi_Y^i-z^i_\nu\xi_Y^\nu
\right]\frac{\partial L}{\partial z^i_\mu}\eta \\
&+\int_{\partial C}(j^1\phi)^*L\xi_Y^\mu d^{n}x_\mu \\
&= \int_C (j^2\phi)^*\left[\frac{\partial L}{\partial
y^i}-\frac{d}{dx^\mu}
\frac{\partial L}{\partial z^i_\mu}\right](\xi_Y^i-z^i_\nu\xi_Y^\nu)\eta \\
&+ \int_{\partial C}
(j^1\phi)^*\left[(\xi^i_Y-z^i_\nu\xi^\nu_Y)\frac{\partial
L}{\partial z^i_\mu}+L\xi^\mu_Y\right]d^{n}x_\mu
\end{align*}
The condition of fixed value at the boundary of $C$ means
$\xi_Y^\mu|_{\partial C}=\xi_Y^i|_{\partial C}=0$, therefore we
have
\begin{equation*}
0 = \int_C(j^2\phi)^*\left[\frac{\partial L}{\partial
y^i}-\frac{d}{dx^\mu} \frac{\partial L}{\partial
z^i_\mu}\right](\xi_Y^i-z^i_\nu\xi_Y^\nu)\eta
\end{equation*}
for arbitrary $\xi_Y^\mu$ and $\xi_Y^i$, whence we obtain the
Euler-Lagrange equations.\hfill$\ \ \ \vrule height 1.5ex
width.8ex depth.3ex \medskip$

\begin{lem}
If $\phi$ is a section of $\pi_{XY}$ and $\xi$ is a $\pi_{YZ}$ vertical
vector field in $Z$, then
\begin{equation*}
(j^1\phi)^*(\iota_\xi\Omega_L)=0
\end{equation*}
\end{lem}

\textit{Proof}. $\xi$ has components $(0,0,w^i_\mu)$, and an easy
computation shows that

\begin{equation*}  \label{eq:inclfield}
\iota_\xi\Omega_L = -w^j_\nu \frac{\partial ^2 L}{\partial z^i_\mu \partial
z^j_\nu}(\theta^i\wedge d^{n}x_\mu) \in \mathcal{I}(\mathcal{C})
\end{equation*}

which vanishes when pulled back by a 1-jet prolongation of a
section of $\pi_{XY}$.\hfill$\ \ \ \vrule height 1.5ex width.8ex
depth.3ex \medskip$

\begin{prop}
\label{intrinsicEL} (\textbf{Intrinsic version of Euler-Lagrange equations})
A section $\phi \in \Gamma (\pi)$ is an extremal of $S$ if and only if
\begin{equation*}
(j^1\phi)^*(\iota_\xi\Omega_L)=0
\end{equation*}
for every vector field $\xi$ on $Z$.
\end{prop}

\textit{Proof}. We have that
\begin{align*}
\int_C (j^1\phi)^*\mathcal{L}_{\xi_Y^{(1)}}\mathcal{L} =\int_C
(j^1\phi)^* \mathcal{L}_{\xi_Y^{(1)}}\Theta_L = -\int_C
(j^1\phi)^*\iota_{\xi_Y^{(1)}}\Omega_L+\int_{\partial
C}(j^1\phi)^*\iota_{\xi_Y^{(1)}}\Theta_L
\end{align*}
Therefore,
\begin{equation*}
-\int_C (j^1\phi)^*\iota_{\xi_Y^{(1)}}\Omega_L = \int_C
(j^2\phi)^*\left[\frac{\partial L}{\partial
y^i}-\frac{d}{dx^\mu}\frac{\partial L}{\partial
z^i_\mu}\right](\xi_Y^i-z^i_\nu\xi_Y^\nu)\eta
\end{equation*}
for every projectable vector field $\xi_Y$ on $Y$. Then, Euler-Lagrange
equations are satisfied in every $C$ if and only if
\begin{equation*}
(j^1\phi)^*\iota_{\xi_Y^{(1)}}\Omega_L = 0
\end{equation*}
for every projectable vector field $\xi_Y$ on $Y$, in every compact $C$ of $X
$. Now different local solutions can be glued together using partitions of
unity, so that we get that
\begin{equation*}
(j^1\phi)^*\iota_{\xi_Y^{(1)}}\Omega_L = 0
\end{equation*}
is the expression for global sections $\phi$.

Finally, any general vector field $\xi_Z$ may be decomposed into a
vector field tangent to $j^1\phi$, the lift of a
$\pi_{XY}$-vertical vector field on $Y$ and a $\pi_{YZ}$-vertical
vector field. Using the preceding lemma, and Proposition
\ref{prop:tangentToSection}, we get the result. \hfill$\ \ \
\vrule height 1.5ex width.8ex depth.3ex \medskip$

\subsection{Regular Lagrangians. De Donder equations}

In some cases, we shall need to assume extra regularity conditions on the
Lagrangian function:

\begin{defn}
For a Lagrangian function $L:Z \longrightarrow \mathbb{R}$ , it is defined
its \textbf{Hessian matrix}
\begin{equation*}
\left( \frac{\partial ^2L}{\partial z^{\alpha}_i\partial z^{\beta}_j}\right)
_{\alpha,\beta,i,j}
\end{equation*}
The Lagrangian is said to be \textbf{regular at} a point whenever such
matrix is regular at that point, and \textbf{regular} whenever it is regular
everywhere.
\end{defn}

When the Lagrangian is regular, the implicit function theorem
allows us to introduce new coordinates for $Z$, called
\textbf{Darboux coordinates} \cite{gent2002,Mar1,Mar2}, namely
$(x^\mu,y^i,\hat{p}_i^\mu)$, which will also be very convenient to
relate the Lagrangian formalism to Hamiltonian formalism.

We introduce the De Donder equations, closely related to the Euler-Lagrange
equations.

\begin{defn}
The following equation on sections $\sigma$ of $\pi_{XZ}$ is called the
\textbf{De Donder equations}:
\begin{align}  \label{eq:dedonder}
\sigma^*(\iota_\xi\Omega_L)=0\qquad\forall\xi\in\mathfrak{X}(Z)
\end{align}
Sections satisfying the De Donder equations and in addition the boundary
condition $\sigma(\partial X) \subseteq B$ are called solutions of the De
Donder equations.
\end{defn}

From proposition \eqref{prop:tangentToSection}, we deduce that De
Donder equations can be equivalently restated in terms of
$\pi_{XZ}$-vertical vector fields. In local coordinates, if
$\sigma(x^\mu)=(x^\mu, \sigma^i(x^\mu),\sigma^i_\nu(x^\mu))$ for
any $\xi = v^i\frac{\partial }{\partial y^i}+w^i_\mu\frac{\partial
}{\partial z^i_\mu}$ the equation is written as
\begin{align*}
0 = &-v^i\left( \frac{\partial L}{\partial y^i} -\frac{\partial
^2L}{\partial x^\nu\partial z^i_\nu}-\frac{\partial
\sigma^j}{\partial x^\mu} \frac{\partial ^2L}{\partial y^j\partial
z^i_\mu} - \frac{\partial \sigma^j_\mu}{\partial
x^\nu}\frac{\partial ^2L}{\partial z^j_\mu\partial z^i_\nu}+
\left(\frac{\partial \sigma^j}{\partial x^\mu}-\sigma^j_\mu\right)
\frac{\partial ^2L}{\partial y^i\partial z^j_\mu}\right) \\
&+ w^i_\mu \left(\left(\frac{\partial \sigma^j}{\partial x^\nu}
-\sigma^j_\nu\right)\frac{\partial ^2L}{\partial z^i_\mu\partial
z^j_\nu} \right),
\end{align*}
or, in other words,
\begin{equation*}
\left.
\begin{array}{r}
\displaystyle{\frac{\partial L}{\partial y^i} -\frac{\partial
^2L}{\partial x^\nu\partial z^i_\nu}-\frac{\partial
\sigma^j}{\partial x^\mu}\frac{
\partial ^2L}{\partial y^j\partial z^i_\mu} - \frac{\partial \sigma^j_\mu}{
\partial x^\nu}\frac{\partial ^2L}{\partial z^j_\mu\partial z^i_\nu}+
\left(\frac{\partial \sigma^j}{\partial x^\mu}-\sigma^j_\mu\right)\frac{\partial
^2L}{\partial y^i\partial z^j_\mu} = 0} \\
\displaystyle{\left(\frac{\partial \sigma^j}{\partial x^\nu}
-\sigma^j_\nu\right)\frac{\partial ^2L}{\partial z^i_\mu\partial
z^j_\nu} = 0}
\end{array}
\right\}
\end{equation*}

From the expression above, we immediately deduce that

\begin{prop}
If the Lagrangian is regular, then if a section $\sigma:X
\longmapsto Z$ of $\pi_{XZ}$ is a solution of the De Donder
equations, then there is a section $\phi: X \longrightarrow Y$ of
$\pi_{XY}$ such that $\sigma = j^1\phi$. Furthermore, $\phi$ is a
solution of the Euler-Lagrange equations.
\end{prop}

Therefore, for regular Lagrangians, the solutions of the De Donder equations
provide the information about the dynamics of the system.

\subsection{The De Donder equations in terms of Ehresmann connections}

Suppose that we have a connection $\Gamma$ in $\pi: Z \longrightarrow X$,
with horizontal projector $\mathbf{h}$. Here, $\Gamma$ is a connection in
the sense of Ehresmann, that is, $\Gamma$ defines a horizontal complement of
the vertical bundle $\mathcal{V}{\pi_{XZ}}$. The horizontal projector has
the following local expression:
\begin{equation*}
\begin{cases}
\displaystyle{\mathbf{h}(\frac{\partial }{\partial x^{\mu}})} & =
\displaystyle{\frac{\partial }{\partial
x^{\mu}}+\Gamma^i_{\mu}\frac{\partial }{\partial
y^i}+\Gamma^i_{\mu\nu }\frac{\partial }{\partial
z^i_{\nu}}} \\
\displaystyle{\mathbf{h}(\frac{\partial }{\partial y^i})} & =
\displaystyle{0} \\
\displaystyle{\mathbf{h}(\frac{\partial }{\partial z^i_{\mu}})} &
= \displaystyle{0}
\end{cases}
\end{equation*}

A direct computation shows that

\begin{align*}
\iota_{\hbox{\textbf{h}}}\Omega_L &= n\Omega_L - \sum_i\left[
\frac{\partial L}{\partial y^i} - \sum_\nu\frac{\partial
^2L}{\partial x^\nu\partial z^i_\nu
}-\sum_{\nu,j}\Gamma^j_\nu\frac{\partial ^2L}{\partial y^j\partial
z^i_\nu}
\right. \\
&\left. - \sum_{\nu,\mu,j}\Gamma^j_{\mu\nu}\frac{\partial ^2L}{\partial
z^j_\mu\partial z^i_\nu} + \sum_{\nu,j}(\Gamma^j_\nu-z^j_\nu)\frac{\partial
^2L}{\partial y^i\partial z^j_\nu}\right] dy^i\wedge d^{n+1}x \\
&-\sum_{\mu,i}\left(\sum_{\nu,j}(\Gamma^j_\nu-z^j_\nu)\frac{\partial
^2L}{\partial z^i_\mu\partial z^j_\nu}\right) dz^i_\mu\wedge
d^{n+1}x
\end{align*}
from where we can state the following.

\begin{prop}
Let $\Gamma$ be a connection with horizontal projector \textrm{\textbf{h}}
verifying
\begin{equation}  \label{eq:hdedonder}
\iota_{\hbox{\rm \textbf{h}}}\Omega_L = n\Omega_L
\end{equation}
If $\sigma$ is a horizontal local integral section of $\Gamma$, then $\sigma$
is a solution of the De Donder equations.
\end{prop}

\textit{Proof}. \textbf{h} satisfies (\ref{eq:hdedonder}) if and only if
\begin{equation*}
\left.
\begin{array}{r}
\displaystyle{\ \frac{\partial L}{\partial y^i} -\frac{\partial
^2L}{\partial x^\nu\partial z^i_\nu}-\Gamma^j_\nu\frac{\partial
^2L}{\partial y^j\partial z^i_\nu} -
\Gamma^j_{\mu\nu}\frac{\partial ^2L}{\partial z^j_\mu\partial
z^i_\nu}+ (\Gamma^j_\nu-z^j_\nu)\frac{\partial ^2L}{\partial
y^i\partial z^j_\nu} = 0} \\
\displaystyle{(\Gamma^j_\nu-z^j_\nu)\frac{\partial ^2L}{\partial
z^i_\mu\partial z^j_\nu} = 0}
\end{array}%
\right\}
\end{equation*}

If $\sigma(x^\mu)=(x^\mu, \sigma^i(x^\mu),\sigma^i_\nu(x^\mu))$ is a
horizontal local integral section of $\Gamma$, then we have that
\begin{equation}  \label{eq:TsigmaEqH}
\hbox{\textbf{h}}(\frac{\partial }{\partial
x^\mu})=T\sigma(\frac{\partial }{\partial x^\mu})
\end{equation}
which means that $\displaystyle{\Gamma_\mu^i=\frac{\partial
\sigma^i}{\partial x^\mu}}$ and
$\displaystyle{\Gamma^i_{\mu\nu}=\frac{\partial
\sigma^i_\nu}{\partial x^\mu}}$, and therefore
(\ref{eq:hdedonder}) becomes the De Donder equations in
coordinates.

Local solutions can be glued together using partitions of
unity.\hfill$\ \ \ \vrule height 1.5ex width.8ex depth.3ex
\medskip$

If we consider boundary conditions, then the connection \textbf{h}
induces a connection $\partial$\textbf{h} in the fibration
$\pi_{\partial X B}: B \longrightarrow \partial X$, since we are
considering sections $\sigma\in\Gamma(\pi_{XZ})$ such that
$\sigma(\partial X)\subseteq B$.

In this way, the equation \eqref{eq:hdedonder} becomes
$\iota_{\hbox{\textbf{h}}}\Omega_L = n\Omega_L$ with the
additional condition that \textbf{h} induces $\partial$\textbf{h}
(or equivalently $\hbox{\bf h} _z(T_zB)\subseteq T_zB$ for all
$z\in B)$.

In the regular case (or for semiholonomic connections, that is $\Gamma^i_\mu
= z^i_\mu)$, two of these solutions differ by a $(1,1)-$tensor field $T$,
locally given by
\begin{equation*}
T=T^i_{\mu\nu}dx^\nu\otimes\frac{\partial }{\partial z^i_\mu}
\end{equation*}
and verifying
\begin{equation*}
T^i_{\mu\nu}\frac{\partial ^2L}{\partial z^i_\mu \partial z^j_\nu}=0
\end{equation*}

\begin{remark}
\textrm{An alternative approach may be considered if we express
(\ref{eq:hdedonder}) for horizontal integrable distributions in
terms of multivector fields generating those distributions. For
further details, see \cite{EMR1,EMR2,EMR4,EMR5,EMR6,EMR7} and
\cite{FPR,PR1,PR2}. }
\end{remark}

\subsection{The singular case}

\label{singularLagrSection}

For a singular Lagrangian $L$, one cannot expect to find globally defined
solutions; in general, if such connection $\mathbf{h}$ exists, it does so
only along a submanifold $Z_f$ of $Z$.

In \cite{LMMS,LMMS2} the authors have developed a constraint
algorithm which extends the Dirac-Bergmann-Gotay-Nester-Hinds
algorithm for Mechanics (see \cite{Got,GN1,GN2}, and also
\cite{Kova,barsa} for more recent developments).

Put $Z_1=Z$. We then consider the subset
\begin{eqnarray*}
Z_2 & = & \{z \in Z \; |\; \exists \hbox{\bf h}_{z} : T_z Z \longrightarrow
T_zZ \quad \hbox{linear such that }\ \hbox{\bf
h}^2_z = \hbox{\bf h}_z, \ker \hbox{\bf h}_z = (\mathcal{V}\pi_{X Z})_z, \\
&& i_{\hbox{\bf h}_z} \Omega_{L}(z) = n \Omega_{L}(z), \hbox{and
for}\,z\in B, \hbox{we also have} \,\hbox{\bf h}_z(T_zB)\subseteq T_zB \}.
\end{eqnarray*}
If $Z_2$ is a submanifold, then there are solutions but we have to include
the tangency condition, and consider a new step (denoting $B_2=B\cap Z_2$,
and in general, $B_r=B\cap Z_r)$:
\begin{eqnarray*}
Z_3 & = & \{z \in Z_2 \; |\; \exists \hbox{\bf h}_{z} : T_zZ \longrightarrow
T_z Z_2 \quad \hbox{linear such that }\ \hbox{\bf
h}^2_z = \hbox{\bf h}_z, \ker \hbox{\bf h}_z = (\mathcal{V}\pi_{X Z})_z, \,
\\
&& i_{\hbox{\bf h}_z} \Omega_{L}(z) = n \Omega_{L}(z) , \hbox{and
for}\,z\in B_2, \hbox{we also have} \, \hbox{\bf
h}_z(T_zB_2)\subseteq T_zB_2\}.
\end{eqnarray*}
If $Z_3$ is a submanifold of $Z_2$, but $\hbox{\bf h}_z(T_zZ)$ is
not contained in $T_zZ_3$ and $\mathbf{h}_z(T_zB)$ is not
contained in $T_zB$ for $z \in B$, we go to the third step, and so
on. In the favourable case, we would obtain a final constraint
submanifold $Z_f$ of non-zero dimension, and a connection for the
fibration $\pi_{XZ}: Z \longrightarrow X$ along the submanifold
$Z_f$ (in fact, a family of connections) with horizontal projector
$\hbox{\bf h}$ which is a solution of equation
(\ref{eq:hdedonder}), and, in addition, it satisfies the boundary
condition .

There is an additional problem, since our connection would be a solution of
the De Donder problem, but not a solution of the Euler-Lagrange equations.
This problem is solved constructing a submanifold of $Z_f$ where such a
solution exists (see \cite{LMMS,LMMS2} for more details).

\subsection{Multisymplectic forms. Brackets}

\begin{defn}
\cite{Gotay} A \textbf{multisymplectic form} $\Omega$ in a manifold $M$ is a
closed $k$-form $(k>1)$ having the following non-degeneracy property:
\begin{equation*}
\iota_v\Omega = 0 \hbox{ if and only if } v=0\qquad\forall v\in T_xM,
\forall x\in M
\end{equation*}
A \textbf{multisymplectic manifold} is a manifold endowed with a
multisymplectic form.
\end{defn}

The properties of multisymplectic manifolds have been widely
explored in \cite{CIL,gent2002,Mar1,Mar2}.

\begin{prop}
\label{prop:OLmultisymp} For $n>0$, the Lagrangian $L$ is regular if and
only if $\Omega_L$ is a multisymplectic form
\end{prop}

\textit{Proof}. As the Lagrangian is regular, we can use Darboux coordinates
$(x^\mu,y^i,\hat{p}_i^\mu)$ (see also Definition \ref{defpp}), and the
expression of $\Omega_L$ in these coordinates was stated shortly after its
definition. From the following computations:
\begin{align*}
\iota_{\partial / \partial x^\nu}\Omega_L &= -\frac{\partial
\hat{p}}{\partial x^\nu} d^{n+1}x + d\hat{p}\wedge d^{n}x_\nu +
d\hat{p}_i^\mu\wedge
dy^i \wedge d^{n-1}x_{\mu\nu} \\
&= \frac{\partial \hat{p}}{\partial y^i}dy^i\wedge d^{n}x_\nu +
\frac{\partial \hat{p}}{\partial
\hat{p}_i^\mu}d\hat{p}_i^\mu\wedge d^{n}x_\nu + d\hat{p}_i^\mu\wedge dy^i \wedge d^{n-1}x_{\mu\nu} \\
\iota_{\partial / \partial y^j}\Omega_L &= \frac{\partial \hat{p}}{\partial
y^j}d^{n+1}x - d\hat{p}_j^\mu\wedge d^{n}x_{\mu} \\
\iota_{\partial / \partial \hat{p}_j^\nu}\Omega_L &=
\frac{\partial \hat{p}}{\partial \hat{p}_j^\nu}d^{n+1}x +
dy^j\wedge d^{n}x_\nu
\end{align*}
if we have $\xi = A^\nu \frac{\partial }{\partial x^\nu} + B^j
\frac{\partial }{\partial y^j} +C_j^\nu \frac{\partial }{\partial
\hat{p}^\nu_j}$ then
\begin{align*}
\iota_\xi\Omega_L &= \left(B^j \frac{\partial \hat{p}}{\partial
y^j}-C_j^\nu \frac{\partial \hat{p}}{\partial
\hat{p}_j^\nu}\right)d^{n+1}x + \left( A^\nu \frac{\partial
\hat{p}}{\partial \hat{p}_j^\mu} - \delta^\nu_\mu
B^j\right)d\hat{p}_j^\mu\wedge d^{n}x_{\nu} \\
&+ \left( A^\nu \frac{\partial \hat{p}}{\partial y^j} -
C^\nu_j\right)dy^j\wedge d^{n}x_{\nu} + A^\nu d\hat{p}_i^\mu\wedge dy^i
\wedge d^{n-1}x_{\mu\nu}
\end{align*}

Therefore, if $\iota_\xi\Omega_L = 0$ and $n>0$, then from the
last term of the expression above, $A^\nu=0$, and we easily get
that the rest of terms $B^j$ and $C^\nu_j$ vanish as well. The
converse is proven in a similar manner. \hfill$\ \ \ \vrule height
1.5ex width.8ex depth.3ex \medskip$

\begin{remark}
\textrm{The case $n=0$ has many differences from the case $n>0$, and
corresponds to the case of the time-dependent Lagrangian mechanics (see \cite%
{LR89}). In this case, the regularity of $L$ implies that $(Z, \Omega_L, dt)$
(where $dt=\eta$ is the volume form) is a cosymplectic manifold. The
connection equation reduces to
\begin{equation*}
\iota_{\hbox{\textbf h}}\Omega_L = 0
\end{equation*}
where if we call $\tau=\frac{\partial }{\partial t}$ (so that
$\langle\eta|\tau\rangle=1)$, then the horizontal projector
\textbf{h} can be written in coordinates as follows
\begin{equation*}
\mathbf{h}(\tau) = \tau + h^i\frac{\partial }{\partial q^i} +
{h^{\prime}}^i \frac{\partial }{\partial v^i}
\end{equation*}
(for $q^i = y^i, v^i=z^i_0)$. Sections of $\pi_{XY}$ are curves on
$Y,$ and $Z$ can be embedded in $TY$. }

\textrm{One obtains from De Donder equations that ${h^{\prime}}^i
= \frac{\partial h^i}{\partial t}$, and that $h(\tau)$ verifies
the time dependent Euler-Lagrange equations on $J^1\pi$.
Furthermore, for a $(1,1)$-tensor field $h$ on $J^1\pi$, being the
horizontal projector of a distribution solution of
\begin{equation*}
\iota_{\hbox{\textbf {h}}}\Omega_L = 0
\end{equation*}
is equivalent to having $\xi=\mathbf{h}(\tau)$ which verifies
\begin{align*}
\iota_\xi\Omega_L&=0 \\
\iota_\xi\eta&=1
\end{align*}
}

\textrm{From now on within this section, we shall suppose that $n>0$. }
\end{remark}

With multisymplectic structures we can define Hamiltonian vector fields and
forms as we did for symplectic structures. However, existence is no longer
guaranteed.

\begin{defn}
Let $\alpha$ be a $n$-form in $Z$. A vector field $X_\alpha$ is called a
\textbf{Hamiltonian vector field} for $\alpha$, and we say that $\alpha$ is
\textbf{Hamiltonian} whenever
\begin{equation*}
d\alpha = \iota_{X_\alpha}\Omega_L
\end{equation*}
\end{defn}

If $L$ is regular, then the non-degeneracy of $\Omega_L$ guarantees that a
Hamiltonian vector field, if it exists, is unique. Otherwise, we cannot
guarantee its existence, and the Hamiltonian vector field is defined up to
an element in the kernel of $\Omega_L$.\newline
Also note that two forms that differ by a closed form have the same
Hamiltonian vector fields.

\begin{defn}
If $\alpha$ and $\beta$ are two Hamiltonian $n$-forms for which there exist
the corresponding Hamiltonian vector fields $X_\alpha$, $X_\beta$, then we
can define the \textbf{bracket operation} as follows:
\begin{equation*}
\{\alpha,\beta\} =\iota_{X_\beta}\iota_{X_\alpha}\Omega_L
\end{equation*}
\end{defn}

We also have the following result:

\begin{prop}
If and $\alpha$, $\beta$ are Hamiltonian $n$-forms which have a
Hamiltonian vector fields $X_\alpha$ and $X_\beta$ respectively,
then $\{\alpha,\beta\}$ is a Hamiltonian $n$-form which has
associated Hamiltonian vector field $[X_\alpha,X_\beta]$. In other
words,
\begin{equation*}
X_{\{\alpha,\beta\}}=[X_\alpha,X_\beta]
\end{equation*}
\end{prop}

\textit{Proof}.
\begin{align*}
\iota_{[X_\alpha, X_\beta]}\Omega_L &=
\mathcal{L}_{X_\alpha}\iota_{X_\beta}\Omega_L-\iota_{X_\beta}\mathcal{L}_{X_\alpha}\Omega_L \\
& =
\mathcal{L}_{X_\alpha}d\beta-\iota_{X_\beta}d\iota_{X_\alpha}\Omega_L-
\iota_{X_\beta}\iota_{X_\alpha}d\Omega_L \\
& = d\iota_{X_\alpha}d\beta-\iota_{X_\beta}dd\alpha \\
& = - d\iota_{X_\alpha}\iota_{X_\beta}\Omega_L \\
& = d\{\alpha,\beta\},
\end{align*}
and, by uniqueness, we obtain the desired result. \hfill$\ \ \ \vrule height
1.5ex width.8ex depth.3ex \medskip$

The properties of this brackets have been widely studied in
\cite{CIL2,FPR,GS}.

\section{Hamiltonian formalism}

\subsection{Dual jet bundle}

At the beginning of our discussion, we briefly listed the different
approaches to the notion of jet bundle, where one of these is to consider it
certain structure of affine bundle over $Y$.

The dual affine bundle of the jet bundle is called \textbf{dual jet bundle},
and it is usually denoted by $(J^1\pi)^*$, that we shall denote by $Z^*$. An
alternative construction of such bundle is given here.

\begin{defn}
Consider the family of spaces of forms
\begin{equation*}
\Lambda^{n+1}_r Y := \{\sigma\in\Lambda^{n+1}Y \;|\;
\iota_{V_1}\ldots\iota_{V_r}\sigma=0, \forall V_i\; \pi-vertical\; 1\leq i
\leq r \}
\end{equation*}

In particular, the elements of $\Lambda^{n+1}_1 Y$ are called \textbf{%
semibasic $(n+1)$-forms}. It is a fiber bundle over $Y$ of rank $(n+1+m+1)$,
and which elements can be locally expressed as $p(x,y)d^{n+1}x$.

Similarly, $\Lambda^{n+1}_2 Y$ is a vector bundle over $Y$ of rank
$(n+1+m+(n+1)m+1)$, having $\Lambda^{n+1}_1 Y$ as subbundle, and
which elements can be locally expressed as
$p(x,y)d^{n+1}x+p^\mu_i(x,y)dy^i\wedge d^nx_\mu$. The natural
projection will be called:
\begin{equation*}
\nu_r : \Lambda^{n+1}_r Y \longrightarrow Y
\end{equation*}
The quotient bundle
\begin{equation*}
Z^* = (J^1\pi)^* := \Lambda^{n+1}_2 Y / \Lambda^{n+1}_1 Y
\end{equation*}
is a vector bundle over $Y$ of rank $n+1+m+(n+1)m$ which elements can be
locally expressed as $p^\mu_i(x,y)dy^i\wedge d^nx_\mu$, and that is called
the \textbf{dual first order jet bundle}. The canonical projection will be
denoted by $\mu: \Lambda^{n+1}_2 Y \longrightarrow Z^*$.

We can define a projection $\pi_{XZ^*}: Z^* \longrightarrow X$, which is
induced by $\nu_2$ into the quotient space $Z^*$, composed with $\pi_{XY}$.
\end{defn}

\begin{defn}
The manifold $\Lambda^{n+1}_2 Y$ is equipped with the following $(n+1)$-form
\begin{equation*}
\Theta_\omega(X_0,\ldots,X_n) := \omega (T\nu_2(X_0),\ldots,T\nu_2(X_n))
\end{equation*}
which is called the \textbf{multimomentum Liouville form}, and has local
expression
\begin{equation*}
\Theta = pd^{n+1}x +p^\mu_idy^i\wedge d^nx_\mu
\end{equation*}
We also define the \textbf{canonical multisymplectic $(n+2)$-form}
on $\Lambda^{n+1}_2 Y$ by
\begin{equation*}
\Omega := -d\Theta
\end{equation*}
\end{defn}

Notice that $\Omega$ is in fact multisymplectic, by a similar argument to
that given in Proposition \ref{prop:OLmultisymp}.

\subsection{Lift of vector fields to the dual jet bundle}

A vector field $\xi_Y$ on $Y$, having flow $\phi_t$, admits a natural lift
to $\Lambda^k Y$ for any $k$, having flow $(\phi_t^{-1})^*$.

If the vector field $\xi_Y$ is projectable, then the flow
preserves $\Lambda^{n+1}_2Y$ and $\Lambda^{n+1}_1Y$, and therefore
we can define on $\Lambda^{n+1}_2Y$ a vector field which projects
onto a vector field on $Z^*$, which we shall denote by
$\xi_Y^{(1*)}$.

In general, if $\alpha$ is the pull-back to $\Lambda^{n+1}_2 Y$ of certain
semibasic $n$-form on $Y$, locally expressed by
\begin{equation*}
\alpha = \alpha^\nu(x^\mu,y^i) d^nx_\nu ,
\end{equation*}
the additional condition $\pounds_{\xi_Y^{\alpha}} \Theta = d\alpha$ imposed
to vector fields on $\Lambda^{n+1} Y$ which project to $\xi_Y$, determines a
vector field on $\Lambda^{n+1} Y$ that can be defined on $\Lambda^{n+1}_2 Y$.

In other words, we have the following definition.

\begin{defn}
If $\alpha$ is the pull-back to $\Lambda^{n+1}_2 Y$ of a
$\pi_{XY}$-semibasic form, then the $\alpha$-lift of a vector
field $\xi_Y$ on $Y$ to $\Lambda^{n+1}_2 Y$ is defined as the
unique vector field $\xi_Y^\alpha$ satisfying:

(1) $\xi_Y^\alpha$ projects onto $\xi_Y$

(2) $\pounds_{\xi_Y^\alpha}\Theta = d\alpha$
\end{defn}

An easy computation shows that the components $dp(\xi_Y^\alpha)=\xi_Y^p$ and
$dp^\mu_i(\xi_Y^\alpha)=\xi_Y^{p^\mu_i}$ are determined by the equations
(see also \cite{gimmsy1,PR1}):
\begin{align*}
\xi_Y^p &= -p \frac{\partial \xi_Y^\mu}{\partial
x^\mu}-p^\mu_i\frac{\partial \xi_Y^i}{\partial
x^\mu}-\frac{\partial \alpha^\mu}{\partial x^\mu}
\\
\xi_Y^{p^\mu_i} &= p^\nu_i\frac{\partial \xi_Y^\mu}{\partial
x^\nu}-p^\mu_j\frac{\partial \xi_Y^j}{\partial
y^i}-p^\mu_i\frac{\partial \xi_Y^\nu}{\partial
x^\nu}-\frac{\partial \alpha^\mu}{\partial y^i}
\end{align*}

When $\xi_Y$ is $\pi_{XY}$-projectable, with flow $\phi_t$, then the flow of
the 0-lift is precisely $(\phi_t^{-1})^*$.

\subsection{Hamilton equations}

\begin{defn}
A \textbf{Hamiltonian} form is a section $h: Z^* \longrightarrow
\Lambda^{n+1}_2 Y$ of the natural projection $\mu: \Lambda^{n+1}_2
Y\longrightarrow Z^*$.

In local coordinates, $h$ is given by
\begin{equation*}
h(x^\mu, y^i, p^\mu_i) = (x^\mu,y^i, p=-H(x^\mu, y^i, p^\mu_i), p^\mu_i)
\end{equation*}
where $H$ is called a \textbf{Hamiltonian function}.
\end{defn}

\begin{defn}
Given a Hamiltonian, we define the following forms in $Z^*$
\begin{equation*}
\Theta_h := h^*\Theta
\end{equation*}
having local expression
\begin{align*}
\Theta_h &= -Hd^{n+1}x +p^\mu_idy^i\wedge d^nx_\mu \\
&= (-Hdx^\mu +p^\mu_idy^i)\wedge d^nx_\mu
\end{align*}
and
\begin{align*}
\Omega_h :&= h^*\Omega = -d\Theta_h \\
&= (-dH\wedge+dx^\mu +dp^\mu_i\wedge dy^i )\wedge d^nx_\mu
\end{align*}
\end{defn}

\begin{defn}
For a given Hamiltonian $h$, a section $\sigma : X \longrightarrow
Z^*$ of $\pi_{XZ^*}$ is said to satisfy the \textbf{Hamilton
equations} if
\begin{equation*}
\sigma^*(\iota_\xi\Omega_h)=0
\end{equation*}
for all vector field $\xi$ on $Z^*$.

If $\sigma$ has local expression
$\sigma(x^\mu)=(x^\mu,\sigma^i(x^\mu), \sigma^\nu_i(x^\mu))$, then
the Hamilton equations are written in coordinates as follows

\begin{equation*}
\frac{\partial \sigma^i}{\partial x^\mu}=\frac{\partial H}{\partial p^\mu_i}
\end{equation*}
\begin{equation*}
\sum_{\mu=1}^m \frac{\partial \sigma^\mu_i}{\partial x^\mu}=-\frac{\partial H%
}{\partial y^i}
\end{equation*}
\end{defn}

As for the Lagrangian case, we can also consider the case of
having a boundary condition given by a subbundle $B^*\subseteq
\partial Z^*$ of $\tilde{\pi}_{\partial X \partial Z}$, which
imposes a restriction on the possible solutions for the Hamilton
equations. The additional requirement for the solutions is
naturally that they must satisfy $\sigma(\partial X)\subseteq
B^*$, and we also need to assume that
\begin{equation*}
i_{B^*}^*\Theta_h = d\Pi^*
\end{equation*}
for certain $n$-form $\Pi^*$ on $B^*$, where $i_{B^*}: B^* \longrightarrow
\partial Z^*$ denotes the canonical inclusion.

There is also another formulation of the Hamilton equations in terms of
connections.

Suppose that we have a connection $\Gamma$ (in the sense of
Ehresmann) in $\pi_{XZ^*}: Z^* \longrightarrow X$, with horizontal
projector $\mathbf{h}$, and having a local expression as follows

\begin{equation*}
\begin{cases}
\displaystyle{\mathbf{h}(\frac{\partial }{\partial x^{\mu}})} & =
\displaystyle{\frac{\partial }{\partial
x^{\mu}}+\Gamma^i_{\mu}\frac{\partial }{\partial
y^i}+\Gamma^\nu_{i\mu} \frac{\partial }{\partial p^\nu_i}} \\
\displaystyle{\mathbf{h}(\frac{\partial }{\partial y^i})} & =
\displaystyle{0} \\
\displaystyle{\mathbf{h}(\frac{\partial }{\partial p^\mu_i})} & =
\displaystyle{0}
\end{cases}
\end{equation*}

A direct computation shows that

\begin{align*}
\iota_{\mathbf{h}}\Omega_h = n\Omega_h &- \left( \frac{\partial H}{\partial
y^i} + \sum_{\mu=1}^m \Gamma^\mu_{i\mu}\right) dy^i\wedge d^{n+1}x \\
&+ \left(\frac{\partial H}{\partial p^\mu_i}-\Gamma^i_{\mu}\right)
dp^\mu_i\wedge d^{n+1}x
\end{align*}

From where we can state the following.

\begin{prop}
Let $\Gamma$ be a connection with horizontal projector \textrm{\textbf{h}}
verifying
\begin{equation}  \label{eq:hHamilton}
\iota_{\hbox{\rm \textbf{h}}}\Omega_h = n\Omega_h
\end{equation}
and also the boundary compatibility condition $\hbox{\bf
h}_\alpha(T_\alpha B^*)\subseteq T_\alpha B^*$ for $\alpha\in Z^*$ (i.e.,
\textrm{\textbf{h}} induces a connection $\partial$\textrm{\textbf{h}} in
the fibration $\pi_{\partial X B^*}: B^* \longrightarrow \partial X)$.

If $\sigma$ is a horizontal integral local section of $\Gamma$, then $\sigma$
is a solution of the Hamilton equations.

Therefore, one can think of the preceding equation as an alternative
approach to the Hamilton equations.
\end{prop}

\subsection{The Legendre transformation}

We shall generalize to field theories the notion of Legendre transformation
in Classical Mechanics.

\begin{defn}
Associated to the Lagrangian function we can define the \textbf{Legendre
transformation} $Leg_L : Z \longrightarrow \Lambda^{n+1}_2 Y$ as follows,
given $\xi_1,\ldots,\xi_n\in (T_{\pi_{YZ}z}) Y$,
\begin{equation*}
(Leg_L(z))(\xi_1,\ldots,\xi_n) =
(\Theta_L)_z(\tilde{\xi}_1,\ldots,\tilde{\xi}_n)
\end{equation*}
where $\tilde{\xi}_i$ is a tangent vector at $z\in Z$ which
projects onto $\xi_i$.

It is well defined, as $\iota_\xi\Theta_L=0$ for $\pi_{YZ}$-vertical vector
fields (see lemma \ref{lemma:verticalOnOmega}), and $\iota_\xi\iota_\zeta
Leg_L(z)=0$ for $\xi,\zeta \in \mathcal{V}\pi$, therefore, $Leg_L(z)\in
\Lambda^{n+1}_2 Y$.

In local coordinates,
\begin{equation*}
Leg_L(x^\mu,y^i,z^i_\mu) = \left( x^\mu, y^i,
p=L-z^i_\mu\frac{\partial L}{\partial
z^i_\mu},p^\mu_i=\frac{\partial L}{\partial z^i_\mu}\right)
\end{equation*}
which shows that $Leg_L$ is a fibered map over $Y$.
\end{defn}

For an expression of the Legendre transformation in terms of affine duals,
see \cite{gimmsy1}.

\begin{defn}
We also define the \textbf{Legendre map} $leg_L := \mu\circ Leg_L: Z
\longrightarrow Z^*$, which in coordinates has the form:
\begin{equation*}
leg_L(x^\mu,y^i,z^i_\mu) = \left( x^\mu, y^i,
p^\mu_i=\frac{\partial L}{\partial z^i_\mu}=\hat{p}^\mu_i\right)
\end{equation*}
\end{defn}

From the local expressions of $\Theta_L$, the following proposition is
obvious.

\begin{prop}
All these facts hold:

(i) The Lagrangian is regular if and only if then the Legendre map $leg_L$
is a local diffeomorphism.

(ii) If we choose a Hamiltonian $h$, then we have the following relations:
\begin{equation*}
(Leg_L)^*\Theta = \Theta_L,\qquad (Leg_L)^*\Omega = \Omega_L\newline
\end{equation*}
\begin{equation*}
(leg_L)^*\Theta_h = \Theta_L,\qquad (leg_L)^*\Omega_h = \Omega_L
\end{equation*}
\end{prop}

\begin{defn}
A Lagrangian $L$ is called \textbf{hyperregular} whenever $leg_L$
is a diffeomorphism (and therefore, it is regular). Also assume
that $leg_L^*(\Pi^*)=\Pi$.
\end{defn}

We also have the following equivalence theorem, which is a straightforward
computation.

\begin{thm}
(\textbf{equivalence theorem}). Suppose that the Lagrangian is regular. Then
if a section $\sigma_1$ of $\pi_{XZ}$ satisfies the De Donder equations
\begin{equation*}
\sigma_1^*(\iota_\xi\Omega_L)=0 \qquad\qquad \forall\xi\in\mathfrak{X}(Z)
\end{equation*}
then $\sigma_2^* := leg\circ\sigma_1$ verifies the Hamilton equations
\begin{equation*}
\sigma_2^*(\iota_\xi\Omega_h)=0 \qquad\qquad \forall\xi\in\mathfrak{X}(Z^*)
\end{equation*}
Reciprocally, if $\sigma_2$ verifies Hamilton equations, then (the locally
defined) $\sigma_1 := leg^{-1}_L\circ\sigma_2$ verifies the De Donder
equations. Therefore, De Donder equations are equivalent to Hamilton
equations.
\end{thm}

\begin{remark}
\textrm{A rutinary computation also shows that, for a regular Lagrangian, if
$\Gamma$ is a connection solution of \eqref{eq:hdedonder} then $%
Tleg_L(\Gamma)$ is a solution for the equation in terms of connections on
the Hamiltonian side. }

\textrm{Furthermore, a boundary condition $B$ on $Z$ automatically
induces a boundary condition $B^*$ in $Z^*$, by $leg_L(B)=B^*$,
which implies that $Tleg_L(T_zB)\subseteq T_{leg_L(z)}B^*$, and in
turn proves that compatible connection projectors relate to each
other via the Legendre map. }
\end{remark}

\subsection{Almost regular Lagrangians}

When the Lagrangian is not regular then to develop a Hamiltonian
counterpart, we need some weak regularity condition on the Lagrangian $L$,
the almost-regularity assumption.

\begin{defn}
A Lagrangian $L : Z\longrightarrow \mathbb{R}$ is said to be \textbf{almost
regular} if $Leg_L(Z)=\tilde{M}_1$ is a submanifold of $\Lambda^{n+1}_2Y$,
and $Leg_L : Z \longrightarrow \tilde{M}_1$ is a submersion with connected
fibers.
\end{defn}

If $L$ is almost regular, we deduce that:

\begin{itemize}
\item $M_1 = leg_L(Z)$ is a submanifold of $Z^*$, and in addition, a
fibration over $X$ and $Y$.

\item The restriction $\mu_1 : \tilde{M}_1 \longrightarrow M_1$ of $\mu$ is
a diffeomorphism.

\item The mapping $leg_L : Z \longrightarrow M_1$ is a submersion with
connected fibers.
\end{itemize}

On the hypothesis of almost regularity, we can define a mapping
$h_1=(\mu_1)^{-1} : {M}_1 \longrightarrow \tilde{M}_1$, and a
$(n+2)$-form ${\Omega}_{M_1}$ on ${M}_1$ by ${\Omega}_{M_1} =
h_1^* (j^*\Omega)$ considering the inclusion map $j:\tilde{M}_1
\hookrightarrow \Lambda^{n+1}_2 Y$. Obviously, we have $leg_1^*
{\Omega}_{M_1} = \Omega_L$, where $j\circ leg_1 = leg_L$ (see
Figure 2).

\unitlength=0.4mm \special{em:linewidth 0.4pt} \linethickness{1pt}
\begin{picture}(400,200)(50,0)
\put(90,115){\makebox(0,0){$Z$}} \bezier{292}(90,119)(200,
215)(330.00,159) \put(330,159){\vector(2,-1){0}}
\bezier{292}(90,105)(200, 30)(325.00,71)
\put(325,71){\vector(2,1){0}}
\put(210,150){\makebox(0,0){$\tilde{M}_1 = Leg_L(Z)$}}
\put(290,160){\makebox(0,0){$\tilde{j}$}}
\put(140,141){\makebox(0,0){$Leg_1$}}
\put(140,90){\makebox(0,0){$leg_1$}}
\put(210,80){\makebox(0,0){$M_1=leg_L(Z)$}}
\put(290,90){\makebox(0,0){$j$}}
\put(340,124){\makebox(0,0){$\mu$}}
\put(220,124){\makebox(0,0){$\mu_1$}}
\put(210,185){\makebox(0,0){$Leg_L$}}
\put(210,50){\makebox(0,0){$leg_L$}}
\put(330,150){\makebox(0,0){$\Lambda^{n+1}_2 Y$}}
\put(330,80){\makebox(0,0){$Z^*$}} \put(105,120){\vector(3,1){60}}
\put(262,150){\vector(1,0){50}} \put(105,110){\vector(3,-1){60}}
\put(262,80){\vector(1,0){50}} \put(330,140){\vector(0,-1){45}}
\put(210,140){\vector(0,-1){45}}
\put(230.00,25.00){\makebox(0,0)[ct]{Figure 2}}
\end{picture}

The Hamiltonian description is now based in the equation
\begin{equation}  \label{connection3}
i_{\tilde{\hbox{\bf h}}} {\Omega}_{M_1} = n {\Omega}_{M_1}
\end{equation}
where $\tilde{\hbox{\bf h}}$ is a connection in the fibration
$\pi_{X M_1} : M_1 \longrightarrow X$, and the additional boundary
condition for $\tilde{\hbox{\bf h}}$.

Proceeding as before, we construct a constraint algorithm as follows. First,
we denote by $B_1^* = B^*\cap M_1$, and will assume it to be a submanifold
of $B^*$ (and in general we shall denote $B_r^* = B^*\cap M_r$, which will
also be assumed to be a submanifold of $B_{r-1}^*)$, and we define
\begin{eqnarray*}
M_2 & = & \{\tilde{z} \in M_1 \; |\; \exists \tilde{\hbox{\bf
h}}_{\tilde{z}} : T_{\tilde{z}} M_1 \longrightarrow
T_{\tilde{z}}M_1 \quad \hbox{linear such that }\ \tilde{\hbox{\bf
h}}^2_{\tilde{z}} = \tilde{\hbox{\bf h}}_{\tilde{z}}, \ker
\tilde{\hbox{\bf
h}}_{\tilde{z}} = (\mathcal{V}\pi_{X M_1})_{\tilde{z}}, \\
&& i_{\tilde{\hbox{\bf h}}_{\tilde{z}}} {\Omega}_{M_1}(\tilde{z})
= n {\Omega}_{M_1}(\tilde{z}), \hbox{and for}\,\tilde{z}\in B^*_1
\hbox{we also have} \, \tilde{\hbox{\bf
h}}_{\tilde{z}}(T_{\tilde{z}}B^*_1)\subseteq T_{\tilde{z}}B^*_1\}.
\end{eqnarray*}
If $M_2$ is a submanifold (possibly with boundary) then there are solutions
but we have to include the tangency conditions, and consider a new step:
\begin{eqnarray*}
M_3 & = & \{\tilde{z} \in M_2 \; |\; \exists \tilde{\hbox{\bf
h}}_{\tilde{z}} : T_{\tilde{z}}M_1 \longrightarrow T_{\tilde{z}}
M_2 \quad \hbox{linear such that }\ \tilde{\hbox{\bf
h}}^2_{\tilde{z}} = \tilde{\hbox{\bf h}}_{\tilde{z}}, \ker
\tilde{\hbox{\bf
h}}_{\tilde{z}} = (\mathcal{V}\pi_{X M_{1}})_{\tilde{z}},\, \\
&& i_{\tilde{\hbox{\bf h}}_{\tilde{z}}} {\Omega}_{M_1}(\tilde{z})
= n {\Omega}_{M_1}(\tilde{z}), \hbox{and for}\,\tilde{z}\in
B^*\cap M_2 \hbox{we also have}
\,\tilde{\hbox{\bf h}}_{%
\tilde{z}}(T_{\tilde{z}}B^*)\subseteq T_{\tilde{z}}B^*\}.
\end{eqnarray*}
If $M_3$ is a submanifold of $M_2$, but $\tilde{\hbox{\bf
h}}_{\tilde{z}}(T_{\tilde{z}}M_1)$ is not contained in $T_{\tilde{z}}M_3$,
and $\tilde{\hbox{\bf
h}}_{\tilde{z}}(T_{\tilde{z}}B^*)$ is not contained in $T_{\tilde{z}}B^*$
for $z \in B^*$, we go to the third step, and so on. Thus, we proceed
further to obtain a sequence of embedded submanifolds
\begin{equation*}
...\hookrightarrow M_3\hookrightarrow M_2\hookrightarrow M_1\hookrightarrow
Z^*
\end{equation*}
with boundaries
\begin{equation*}
...\hookrightarrow B^*_3\hookrightarrow B^*_2\hookrightarrow
B^*_1\hookrightarrow B^*
\end{equation*}
If this constraint algorithm stabilizes, we shall obtain a final
constraint submanifold $M_f$ of non-zero dimension and a
connection in the fibration $\pi_{X M_1} : M_{1} \longrightarrow
X$ along the submanifold $M_f$ (in fact, a family of connections)
with horizontal projector $\tilde{\hbox{\bf h}}$ verifying the
boundary compatibility condition, and which is a solution of
equation (\ref{connection3}) and satisfies the boundary condition.
$M_f$ projects onto an open submanifold of $X$ (and $B^*_f$
projects also onto an open submanifold of $\partial X)$.

If $M_f$ is the final constraint submanifold and $j_{f1}:
M_f\longrightarrow M_1$ is the canonical immersion then we may
consider the $(n+2)$-form $\Omega_{M_f}=j_{f1}^* \Omega_{M_1}$,
and the $(n+1)$-form $\Theta_{M_f}=i_{f1}^*\Theta_{M_1}$, where
$\Omega_{M_f} = -d\Theta_{M_f}$.

Denoting by $leg_i := leg_L|_{Z_i}$, a direct computation shows
that $leg_1(Z_{a})=M_{a}$ for each integer.
\begin{equation*}
\begin{array}{lclcc}
Z_1=Z & \overset{leg_1}{\begin{picture}(70,0) \put(0,0){\vector(1,0){70}}
\end{picture} } & leg_L(Z)=M_1 & \overset{j}{\begin{picture}(45,0)
\bezier{300}(0,5)(2.5,0)(5,0) \put(5,0){\vector(1,0){40}} \end{picture} } &
Z^* \\
\uparrow i_1 &  & \uparrow j_1 &  &  \\
Z_2 & \overset{leg_2}{\begin{picture}(70,0) \put(0,0){\vector(1,0){70}}
\end{picture} } & M_2 &  &  \\
\uparrow i_2 &  & \uparrow j_2 &  &  \\
Z_3 & \overset{leg_3}{\begin{picture}(70,0) \put(0,0){\vector(1,0){70}}
\end{picture} } & M_3 &  &  \\
\uparrow i_3 &  & \uparrow j_3 &  &  \\
\vdots &  & \vdots &  &  \\
\uparrow i_{k-2} &  & \uparrow j_{k-2} &  &  \\
Z_{k-1} & \overset{leg_{k-1}}{\begin{picture}(70,0)
\put(0,0){\vector(1,0){70}} \end{picture} } & M_{k-1} &  &  \\
\uparrow i_{k-1} &  & \uparrow j_{k-1} &  &  \\
Z_k & \overset{leg_k}{\begin{picture}(70,0) \put(0,0){\vector(1,0){70}}
\end{picture} } & Z_k &  &
\end{array}
\end{equation*}

In consequence, both algorithms have the same behaviour; in
particular, if one of them stabilizes, so does the other, and at
the same step. In particular, we have $leg_1(Z_f)=M_f$.  In such a
case, the restriction $leg_f : Z_f \longrightarrow M_f$ is a
surjective submersion (that is, a fibration) and
$leg^{-1}_{f}(leg_{f}(z)) = leg_{1}^{-1}(leg_{1}(z))$, for all $z
\in Z_{f}$ (that is, its fibres are the ones of $leg_1)$.

Therefore, the Lagrangian and Hamiltonian sides can be compared
through the fibration $leg_f :Z_f \longrightarrow {M}_f$. Indeed,
if we have a connection in the fibration $\pi_{XZ} : Z
\longrightarrow X$ along the submanifold $Z_f$ with horizontal
projector $\hbox{\bf h}$ which is a solution of equation
(\ref{eq:hdedonder}) (the De Donder equations) and satisfies the
boundary condition and, in addition, the connection is projectable
via $Leg_f$ to a connection in the fibration $\pi_{X \tilde{Z}} :
\tilde{Z} \longrightarrow X$ along the submanifold $M_f$, then the
horizontal projector of the projected connection is a solution of
equation (\ref{eq:hHamilton}) (the Hamilton equations) and
satisfies the boundary contion, too. Conversely, given a
connection in the fibration $\pi_{X \tilde{Z}} : \tilde{Z}
\longrightarrow X$ along the submanifold $M_f$, with horizontal
projector $\tilde{\hbox{\bf h}}$ which is a solution of equation
(\ref{eq:hHamilton}) satisfying the boundary condition, then every
connection in the fibration $\pi_{XZ} : Z \longrightarrow X$ along
the submanifold $Z_f$ that projects onto $\tilde{\hbox{\bf h}}$ is
a solution of the De Donder equations (\ref{eq:hdedonder}) and
satisfies the boundary condition.

\section{Cartan formalism in the space of Cauchy data}

\subsection{Cauchy surfaces. Initial value problem}

\begin{defn}
A \textbf{Cauchy surface} is a pair $(M,\tau)$ formed by a compact
oriented $n$-manifold $M$ embedded in the base space $X$ by $\tau:
M \longrightarrow X$, such that $\tau(\partial M) \subseteq
\partial X$, and the interior of $M$ is included in the interior
of $X$. Two of such Cauchy surfaces are considered the same up to
an orientation and volume preserving diffeomorphism of
$M$.\newline In what follows, we shall fix $M$, and consider
certain space $\tilde{X}$ of such embeddings. We shall rather call
\textbf{Cauchy surfaces} to such embeddings.
\end{defn}

The choice of $M$ and $\tilde{X}$ depends on the physical theory which we
aim to describe with this model.

\begin{defn}
A \textbf{space of Cauchy data} is the manifold of embeddings $\gamma: M \to
Z$ such that there exists a section $\phi$ of $\pi_{XY}$ satisfying
\begin{equation*}
\gamma = (j^1\phi)\circ\tau
\end{equation*}
where $\tau := \pi_{XZ}\circ\gamma \in \tilde{X}$, and
$\gamma(\partial M)\subseteq B$.\newline The space of such
embeddings shall be denoted by $\tilde{Z}$, and we shall denote by
$\pi_{\tilde{X}\tilde{Z}}$ the projection
$\pi_{\tilde{X}\tilde{Z}}(\gamma)=\pi_{XZ}\circ\gamma$. We shall
also require this projection to be a locally trivial fibration.
\end{defn}

\begin{defn}
The \textbf{space of Dirichlet data} is the manifold $\tilde{Y}$
of all the embeddings $\delta: M \longrightarrow Y$ of the form
$\delta = \pi_{YZ} \circ \gamma$ for $\gamma \in \tilde{Z}$. We
also define $\pi_{\tilde{Y} \tilde{Z}} : \tilde{Z} \longrightarrow
\tilde{Y}$ as $\pi_{\tilde{Y}\tilde{Z}} (\gamma) = \pi_{YZ} \circ
\gamma$.\newline We denote by $\pi_{\tilde{X}\tilde{Y}}$ the
unique mapping from $\tilde{Y}$ to $\tilde{X}$ such that
$\pi_{\tilde{X}\tilde{Z}}=\pi_{\tilde{X}\tilde{Y}}\circ\pi_{\tilde{Y}\tilde{Z}}$
(see Figure 3)
\end{defn}

A tangent vector $v$ at $\gamma\in\tilde{Z}$ can be seen as a
vector field along $\gamma$, that is, $v: M\longrightarrow TZ$
such that $\tau_Z\circ v=\gamma$, where $\tau_Z: TZ\longrightarrow
Z$ is the canonical projection. Therefore, we identify vectors in
$T_{\gamma}\tilde{Z}$ with vector fields on $\gamma (M)$. Thus, a
vector field $\xi_Z$ on $Z$ induces a vector field
$\xi_{\tilde{Z}}$ on $\tilde{Z}$, where for every
$\gamma\in\tilde{Z}$, its representative tangent vector at
$\gamma\in\tilde{Z}$ is given by
\begin{equation*}
\xi_{\tilde{Z}}(\gamma)(u) = \xi_Z(\gamma(u))
\end{equation*}
for $u\in M$. And conversely, forms on $Z$ can be considered to act upon
tangent vectors of $\tilde{Z}$, for if $z=\gamma(u)$, $\alpha$ is a $r$-form
on $Z$ and $v\in T_\gamma\tilde{Z}$, then $\iota_v\alpha$ is a $(r-1)$-form
on $Z$ defined by
\begin{equation*}
(\iota_v\alpha)_z := \iota_{v(u)}\alpha_z
\end{equation*}
In practice, no distinction between them will be made.

\unitlength=1mm\special{em:linewidth 0.4pt} \linethickness{1pt}
\begin{picture}(70.00,90.00)
\put(20.00,85.00){\makebox(0,0)[cb]{$Z$}}
\put(40.00,50.00){\makebox(0,0)[cc]{$Y$}}
\put(20.00,10.00){\makebox(0,0)[ct]{$X$}}
\put(20.00,82.00){\vector(0,-1){68.33}}
\put(22.00,83.33){\vector(1,-2){15.33}}
\put(38.67,48.00){\vector(-1,-2){16.00}}
\put(17.67,50.67){\makebox(0,0)[rc]{$\pi_{XZ}$}}
\put(29.67,66.00){\makebox(0,0)[rb]{$\pi_{YZ}$}}
\put(31.33,34.00){\makebox(0,0)[rb]{$\pi_{XY}$}}
\put(70.00,50.00){\makebox(0,0)[cc]{$M$}} \linethickness{0.4pt}
\put(67.33,54){\vector(-4,3){41}}
\put(66.67,53){\vector(-4,3){41}}
\put(65.67,52){\vector(-4,3){41}}
\put(64.33,51.00){\vector(-1,0){20.67}}
\put(64.33,49.67){\vector(-1,0){20.67}}
\put(64.33,48.33){\vector(-1,0){20.67}}
\put(65.67,46.67){\vector(-4,-3){40.67}}
\put(67.00,45.67){\vector(-4,-3){40.67}}
\put(68.00,44.67){\vector(-4,-3){40.67}}
\put(52.33,70.67){\makebox(0,0)[lb]{$\tilde{Z}$}}
\put(51.67,53.33){\makebox(0,0)[cb]{$\tilde{Y}$}}
\put(50.33,26.67){\makebox(0,0)[lb]{$\tilde{X}$}}
\linethickness{1pt}
\put(110.00,75.00){\makebox(0,0)[cb]{$\tilde{Z}$}}
\put(140.00,50.00){\makebox(0,0)[cc]{$\tilde{Y}$}}
\put(110.00,20.00){\makebox(0,0)[ct]{$\tilde{X}$}}
\put(110.00,72.00){\vector(0,-1){50.33}}
\put(112.00,73){\vector(1,-1){23.33}}
\put(136.67,46.00){\vector(-1,-1){25.00}}
\put(118.67,50.67){\makebox(0,0)[rc]{$\pi_{\tilde{X}\tilde{Z}}$}}
\put(131.67,62.00){\makebox(0,0)[rb]{$\pi_{\tilde{Y}\tilde{Z}}$}}
\put(134.33,32.00){\makebox(0,0)[rb]{$\pi_{\tilde{X}\tilde{Y}}$}}
\put(80.00, 4){\makebox(0,0)[cc]{Figure 3}}
\end{picture}

Integration gives a standard method for obtaining $k$-forms on $\tilde{Z}$
from $(k+n)$-forms on $Z$ as follows.

\begin{defn}
If $\alpha$ is a $(k+n)$-form in $Z$ such that $i_B^*\alpha=d\beta$, we
define the $k$-form $\tilde{\alpha}$ on $\tilde{Z}$ by
\begin{align}  \label{integration}
\iota_{\tilde{\zeta}_1}\ldots\iota_{\tilde{\zeta}_k}\tilde{\alpha}_\gamma =
\int_M \gamma^*\iota_{\zeta_1}\ldots\iota_{\zeta_k}\tilde{\alpha}_\alpha -
(-1)^k \int_{\partial M} \gamma^*\iota_{\zeta_1}\ldots\iota_{\zeta_k}\beta
\end{align}
for $\tilde{\zeta}_1,\ldots, \tilde{\zeta}_k\in T_\gamma\tilde{Z},
\gamma\in\tilde{Z}$.
\end{defn}

In particular, the Poincar\'e-Cartan $(n+1)$-form $\Theta_L$ and
$(n+2)$-form $\Omega_L$ also induce a 1-form
$\widetilde{\Theta_L}$ and a 2-form $\widetilde{\Omega_L}$ on
$\tilde{Z}$, given by:
\begin{equation*}
(\widetilde{\Theta_L})_\gamma(\tilde{\xi})=\int_M
\gamma^*(\iota_\xi\Theta_L)+\int_{\partial M} \gamma^*(\iota_\xi\Pi)
\end{equation*}
and also
\begin{equation*}
\widetilde{\Omega_L}(\tilde{\xi}_1,\tilde{\xi}_2)=\int_M \gamma^*
(\iota_{\xi_2}\iota_{\xi_1}\Omega_L).
\end{equation*}

\begin{lem}
\label{lema4_1} If $\tilde{\xi}$ is a vector field on $\tilde{Z}$ defined
from a vector field $\xi$ on $Z$, and $\alpha$ is an $n$-form on $Z$ such
that $i_B^*\alpha = d\beta$ then
\begin{equation*}
d\tilde{\alpha}(\tilde{\xi})_\gamma=(\pounds_{\tilde{\xi}}\tilde{\alpha}
)_\gamma = \int_M \gamma^*(\pounds_\xi\alpha)-\int_{\partial M}
\gamma^*(\pounds_\xi\beta)
\end{equation*}
\end{lem}

\textit{Proof}. First observe that $\tilde{\alpha}$ is a function. In this
case, if $c_{\tilde{Z}}(t)$ is a curve such that $c_{\tilde{Z}}(0)=\gamma$
and $\dot{c}_{\tilde{Z}}(0)=\xi(\gamma)$, then
\begin{align*}
d\tilde{\alpha} (\tilde{\xi})_\gamma &=
\tilde{\xi}_\gamma(\tilde{\alpha}) =
\frac{d}{dt}(\tilde{\alpha}\circ c_{\tilde{Z}}(t))_{|t=0} =
\frac{d}{dt}\left[\int_M
\left(c_{\tilde{Z}}(t)^*\alpha\right)-\int_{\partial M}
\left(c_{\tilde{Z}}(t)^*\beta\right)\right]_{|t=0} \\
&= \int_M \frac{d}{dt} \left(c_{\tilde{Z}}(t)^*\alpha\right)_{|t=0}-
\int_{\partial M} \frac{d}{dt} \left(c_{\tilde{Z}}(t)^*\beta\right)_{|t=0} =
\int_M \gamma^*(\pounds_{{\xi}}\alpha)-\int_{\partial M}
\gamma^*(\pounds_\xi\beta).
\end{align*}
\vspace{-1.7cm}

$\,$\hfill\ \ \ \vrule height 1.5ex width.8ex depth.3ex \medskip

\

The previous result can be also extended for forms of higher
degree, and for arbitrary fibrations over $X$.

Let $\xi$ be a complete vector field on a fibration $W$ over $X$,
and let us denote by $\tilde{W}$ certain space of embeddings in
$W$, and by $\tilde{\xi} $ the vector field defined on $\tilde{W}$
from $\xi$ (that is, $\tilde{\xi}(\gamma)(u)=\xi(\gamma(u)))$.

Fix $\gamma\in\tilde{W}$. For every $u\in M$, consider an integral curve $c^u
$ of $\xi$ through $\gamma(u)$, that is
\begin{align*}
c^u(0) &= \gamma(u) \\
\dot{c}^u(0) &= \xi(\gamma(u))
\end{align*}

Let us define a curve $\tilde{c}$ on $\tilde{W}$ by
\begin{equation*}
\tilde{c}(t)(u) = c^u(t).
\end{equation*}

Then we have that

\begin{prop}
$\tilde{c}$ is an integral curve of $\tilde{\xi}$ through $\gamma$.
\end{prop}

\textit{Proof.} To see this, we just have to compute
\begin{equation*}
\tilde{c}(0)(u) = c^u(0)=\gamma(u)
\end{equation*}
and
\begin{equation*}
\dot{\tilde{c}}(0)(u) = \frac{d}{dt}(\tilde{c}(t))|_{t=0}(u) =
\frac{d}{dt} (\tilde{c}(t)(u))|_{t=0} = \frac{d}{dt} c^u(t)|_{t=0}
= \dot{c}^u(t) = \xi(\gamma(u)) = \tilde{\xi}(\gamma)(u).\ \ \
\vrule height 1.5ex width.8ex depth.3ex \medskip
\end{equation*}

$\tilde{c}$ will be said to be the associated curve to the flow given by the
$c^u$'s.

In particular, if we also have a diffeomorphism $F: W \longrightarrow W$, it
is easy to see that the curve (denoted by $\widetilde{F\circ c})$ associated
to the family $F\circ c^u$ is precisely $\tilde{F}\circ\tilde{c}$.

To see this, and using the preceding notation, note first that
\begin{equation*}
\widetilde{F\circ c}(t)(u)=(F\circ c)^u(t)=(F\circ
c^u)(t)=F(c^u(t))=F(\tilde{c}(t)(u))=(\tilde{F}\circ\tilde{c}(t))(u),
\end{equation*}
from which we deduce

\begin{cor}
If $F: W \longrightarrow W$ is a diffeomorphism, then
$T\tilde{F}(\tilde{\xi})=\widetilde{TF(\xi)}$.
\end{cor}

The next step is to study the pullback of forms.

\begin{prop}
If $F: W \longrightarrow W$ is a diffeomorphism, and $\alpha$ is a
$(n+k)$-form on $W$, such that $i_B^*\alpha=d\beta$, then
\begin{equation*}
\tilde{F}^*\tilde{\alpha} = \widetilde{F^*\alpha}
\end{equation*}
\end{prop}

\textit{Proof.} Let $\widetilde{V_1},\ldots,\widetilde{V_k}\in
T_{\tilde{F}^{-1}(\gamma)}\tilde{W}$. We have that
\begin{align*}
\iota_{\widetilde{V_1}}\ldots\iota_{\widetilde{V_k}}\tilde{F}^*\tilde{\alpha}
&=
\tilde{\alpha}(T\tilde{F}(\widetilde{V_1}),\ldots,T\tilde{F}(\widetilde{V_k}))
= \tilde{\alpha}(\widetilde{TF(V_1)},\ldots,\widetilde{TF(V_k)}) \\
&= \int_M \gamma^* \iota_{TF(V_1)}\ldots\iota_{TF(V_k)}\alpha -
(-1)^k\int_{\partial M} \gamma^* \iota_{TF(V_1)}\ldots\iota_{TF(V_k)}\beta \\
&= \int_M(F^{-1}\circ\gamma)^*F^*\iota_{TF(V_1)}\ldots\iota_{TF(V_k)}\alpha
- (-1)^k\int_{\partial
M}(F^{-1}\circ\gamma)^*F^*\iota_{TF(V_1)}\ldots\iota_{TF(V_k)}\beta \\
&= \int_M(F^{-1}\circ\gamma)^*\iota_{V_1}\ldots\iota_{V_k}F^*\alpha -
(-1)^k\int_{\partial M}
(F^{-1}\circ\gamma)^*\iota_{V_1}\ldots\iota_{V_k}F^*\beta \\
&= \iota_{\widetilde{V_1}}\ldots\iota_{\widetilde{V_k}}\widetilde{F^*\alpha}.
\end{align*}
\vspace{-1.7cm}

$\,$\hfill\ \ \ \vrule height 1.5ex width.8ex depth.3ex \medskip

\

Finally,

\begin{prop}
If $\xi$ is a vector field on $\tilde{W}$, then
\begin{equation*}
\pounds_{\tilde{\xi}}\tilde{\alpha}=\widetilde{\pounds_\xi\alpha}
\end{equation*}
\end{prop}

\textit{Proof.} Let $\widetilde{V_1},\ldots,\widetilde{V_k}\in
T_\gamma\tilde{W}$, and denote by $\phi_t$ the flow of $\xi$. Then
we have that
\begin{align*}
\iota_{\widetilde{V_1}}\ldots\iota_{\widetilde{V_k}}\pounds_{\tilde{\xi}}
\tilde{\alpha}&=
\iota_{\widetilde{V_1}}\ldots\iota_{\widetilde{V_k}}\frac{d}{dt}\widetilde{\phi_t}^*\tilde{\alpha}|_{t=0}=
\iota_{\widetilde{V_1}}\ldots\iota_{\widetilde{V_k}}\frac{d}{dt}\widetilde{\phi_t^*\alpha}|_{t=0}
\\
&=\frac{d}{dt}\left(\iota_{\widetilde{V_1}}\ldots\iota_{\widetilde{V_k}}
\widetilde{\phi_t^*\alpha}\right)|_{t=0}
=\frac{d}{dt}\left(\int_M\iota_{V_1}\ldots\iota_{V_k}\phi_t^*\alpha
- (-1)^k\int_{\partial
M}\iota_{V_1}\ldots\iota_{V_k}\phi_t^*\beta\right)|_{t=0} \\
&=\int_M\iota_{V_1}\ldots\iota_{V_k}\frac{d}{dt}\left(\phi_t^*\alpha\right)|_{t=0}
- (-1)^k\int_{\partial M}\iota_{V_1}\ldots\iota_{V_k}\frac{d}{dt}\left(\phi_t^*\beta\right)|_{t=0} \\
&=\int_M\iota_{V_1}\ldots\iota_{V_k}\pounds_\xi\alpha - (-1)^k\int_{\partial
M}\iota_{V_1}\ldots\iota_{V_k}\pounds_\xi\beta \\
&=\iota_{\widetilde{V_1}}\ldots\iota_{\widetilde{V_k}}\widetilde{\pounds_\xi\alpha}.
\end{align*}
where for the last bit just notice that
$i_B^*\pounds_\xi\alpha=\pounds_\xi i_B^*\alpha = \pounds_\xi
d\beta = d\pounds_\xi\beta.\hfill\ \ \ \vrule height 1.5ex
width.8ex depth.3ex \medskip$

Back to the fibration $Z \longrightarrow X$, the consistency of our
definition of forms respect to the exterior derivative is ensured by the
following proposition

\begin{prop}
If $\alpha$ is an $n$-form or an $(n+1)$-form, then
\begin{equation*}
\widetilde{d\alpha} = d\tilde{\alpha}
\end{equation*}
In particular,
\begin{equation*}
\widetilde{\Omega_L}:=-d\widetilde{\Theta_L}
\end{equation*}
\end{prop}

\textit{Proof}. For $n$-forms we use the previous lemma
\begin{eqnarray*}
(d\tilde{\alpha})_\gamma(\tilde{\xi})&=& \int_M \gamma^*
\pounds_\xi\alpha-\int_{\partial M}\gamma^* \pounds_\xi\beta = \int_M
\gamma^* \iota_\xi d\alpha + \int_M \gamma^*d\iota_\xi\alpha-\int_{\partial
M}\gamma^*(i_\xi d\beta+di_{\xi}\beta) \\
& =& \int_M \gamma^* \iota_\xi d\alpha = (\widetilde{d\alpha})_\gamma(\xi)
\end{eqnarray*}
For $(n+1)$-forms:
\begin{align*}
d\widetilde{\alpha}(\xi,\zeta)_\gamma &= \{\xi(\widetilde{\alpha}(\zeta)) -
\zeta(\widetilde{\alpha}(\xi)) - \widetilde{\alpha}([\zeta,\xi])\}_\gamma \\
&=\int_M
\gamma^*\{\pounds_\xi(\iota_\zeta\alpha)-\pounds_\zeta(\iota_\xi\alpha)-
\iota_{[\xi,\zeta]}\alpha\} \\
&+\int_{\partial M}
\gamma^*\{\pounds_\xi(\iota_\zeta\beta)-\pounds_\zeta(\iota_\xi\beta)-
\iota_{[\xi,\zeta]}\beta\} \\
&=\int_M \gamma^*\{\iota_\zeta\iota_\xi d\alpha -
d\iota_\zeta\iota_\xi\alpha\} \\
&+\int_{\partial M} \gamma^*\{\iota_\zeta\iota_\xi d\beta -
d\iota_\zeta\iota_\xi\beta\} \\
&=\int_M \gamma^*(\iota_\zeta\iota_\xi d\alpha) - \int_{\partial
M}\gamma^*(\iota_\zeta\iota_\xi(d\beta-\alpha)) \\
&=\int_M \gamma^*(\iota_\zeta\iota_\xi d\alpha) \\
&=\widetilde{d\alpha}(\xi,\zeta)_\gamma.
\end{align*}
\hfill$\ \ \ \vrule height 1.5ex width.8ex depth.3ex \medskip$

\subsection{The De Donder equations in the space of Cauchy data}

The De Donder equations of Field Theories have a presymplectic
counterpart in the spaces of Cauchy data. The relationship between
both can be found in \cite{BSF} (see also \cite{gimmsy1}), and
requires the definition of a slicing of the base manifold $X$.

\begin{defn}
We say that a curve $c_{\tilde{X}}$ in $\tilde{X}$ defined on a
domain $I \subseteq \mathbb{R}$ \textbf{splits} $X$ if the mapping
$\Phi: I \times M \longrightarrow X$, such that
$\Phi(t,u)=c_{\tilde{X}}(t)(u)$, is a diffeomorphism. In
particular, the partial mapping $\Phi(t,\cdot)$ (defined by
$\Phi(t,\cdot)(u)=\Phi(t,u))$ is an element of $\tilde{X}$ for all
$t\in I $. In this case, $c_{\tilde{X}}$ is said to be a
\textbf{slicing}.

In this situation, we can rearrange coordinates in $X$ such that
if $\frac{\partial }{\partial t}$ generates the tangent space to
$I$, then $T\Phi(\frac{\partial }{\partial t})=\frac{\partial
}{\partial x^0}$, and we
consider $\frac{\partial }{\partial x^1},\ldots,\frac{\partial }{\partial x^n%
}$ as local tangent vector fields on $M$ or $X$.
\end{defn}

\begin{defn}
We can also define the concept of \textbf{infinitesimal slicing}
at $\tau\in\tilde{X}$ as a tangent vector $v\in T_\tau\tilde{X}$
such that for every $u\in M$, $v(u)$ is transverse to $Im\;\tau$.
\end{defn}

If $c_{\tilde{Z}}$ is a curve in $\tilde{Z}$ such that its
projection $c_{\tilde{X}}$ to $\tilde{X}$ splits $X$, then it
defines a local section $\sigma$ of $\pi_{XZ}$ by
\begin{equation}  \label{eq:liftcurve}
\sigma(c_{\tilde{X}}(t)(u))=c_{\tilde{Z}}(t)(u)
\end{equation}

Conversely, if $\sigma$ is a section of $\pi_{XZ}$, and
$c_{\tilde{X}}$ is a curve on $\tilde{X}$ (not necessarily a
slicing), we define a curve $c_{\tilde{Z}}$ on $Z$ by using
\eqref{eq:liftcurve}. The following result relating equations in
$Z$ and $\tilde{Z}$ can be found in \cite{BSF}.

\begin{thm}
If $\sigma$ satisfies the De Donder equations, then $c_{\tilde{Z}}$ defined
as above verifies
\begin{equation}  \label{DeDonderTilde}
\iota_{\dot{c}_{\tilde{Z}}}\widetilde{\Omega_L}=0
\end{equation}
Conversely, if $c_{\tilde{Z}}$ is a curve on $Z$ satisfying
\eqref{DeDonderTilde}, and its projection $c_{\tilde{X}}$ to
$\tilde{X}$ splits $X$, then the section $\sigma$ of $\pi_{XZ}$
defined by \eqref{eq:liftcurve} verifies the De Donder equations.
\end{thm}

\textit{Proof}. Assume that $\sigma$ verifies the De Donder
equations. From \eqref{eq:liftcurve} we obtain that
$\dot{c}_{\tilde{Z}}=\sigma_*\dot{c}_{\tilde{X}}$, whence

\begin{equation*}
c_{\tilde{Z}}(t)^*(\iota_{\dot{c}_{\tilde{Z}}}\iota_\xi\Omega_L) =
c_{\tilde{X}}(t)^* \sigma^*
(\iota_{\dot{c}_{\tilde{Z}}}\iota_\xi\Omega_L) =
c_{\tilde{X}}(t)^*
(\iota_{\dot{c}_{\tilde{X}}}\sigma^*\iota_\xi\Omega_L) = 0
\end{equation*}
for all $\xi$. Now integrate over $M$ to obtain the desired result. For the
converse, consider the integral
\begin{equation*}
0 = \int_M c_{\tilde{X}}(t)^*
(\iota_{\dot{c}_{\tilde{X}}}\sigma^*\iota_\xi\Omega_L) = 0
\end{equation*}
since this is true for every $\xi$, from the Fundamental Theorem of Calculus
of Variations, we deduce
\begin{equation*}
c_{\tilde{X}}(t)^* (\iota_{\dot{c}_{\tilde{X}}}\sigma^*\iota_\xi\Omega_L) =
0
\end{equation*}

Now if $c_{\tilde{X}}$ splits $X$, then $\dot{c}_{\tilde{X}}(t)$ is
transverse to $c_{\tilde{X}}(t)(M)$, which implies the De Donder
equations.\hfill $\ \ \ \vrule height 1.5ex width.8ex depth.3ex \medskip$

Note that, in particular, if \textbf{h} is the horizontal projector of a
connection which is a solution of the De Donder equations for a connection
\begin{align}  \label{DeDonderh}
\iota_{\hbox{\textbf{h}}}\Omega_L = n \Omega_L
\end{align}
and if $\sigma$ is a horizontal local section of \textbf{h}, the results
above show that the solution to \eqref{DeDonderTilde} is the horizontal lift
of $\dot{c}_{\tilde{X}}$ through \textbf{h}. Or more generally, the
solutions are obtained as horizontal lifts of infinitesimal slicings through
the connection solution to \eqref{DeDonderh}.

\subsection{The singular case}

For a singular Lagrangian, we cannot guarantee the existence of a
curve $c_{\tilde{Z}}$ in $\tilde{Z}$ as a solution of the De
Donder equations in $\tilde{Z}$.

Therefore, we propose an algorithm similar to that of a general
presymplectic space (developed in \cite{Got,GN1,GN2}; see also
\cite{canarias2,canarias3,canarias1} for the time dependent case),
where to the condition that defines the manifold obtained in each
step (which is the existence of a tangent vector verifying the De
Donder equations), we add the fact that this tangent vector must
project onto an infinitesimal slicing.

Naming $\tilde{Z}_1 := \tilde{Z}$, we define $\tilde{Z}_2$ and the
subsequent subsets (requiring them to be submanifolds) as follows
\begin{align*}
\tilde{Z}_2 &:= \{ \gamma\in\tilde{Z}_1 | \exists v\in T_\gamma
\tilde{Z}_1 \;\hbox{such that}\; T\pi_{\tilde{X}\tilde{Z}}(v) \;
\hbox{is an infinitesimal slicing and
}\;\iota_v\widetilde{\Omega_L}|_\gamma = 0 \} \\
\tilde{Z}_3 &:= \{ \gamma\in\tilde{Z}_2 | \exists v\in T_\gamma
\tilde{Z}_2 \;\hbox{such that}\; T\pi_{\tilde{X}\tilde{Z}}(v) \;
\hbox{is an infinitesimal slicing and
}\;\iota_v\widetilde{\Omega_L}|_\gamma = 0 \} \\
&\ldots
\end{align*}
In the favourable case, the algorithm will stop at certain final non-zero
dimensional constraint submanifold $\tilde{Z}_f$.

This algorithm is closely related to the algorithm in the finite dimensional
spaces. We turn now to state the link between them.

\begin{prop}
Suppose that we have $v\in T_\gamma \tilde{Z}_1$ such that
$T\pi_{\tilde{X} \tilde{Z}}(v)$ is an infinitesimal slicing and
$\iota_v\widetilde{\Omega_L} |_\gamma = 0$. Then, for every $u\in
M$ we have that
\begin{equation*}
H_{\gamma(u)} := T_u\gamma (T_uM) \oplus \langle v(u)\rangle
\end{equation*}
is a horizontal subspace of $T_{\gamma(u)}Z$ which horizontal
projector \textbf{h} verifies the De Donder equations for
connections satisfying \eqref{DeDonderh} at $\gamma(u)$:
\begin{equation*}
\iota_{\hbox{\rm \textbf{h}}}\Omega_L|_{\gamma(u)} = n \Omega_L|_{\gamma(u)}
\end{equation*}
\end{prop}

\textit{Proof.} The fact that $v$ projects onto an infinitesimal slicing
guarantees that $H_{\gamma(u)}$ is indeed horizontal.

The other hypothesis states that
\begin{equation*}
\gamma^*(\iota_\xi\iota_{v_{\gamma(u)}}\Omega_L)=0
\end{equation*}
for every $\xi\in T_{\gamma(u)}Z$, that is, if $\langle v_1, v_2, \ldots,
v_n\rangle$ is a basis for $T_uM$, then
\begin{equation*}
\iota_\xi\iota_{v_{\gamma(u)}}\Omega_L(T_u\gamma(v_1),
T_u\gamma(v_2),\ldots,T_u\gamma(v_n))=0
\end{equation*}
or in other words,
\begin{equation*}
\Omega_L (\xi, H_1, H_2, \ldots, H_{n+1}) = 0
\end{equation*}
for every $\xi\in T_{\gamma(u)}Z$ and every collection $H_1, H_2, \ldots,
H_{n+1}$ of horizontal tangent vectors.

We want to prove that
$\iota_{\hbox{\textbf{h}}}\Omega_L|_{\gamma(u)} = n
\Omega_L|_{\gamma(u)}$, or equivalently,
$\iota_\xi\iota_{\hbox{\textbf{h}}}\Omega_L|_{\gamma(u)} =
n\iota_\xi \Omega_L|_{\gamma(u)}$, for every $\xi\in
T_{\gamma(u)}Z$.

From the previous remarks, we see that the condition results to be true when
it is evaluated on $n+1$ horizontal vector fields.

Suppose that $V_1$ is a vertical tangent vector to $\gamma(u)$. Then (as
\textbf{h}$(V_1)=0)$,
\begin{eqnarray*}
\iota_{\hbox{\textbf{h}}}\Omega_L (\xi, V_1, H_1, \ldots, H_n) &= &\Omega_L
( \hbox{\textbf{h}}(\xi), V_1, H_1, \ldots, H_n) + n\Omega_L (\xi, V_1, H_1,
\ldots, H_n)
\end{eqnarray*}
where the first term vanishes due to the previous remarks. Thus, the
expression holds when applied to any two tangent vector, and to any $n$
horizontal tangent vectors.

For the next step, having two vertical vectors, remember that $\Omega_L$ is
annihilated by three vertical tangent vectors. Therefore,
\begin{align*}
\iota_{\hbox{\textbf{h}}}\Omega_L(&\xi, V_1, V_2, H_1, \ldots, H_{n-1}) =
\Omega_L(\hbox{\textbf{h}}(\xi), V_1,V_2, H_1, \ldots, H_{n-1}) \\
&+ (n-1)\Omega_L(\xi, V_1,V_2, H_1, \ldots, H_{n-1}) \\
&=\Omega_L(\xi, V_1,V_2, H_1, \ldots, H_{n-1}) + (n-1)\Omega_L(\xi, V_1,V_2,
H_1, \ldots, H_{n-1}) \\
&=n\Omega_L(\xi, V_1,V_2, H_1, \ldots, H_{n-1})
\end{align*}
Finally, from the mentioned properties of $\Omega_L$, the expression also
holds for a higher number of vertical tangent vectors, and so the expression
holds in general. \hfill$\ \ \ \vrule height 1.5ex width.8ex depth.3ex
\medskip$

As an immediate result, we have that

\begin{cor}
If $\gamma\in\tilde{Z}_2$, then $Im\gamma\subseteq Z_2$.
\end{cor}

and in general,

\begin{prop}
If $\gamma\in\tilde{Z}_i$, then $Im\gamma\subseteq Z_i$.
\end{prop}

\textit{Proof}. If $\gamma\in \tilde{Z}_i$ (which implies that
there exists $v\in T\tilde{Z}_i$ such that
$\iota_v\widetilde{\Omega_L}|_\gamma = 0)$, then for every $u\in
M$ we define $H_{\gamma(u)} := T\gamma_u (T_uM) \oplus \langle
v(u) \rangle$.

We need to justify in each step that $H_{\gamma(u)} \subseteq
T_{\gamma(u)}Z_i$, which amounts to prove that $T\gamma_u (T_uM)\subseteq
T_{\gamma(u)}Z_i$ and $v(u)\in T_{\gamma(u)}Z_i$. The first assertion is
true by construction of the subsets.

To see that $v(u)\in T_{\gamma(u)}Z_i$, we proceed inductively,
starting on $i=2$, for which the result is true because of the
preceding corollary.

We assume it to be true for all the steps until the $i$-th, and we prove
that $v(u)\in T_{\gamma(u)}Z_{i+1}$.

As $\gamma\in\tilde{Z}_{i+1}$, there exists $v\in
T_\gamma\tilde{Z}_i$ such that $\iota_v\widetilde{\Omega_L}=0$.
Thus, there exists a curve $c:
(-\varepsilon,\varepsilon)\longrightarrow \tilde{Z}_i$ (and thus
$Im(c)(t)\subseteq Z_i$) such that $c(0)=\gamma$ and
$\dot{c}(o)=v$. We deduce that $v(u)\in T_{\gamma(u)}Z_i.\hfill\ \
\ \vrule height 1.5ex width.8ex depth.3ex \medskip$

\begin{remark}
\textrm{Suppose now that $\tilde{X}$ admits an slicing. In the case in which
$z\in Z_i$ is such that $\pi_{XZ}(z)$ belongs to the image of the slicing,
and \textbf{h$_z$} is integrable, then there exists $\gamma\in\tilde{Z}_i$,
and $u\in M$ such that $\gamma(u)=z$. }

\textrm{As before, we prove first the case $i=2$. If $\sigma$ is
an horizontal local section of \textbf{h} at $z$, then we use the
slicing to define the curve $c_{\tilde{Z}}(t)$, which verifies the
De Donder equations in $\tilde{Z}$, and projects onto the slicing,
therefore we can take $\gamma=c_{\tilde{Z}}(t)$ for some $t$. }

\textrm{For the case $i>1$, simply observe that if $H_{\gamma(u)}\subseteq
Z_i$, then $\dot{c}_{\tilde{Z}}(t)(u^{\prime})$ must be tangent to $Z_2$ for
all $u^{\prime}\in M$, and a very similar argument to that of the preceding
section proves that $\gamma=c_{\tilde{Z}}(t)\in\tilde{Z}_2$. }
\end{remark}

\subsection{Brackets}

Notice that, in general, the only fact over $\widetilde{\Omega_L}$
that we can guarantee is that it is presymplectic, as we cannot
guarantee nor the existence neither the uniqueness of Hamiltonian
vector fields associated to functions defined on $\tilde{Z}$. For
further details see \cite{Le-Da1} and \cite{Le-Da2}.

\begin{defn}
Given a function $f$ in $\tilde{Z}$ and a vector field
$\tilde{\xi}$ on $\tilde{Z}$, we shall say that $f$ is a
\textbf{Hamiltonian function}, and that $\tilde{\xi}$ is a
\textbf{Hamiltonian vector field} for $f$ if
\begin{equation*}
\iota_{\tilde{\xi}}\widetilde{\Omega_L}=df
\end{equation*}
\end{defn}

\begin{prop}
If $\alpha$ is a Hamiltonian $n$-form in $Z$ for $\Omega_L$ which
is exact on $\partial Z$, say $\tilde{\alpha}_{|\partial Z} =
d\tilde{\beta}$, then $\tilde{\alpha}$ is a Hamiltonian function
on $\tilde{Z}$ for $\widetilde{\Omega_L}$. More precisely, if
$X_\alpha$ is a Hamiltonian vector field for $\alpha$, then
$X_{\tilde{\alpha}}$ defined on $\tilde{Z}$ by
\begin{equation*}
[X_{\tilde{\alpha}}(\gamma)](u) = X_\alpha(\gamma(u))
\end{equation*}
is a Hamiltonian vector field for $\tilde{\alpha}$
\end{prop}

\textit{Proof}. Take a tangent vector $\tilde{\xi}$ to $\tilde{Z}$, then by
lemma \eqref{lema4_1}
\begin{align*}
(d\tilde{\alpha})(\tilde{\xi})|_\gamma &= \int_M \gamma^*(\pounds_\xi\alpha)
- \int_{\partial M} \gamma^*(\pounds_\xi\beta) \\
&= \int_M \gamma^*\iota_\xi d\alpha + \int_M \gamma^* d \iota_\xi \alpha -
\int_{\partial M} \gamma^* \iota_\xi d\beta \\
&= \int_M \gamma^*\iota_\xi d\alpha = \int_M
\gamma^*\iota_\xi\iota_{X_\alpha}\Omega_L =
\iota_{\tilde{X}_\alpha}\widetilde{\Omega_L}(\tilde{\xi})|_\gamma.
\end{align*}
which proves that $d\tilde{\alpha} =
\iota_{X_{\tilde{\alpha}}}\widetilde{\Omega_L}. $ \hfill$\ \ \
\vrule height 1.5ex width.8ex depth.3ex \medskip$

If $f$ is a Hamiltonian function on $\tilde{Z}$, then its
associated Hamiltonian vector field is defined up to an element in
the kernel of $\widetilde{\Omega_L}$, therefore we can define the
bracket operation for these functions as follows.

\begin{defn}
If $f$ and $g$ are Hamiltonian functions on $\tilde{Z}$, with associated
Hamiltonian vector fields $X_f$ and $X_g$, then we define:
\begin{equation*}
\{f,g\} := \widetilde{\Omega_L}(X_f,X_g)
\end{equation*}
\end{defn}

Notice that $i_B^*\Omega_L=0$, thus if $\alpha_1$ and $\alpha_2$
are Hamiltonian forms which are exact on the boundary, then
$i_B^*\{\alpha_1,\alpha_2\}=0$.

\begin{prop}
If $\alpha_1$ and $\alpha_2$ are Hamiltonian $n$-forms which are
exact on $\partial Z$, then
\begin{equation*}
\{\tilde{\alpha_1},\tilde{\alpha_2}\} = \widetilde{\{\alpha_1,\alpha_2\}}
\end{equation*}
\end{prop}

\textit{Proof}.
\begin{equation*}
\{\tilde{\alpha_1},\tilde{\alpha_2}\} =
\widetilde{\Omega_L}(X_{\tilde{\alpha_1}},X_{\tilde{\alpha_2}}) =
\int_M \gamma^* \iota_{X_{\alpha_2}} \iota_{X_{\alpha_1}} \Omega_L
= \int_M \gamma^* \{\alpha_1,\alpha_2\} =
\widetilde{\{\alpha_1,\alpha_2\}}.
\end{equation*}
\hfill$\ \ \ \vrule height 1.5ex width.8ex depth.3ex \medskip$

In \cite{CIL2,FPR} and \cite{GS} the authors explore the properties of a
generalisation of this bracket, which satisfies the graded versions of
several properties, such as skew-symmetry and Jacobi identity.

\begin{remark}
\textrm{We could alternatively use the space of Cauchy data $\tilde{Z^*}$,
defined in the obvious way. But nothing different or new would be obtained.
In fact, assume for simplicity that $L$ is hyperregular. Then we would have
a diffeomorphism $\widetilde{leg_L} : \tilde{Z} \longrightarrow \tilde{Z^*}$
defined by composition:
\begin{equation*}
\widetilde{leg_L} (\gamma) = leg_L\circ\gamma
\end{equation*}
for all $\gamma \in \tilde{Z}$. }

\textrm{If the Lagrangian is not regular, but at least is almost regular, we
invite to the reader to develop the corresponding scheme. The only delicate
point is that we have to consider the second order problem in the Lagrangian
side, so that $\widetilde{leg_L} : \tilde{Z} \longrightarrow \tilde{Z^*}$
becomes a fibration. }

\textrm{In what follows, we shall emphasize the discussion in the Lagrangian
side, since, as we have shown, the equivalence with the Hamiltonian side is
obvious. }
\end{remark}

\section{Symmetries. Noether's theorems}

We are now interested in studying the presence of symmetries which would
eventually produce preserved quantities, and allow us to reduce the
complexity of the dynamical system and to obtain valuable information about
its behaviour. For every type of symmetry, there will be a form of the
Noether's theorem, which will show up the preserved quantity obtained from
it (see \cite{OLVER}).

We shall suppose that we are in the regular Lagrangian case, unless stated
otherwise.

In our framework for field theory, we define a preserved quantity in the
following manner:

\begin{defn}
A \textbf{preserved quantity for the Euler-Lagrange equations} is
an $n$-form $\alpha$ on $Z$ such that $(j^1\phi)^*d\alpha=0$ for
every solution $\phi$ of the Euler-Lagrange equations. If $\alpha$
is a preserved quantity, then $\tilde{\alpha}$ is called its
associated \textbf{momentum}.
\end{defn}

Notice that if $\alpha$ is a preserved quantity, and $\Lambda$ is
a closed form, then $\alpha+\Lambda$ is also a preserved quantity.
Similarly, if $\gamma$ is an $n$-form which belongs to the
differential ideal $\mathcal{I}(\mathcal{C})$, then
$\alpha+\gamma$ is also a preserved quantity (see \cite{OLVER} for
a further discussion).

We turn now to obtain preserved quantities from symmetries.

\subsection{Symmetries of the Lagrangian}

We shall define the notion of symmetry based on the the variation of the
Poincar\'e-Cartan $(n+1)$-form along prolongations of vector fields. Suppose
that $\xi_Y$ is a vector field defined on $Y$, and abbreviate by $F$ the
function such that

\begin{equation*}
\pounds_{\xi_Y^{(1)}}\mathcal{L} - F\eta \in \mathcal{I}(\mathcal{C})
\end{equation*}
having local expression

\begin{equation}  \label{eq:F}
F = \xi^{(1)}_Y(L) + \left( \frac{\partial \xi_Y^\mu}{\partial x^\mu} +
z^i_\nu\frac{\partial \xi_Y^\nu}{\partial y^i}\right)L .
\end{equation}

After a lengthy computation we get that

\begin{align}  \label{eq:lYthetaL}
\pounds_{\xi_Y^{(1)}}\Theta_L &= F\eta + \frac{\partial
F}{\partial z^i_\mu}\theta^i\wedge d^nx_\mu  \notag \\
&+z^j_\nu\left(\frac{\partial \xi_Y^\nu}{\partial
y^j}\frac{\partial L} {\partial z^i_\mu}-\frac{\partial
\xi_Y^\mu}{\partial y^j}\frac{\partial L}{\partial z^i_\nu}\right)\theta^i\wedge d^nx_\mu \\
&- \frac{\partial \xi_Y^\nu}{\partial y^j}\frac{\partial
L}{\partial z^i_\mu} \theta^i\wedge dy^j\wedge d^{n-1}x_{\nu\mu}
\notag
\end{align}

\begin{defn}
\label{def:SimLag} A vector field $\xi_Y$ on $Y$ is said to be an
\textbf{infinitesimal symmetry of the Lagrangian} or a
\textbf{variational symmetry} if $\pounds_{\xi_Y^{(1)}}\Theta_L
\in \mathcal{I}(\mathcal{C})$ (the differential ideal generated by
the contact forms), and $\xi_Y^{(1)}$ is also tangent to $B$ and
verifies $\pounds_{\xi_Y^{(1)}|_B}\Pi=0$
\end{defn}

We shall only deal with infinitesimal symmetries of the Lagrangian, so for
brevity they will be referred simply as symmetries of the Lagrangian.

From the definition and the expression \eqref{eq:lYthetaL}, it is obvious to
see that

\begin{prop}
If a vector field $\xi_Y$ on $Y$ is a symmetry of the Lagrangian, then $F=0$
(where $F$ was defined in \eqref{eq:F}).
\end{prop}

\begin{remark}
\textrm{In our construction, we choose as definition of the
Poincar\'e-Cartan $(n+1)$-form:
\begin{equation*}
\Theta_L=\mathcal{L}+(S_{\eta})^*(dL)
\end{equation*}
or, in fibred coordinates
\begin{equation*}
\Theta_L=L\, d^{n+1} x+\frac{\partial L}{\partial z^i_{\mu}}\theta^i\wedge
d^nx_{\mu}
\end{equation*}
If $n>0$ it is possible to generalize the construction of the
Poincar\'e-Cartan $(n+1)$-form in several different ways. The
unique requirement is that the resulting $\pi_{YZ}$-semibasic
$(n+1)$-form be \emph{Lepage}-equivalent to ${\mathcal L}$, that
is,
\[
\Theta-{\mathcal L}\in {\mathcal I}({\mathcal C})
\]
and $i_V d\Theta\in {\mathcal I}({\mathcal C})$ where $V$ is an arbitrary $\pi_{YZ}$-vertical
vector field. Locally,
\begin{equation}  \label{asqq}
\Theta=\Theta_L+ \cdots
\end{equation}
where the dots signify terms which are at least two-contact (see
\cite{Betounes,Crampin-Saunders,Ka,Kova}). Obviously, all them
gives us identically the same Euler-Lagrange equations. }

\textrm{Therefore, we may substitute in Definitions
\ref{def:SimLag}, \ref{def:SimNoe} and \ref{def:SimCar} the
Poincar\'e-Cartan $(n+1)$-form by any $(n+1)$-form which is
\emph{Lepage}- equivalent to $\Theta_L$. Obviously, the symmetries
of the Euler-Lagrange equations are independent of the class of
\emph{Lepagian} $(n+1)$-form appearing in their definition. }
\end{remark}

We also have the following two special cases, which are easily computed from
the expression of $F$.

\begin{prop}
\label{prop:propertiesF} If $\xi_Y$ is a projectable symmetry of
the Lagrangian $(T\pi_{XY}(\xi_Y)$ is a well defined vector field,
or locally $\frac{\partial \xi_Y^\mu}{\partial y^i}=0)$, or if
$dimX=1$ $(n=0)$, then
\begin{equation*}
\pounds_{\xi_Y^{(1)}}\Theta_L = 0
\end{equation*}
or, equivalently,
\begin{equation*}
\pounds_{\xi_Y^{(1)}}\mathcal{L} = 0
\end{equation*}
Therefore,
\begin{equation*}
\xi_Y^{(1)}(L)= - \sum_\mu\frac{d\xi_Y^\mu}{dx^\mu}L
\end{equation*}
\end{prop}

And as a direct consequence of Proposition \ref{prop:liftLieBracket}, we have

\begin{prop}
The symmetries of the Lagrangian form a Lie subalgebra of $\mathfrak{X}(Y)$.
\end{prop}

\begin{thm}
\textbf{(Noether's theorem)}. If $\xi_Y$ is a symmetry of the
Lagrangian, then $\iota_{\xi_Y^{(1)}}\Theta_L$ is a preserved
quantity, which is exact on the boundary.
\end{thm}

\textit{Proof}. We have that
\begin{equation*}
\pounds_{\xi_Y^{(1)}}\Theta_L = -\iota_{\xi_Y^{(1)}}\Omega_L +
d\iota_{\xi_Y^{(1)}}\Theta_L
\end{equation*}

If $\phi$ is a solution of the Euler-Lagrange equations, then

\begin{equation*}
0 = (j^1\phi)^*\pounds_{\xi_Y^{(1)}}\Theta_L =
-(j^1\phi)^*\iota_{\xi_Y^{(1)}}\Omega_L +
(j^1\phi)^*d\iota_{\xi_Y^{(1)}}\Theta_L,
\end{equation*}
where the first term vanishes by the intrinsic Euler-Lagrange equations (see
Proposition \ref{intrinsicEL}).

Finally, to see that it is exact on the boundary, notice that from
the boundary property of a symmetry of the Lagrangian we infer
that $\iota_{{\xi_Y^{(1)}}_{|B}}d\Pi =
-d\iota_{{\xi_Y^{(1)}}_{|B}}\Pi$, and from this we get
\begin{equation*}
i_B^*(\iota_{\xi_Y^{(1)}}\Theta_L) = \iota_{{\xi_Y^{(1)}}_{|B}}d\Pi =
-d\iota_{{\xi_Y^{(1)}}_{|B}}\Pi
\end{equation*}
\hfill$\ \ \ \vrule height 1.5ex width.8ex depth.3ex \medskip$

Observe that without the boundary condition, we obtain that
$(j^1\phi)^*d\iota_{\xi_Y^{(1)}}\Theta_L=0$, but we cannot be sure
that it is exact on the boundary.

The preserved quantity can be written in local coordinates as
\begin{equation*}
\left(\left[L-z^i_{\mu}\frac{\partial L}{\partial
z^i_{\mu}}\right]\xi_X^{\nu}+\frac{\partial L}{\partial
z^i_{\nu}}\xi_Y^i\right)d^nx_{\nu}-\frac{\partial L}{\partial
z^i_{\mu}}\xi_X^{\nu}dy^i\wedge d^{n-1}x_{\mu\nu}
\end{equation*}

\subsection{Noether symmetries}

\begin{defn}
\label{def:SimNoe} A vector field $\xi_Y$ on $Y$ is said to be a
\textbf{Noether symmetry} or a \textbf{divergence symmetry} if
there exists an $n$-form on $Y$ whose pullback $\alpha$ to $Z$
(that must be exact $\alpha=d\beta$ on $B$) verifies
$\pounds_{\xi_Y^{(1)}}\Theta_L - d\alpha\in
\mathcal{I}(\mathcal{C})$, and $\xi_Y^{(1)}$ is tangent to $B$ and
verifies $\pounds_{\xi_Y^{(1)}|_B}\Pi=0$
\end{defn}

The relation $dy^i = \theta^i + z^i_\mu dx^\mu$ allows us to write $\alpha$
locally as follows

\begin{equation*}
\alpha = \alpha_\mu dx^0\wedge\ldots\wedge\widehat{dx^\mu}\wedge\ldots\wedge
dx^n + \theta
\end{equation*}
for $\theta \in \mathcal{I}(\mathcal{C})$ and

\begin{equation*}
d\alpha - \sum_\mu(\frac{\partial \alpha^\mu}{\partial
x^\mu}+z^i_\mu\frac{\partial \alpha^\mu}{\partial y^i})\eta \in
\mathcal{I}(\mathcal{C})
\end{equation*}

Therefore, if we define:

\begin{equation*}
\tilde{F} = F + \sum_\mu\left(\frac{\partial \alpha^\mu}{\partial
x^\mu}+z^i_\mu\frac{\partial \alpha^\mu}{\partial y^i}\right)
\end{equation*}
then

\begin{prop}
If a vector field $\xi_Y$ on $Y$ is a Noether symmetry then $\tilde{F}=0$.
\end{prop}

Similarly,

\begin{prop}
(1) If $\xi_Y$ is a $\pi_{XY}-$projectable Noether symmetry, then
\begin{equation*}
\pounds_{\xi_Y^{(1)}}\Theta_L = d\alpha
\end{equation*}
Furthermore,
\begin{equation*}
\xi_Y^{(1)}(L)= - \sum_\mu\left(\frac{d\xi_Y^\mu}{dx^\mu}L +
\frac{d\alpha^\mu}{dx^\mu}\right)
\end{equation*}
(2) If $dimX=1$ and $\xi_Y$ is a Noether symmetry then
\begin{equation*}
\pounds_{\xi_Y^{(1)}}\Theta_L = d\alpha
\end{equation*}
\end{prop}

\begin{prop}
\label{poi} Noether symmetries form a Lie subalgebra of $\mathfrak{X}(Y)$,
containing the Lie algebra of the symmetries of the Lagrangian.
\end{prop}

\textit{Proof}.
\begin{align*}
\pounds_{[\xi_Y^{(1)},\zeta_Y^{(1)}]}\Theta_L &=
\pounds_{\xi_Y^{(1)}}\pounds_{\zeta_Y^{(1)}}\Theta_L -
\pounds_{\zeta_Y^{(1)}}\pounds_{\xi_Y^{(1)}}\Theta_L =
\pounds_{\xi_Y^{(1)}}(d\alpha_2 +\theta_2) -
\pounds_{\zeta_Y^{(1)}}(d\alpha_1 + \theta_1) \\
&= d(\pounds_{\xi_Y^{(1)}}\alpha_2 - \pounds_{\zeta_Y^{(1)}}\alpha_1) +
\pounds_{\xi_Y^{(1)}}\theta_2 - \pounds_{\zeta_Y^{(1)}}\theta_1
\end{align*}
and $\pounds_{\xi_Y^{(1)}}\theta_2 - \pounds_{\zeta_Y^{(1)}}\theta_1 \in
\mathcal{I}(\mathcal{C})$.

Finally, since $\xi_Y^{(1)}$ and $\zeta_Y^{(1)}$ are tangent to
$B$, then $[\xi_Y^{(1)},\zeta_Y^{(1)}]$ is also tangent to $B$. We
also have that
$\pounds_{[\xi_Y^{(1)},\zeta_Y^{(1)}]_{|B}}\Pi=\pounds_{{\xi_Y^{(1)}}_{|B}}
\pounds_{{\zeta_Y^{(1)}}_{|B}}\Pi-\pounds_{{\zeta_Y^{(1)}}
_{|B}}\pounds_{{\xi_Y^{(1)}}_{|B}}\Pi=0$ on $B$, and that if
$\alpha_1$ and $\alpha_2$ are exact on $B$, so is
$\pounds_{{\xi_Y^{(1)}}_{|B}}\alpha_2 -
\pounds_{{\zeta_Y^{(1)}}_{|B}}\alpha_1$. $\hfill \ \ \ \vrule
height 1.5ex width.8ex depth.3ex \medskip$

The following Noether's theorem

\begin{thm}
\textbf{(Noether's theorem)}. If $\xi_Y$ is a Noether symmetry,
then $\iota_{\xi_Y^{(1)}}\Theta_L -\alpha$ is a preserved quantity
which is exact on the boundary.
\end{thm}

is proved analogously as we did for the symmetries of the
Lagrangian. We just remark a slight modification introduced to see
that it is exact on the boundary:
\begin{equation*}
i_B^*(\iota_{\xi_Y^{(1)}}\Theta_L-\alpha) = \iota_{{\xi_Y^{(1)}}_{|B}}d\Pi
-d\beta= d(-\iota_{{\xi_Y^{(1)}}_{|B}}\Pi-\beta)
\end{equation*}

\subsection{Cartan symmetries}

\begin{defn}
\label{def:SimCar} A vector field $\xi_Z$ on $Z$ is said to be a
\textbf{Cartan symmetry} if its flow preserves the differential
ideal $\mathcal{I}(\mathcal{C})$ (in other words,
$\psi_{Z,t}^*\theta^i\in\mathcal{I}(\mathcal{C})$, or locally,
$\pounds_{\xi_Z}\mathcal{I}(\mathcal{C})\subseteq\mathcal{I}
(\mathcal{C}))$, and there exists an $n$-form $\alpha$ on $Z$
(that must be exact $\alpha=d\beta$ on $B$) such that
$\pounds_{\xi_Z}\Theta_L - d\alpha \in \mathcal{I}(\mathcal{C})$,
$\xi_Z$ is tangent to $B$ and verifies $\pounds_{\xi_Z|_B}\Pi=0$.
\end{defn}

If $\xi_Y$ is a Noether symmetry, then its 1-jet prolongation is a Cartan
symmetry. Conversely, it is obvious that a projectable Cartan symmetry is
the 1-jet prolongation of its projection, which is therefore a Noether
symmetry.

\begin{prop}
The Cartan symmetries form a subalgebra of $\mathfrak{X}(Z)$.
\end{prop}

We also have

\begin{thm}
\textbf{(Noether's theorem)}. If $\xi_Z$ is a Cartan symmetry,
then $\iota_{\xi_Z}\Theta_L-\alpha$ is a preserved quantity which
is exact on the boundary.
\end{thm}

We also have the obvious relations between the different types of symmetries
that we have exposed here. Every symmetry of the Lagrangian is a Noether
symmetry. And the 1-jet prolongation of any Noether symmetry is a Cartan
symmetry.

And finally,

\begin{prop}
The flow of Cartan symmetries maps solutions of the Euler-Lagrange equations
into solutions of the Euler-Lagrange equations.
\end{prop}

\textit{Proof}. Let $\psi_Z^t$ be the flow of a Cartan symmetry $\xi_Z$.

For any section $\phi\in\Gamma(\pi)$, we can locally define
\begin{equation*}
\psi_{\phi,X}^t := \pi_{XZ}\circ\psi_Z^t\circ j^1\phi
\end{equation*}
$\psi_{\phi,X}^0 = Id_X$, whence for small $t^{\prime}s$, $\psi_{\phi,X}^t$
is a diffeomorphism. Analogously, we define
\begin{equation*}
\psi_{\phi,Y}^t := \pi_{YZ}\circ\psi_Z^t\circ j^1\phi\circ\pi_{XY}
\end{equation*}
With the same argument we see that for small $t^{\prime}s$, $\psi_{\phi,Y}^t$
is as well a diffeomorphism.

If $\phi$ is a solution of the Euler-Lagrange equation, then the flow
transforms $\phi$ into
\begin{equation*}
\psi_{\phi,Y}^t \circ\phi\circ(\psi_{\phi,X}^t)^{-1}
\end{equation*}
Now, for $\theta\in\mathcal{C}$,
\begin{equation*}
(\psi_Z^t\circ j^1\phi\circ(\psi_{\phi,X}^t)^{-1})^*\theta =
((\psi_{\phi,X}^t)^{-1})^*(j^1\phi)^*(\psi_Z^t)^*\theta = 0
\end{equation*}
as $\xi_Z$ is a Cartan symmetry. This means that $\psi_Z^t\circ
j^1\phi\circ(\psi_{\phi,X}^t)^{-1}$ is the 1-jet prolongation of its
projection to $Y$,
\begin{equation*}
\pi_{YZ}\circ\psi_Z^t\circ j^1\phi\circ(\psi_{\phi,X}^t)^{-1} =
\psi_{\phi,Y}^t \circ\phi\circ(\psi_{\phi,X}^t)^{-1}
\end{equation*}
In other words,
\begin{equation*}
j^1(\psi_{\phi,Y}^t \circ\phi\circ(\psi_{\phi,X}^t)^{-1})= \psi_Z^t\circ
j^1\phi\circ(\psi_{\phi,X}^t)^{-1}
\end{equation*}
Now we need to see that the transformed solution verifies the
Euler-Lagrange equations. The preceding equation shows that, being
the symmetry tangent to $B$, the boundary condition will be
satisfied.

In addition, for every compact ($n+1$)-dimensional submanifold $C$, and
every vertical vector field $\xi \in \mathcal{V}(\pi)$, which annihilates at
$\partial C$ (and therefore, so does $\xi^{(1)})$,
\begin{align*}
&\int_{(\psi_{\phi,X}^t)(C)}(j^1(\psi_{\phi,Y}^t
\circ\phi\circ(\psi_{\phi,X}^t)^{-1}))^*\pounds_{\xi^{(1)}}\Theta_L \\
= &\int_{(\psi_{\phi,X}^t)(C)}(\psi_Z^t\circ
j^1\phi\circ(\psi_{\phi,X}^t)^{-1})^*\pounds_{\xi^{(1)}}\Theta_L \\
= &\int_C(\psi_Z^t\circ j^1\phi)^*\pounds_{\xi^{(1)}}\Theta_L = \int_C
(j^1\phi)^*(\psi_Z^t)^*\pounds_{\xi^{(1)}}\Theta_L
\end{align*}
by means of a change of variable. The annihilation of the preceding
expression is infinitesimally equivalent to the annihilation of
\begin{align*}
&\int_C (j^1\phi)^* \pounds_{\xi_Z}\pounds_{\xi^{(1)}}\Theta_L \\
= &\int_C (j^1\phi)^* \pounds_{[\xi_Z,\xi^{(1)}]}\Theta_L - \int_C
(j^1\phi)^* \pounds_{\xi^{(1)}}\pounds_{\xi_Z}\Theta_L
\end{align*}
and we conclude by seeing that
\begin{equation*}
\int_C (j^1\phi)^* \pounds_{[\xi_Z,\xi^{(1)}]}\Theta_L = - \int_C(j^1\phi)^*
\iota_{[\xi_Z,\xi^{(1)}]}\Omega_L + \int_C (j^1\phi)^*
d\iota_{[\xi_Z,\xi^{(1)}]}\Theta_L = 0
\end{equation*}
where the first term vanishes because $\phi$ is a solution of Euler-Lagrange
equations, and second term vanishes due to the boundary condition on $\xi$;
and
\begin{align*}
\int_C (j^1\phi)^* \pounds_{\xi^{(1)}}\pounds_{\xi_Z}\Theta_L &= \int_C
(j^1\phi)^* \pounds_{\xi^{(1)}}(d\alpha + \theta) \\
&= \int_{\partial C} (j^1\phi)^*\pounds_{\xi^{(1)}}\alpha + \int_C
(j^1\phi)^*\pounds_{\xi^{(1)}}\theta = 0
\end{align*}
where the first term vanishes again by the boundary condition on $\xi$.
\hfill$\ \ \ \vrule height 1.5ex width.8ex depth.3ex \medskip$

\subsection{Symmetries for the De Donder equations}

In the discussion of the preceding section, we have used on Noether's
theorem the fact that, for a solution $\phi$ of the Euler-Lagrange
equations, we have
\begin{equation*}
(j^1\phi)^*\theta=0
\end{equation*}
for elements $\theta$ of the differential ideal generated by the contact
forms. However, this result is no longer true for general solutions of the
De Donder equations (more specifically, when the Lagrangian is not regular).
In other words, if $\sigma$ is a solution of the De Donder equations, then
\textbf{not necessarily}
\begin{equation*}
\sigma^*\theta=0
\end{equation*}
for $\theta\in\mathcal{I}(\mathcal{C})$.

Therefore, our definition of symmetry must be more restrictive when we are
dealing with solutions of the De Donder equations.

\begin{defn}
A \textbf{preserved quantity for the De Donder equations} is a
$n$-form $\alpha$ on $Z$ such that $\sigma^*d\alpha=0$ for every
solution $\sigma$ of the De Donder equations. If $\alpha$ is a
preserved quantity, then $\tilde{\alpha}$ is called its associated
\textbf{momentum}.
\end{defn}

Also note that if $\alpha$ is a preserved quantity and $\beta$ is a closed $n
$-form, then $\alpha+\beta$ is also a preserved quantity.

From equation \eqref{eq:TsigmaEqH} we can easily deduce the following.

\begin{prop}
Let \hbox{\rm \textbf{h}} be a solution of the connection equation
\eqref{eq:hdedonder}. Then $\alpha$ is a preserved quantity for
the De Donder equations if and only if $d\alpha$ is annihilated by
any $n$ horizontal tangent vectors at each point.
\end{prop}

\begin{defn}
We have the following definitions of symmetries for the De Donder equations:

(1) A vector field $\xi_Y$ on $Y$ is said to be a \textbf{symmetry of the
Lagrangian}, or a \textbf{variational symmetry} if
\begin{equation*}
\pounds_{\xi_Y^{(1)}}\Theta_L = 0
\end{equation*}
and $\xi_Y^{(1)}$ is tangent to $B$ and verifies
$\pounds_{\xi_Y^{(1)}|_B}\Pi=0$.

(2) A vector field $\xi_Y$ on $Y$ is said to be a \textbf{Noether symmetry},
or a \textbf{divergence symmetry} if
\begin{equation*}
\pounds_{\xi_Y^{(1)}|_B}\Theta_L = d\alpha
\end{equation*}
where $\alpha$ is the pullback to $Z$ of a $n$-form on $Y$ (that
must be exact $\alpha=d\beta$ on $B$), $\xi_Y^{(1)}$ is tangent to
$B$ and verifies $\pounds_{{\xi_Y^{(1)}}_{|B}}\Pi=0$.

(3) A vector field $\xi_Z$ on $Z$ is a \textbf{Cartan symmetry} if
\begin{equation*}
\pounds_{\xi_Z}\Theta_L = d\alpha
\end{equation*}
where $\alpha$ is a $n$-form on $Z$ (that is exact $\alpha=d\beta$ on $B$)
(or, equivalently, if there is a $n$-form $\alpha^{\prime}$ such that
\begin{equation*}
\iota_{\xi_Z}\Omega_L = d\alpha^{\prime}
\end{equation*}
we can put $\alpha^{\prime}= \alpha +\iota_{\xi_Z}\Theta_L)$, in other
words, if $\xi_Z$ is a Hamiltonian vector field), $\xi_Z$ is tangent to $B$
and verifies $\pounds_{\xi_Z|_B}\Pi=0$.
\end{defn}

There is an obvious relation between these types of symmetries,
completely analogous to those between the symmetries for the
Euler-Lagrange equations, that is, a symmetry of the Lagrangian
(resp. a Noether symmetry, Cartan symmetry) for the De Donder
equations is a symmetry of the Lagrangian (resp. a Noether
symmetry, Cartan symmetry) for the Euler-Lagrange equations.

Also note that a small computation shows that, in the case of of a Noether
symmetry, $\alpha$ must be necessarily the pullback of a semibasic $n$-form
on $Y$, locally expressed by
\begin{equation*}
\alpha(x,y,z)= \alpha^\mu(x,y) {d^nx_\mu}
\end{equation*}

Note from the definition of Cartan symmetry that using Cartan's formula we
obtain
\begin{equation*}
\iota_{\xi_Z}\Omega_L = d(\iota_{\xi_Z}\Theta_L+\alpha)
\end{equation*}
and therefore $d\iota_{\xi_Z}\Omega_L=0$, from where
\begin{equation*}
\pounds_{\xi_Z}\Omega_L=0
\end{equation*}

\begin{thm}
\textbf{(Noether's theorem)} If $\xi_Z$ is a Cartan symmetry, such
that $\pounds_{\xi_Z}\Theta_L = d\alpha$, then
$\iota_{\xi_Z}\Theta_L-\alpha$ is a preserved quantity which is
exact on the boundary.
\end{thm}

For the proof, repeat that of the Noether's theorem for Euler-Lagrange
equations, where
\begin{equation*}
\pounds_{\xi_Z}\Theta_L - d\alpha
\end{equation*}
now vanishes by definition.

In the case of a regular Lagrangian, and $n>0$, a computation
similar to that in Proposition \ref{prop:OLmultisymp} for the
expression $\pounds_{\xi_Z}\Omega_L=0$ produces two terms
\begin{equation*}
\frac{\partial^2 L}{\partial z^i_{\mu} \partial z^j_{\nu}}\frac{\partial
\xi_X^{\kappa}}{\partial y^k}\,dz^j_{\nu}\wedge dy^i\wedge dy^k\wedge
d^{n-1}x_{\mu\kappa}
\end{equation*}
and
\begin{equation*}
\frac{\partial^2 L}{\partial z^i_{\mu} \partial z^j_{\nu}}\frac{\partial
\xi_X^{\kappa}}{\partial z^k_{\lambda}}\,dz^j_{\nu}\wedge dy^i\wedge
dz^k_{\lambda}\wedge d^{n-1}x_{\mu\kappa},
\end{equation*}
which show that Cartan symmetries are automatically projectable. For this
reason, and because projectable symmetries are typical of examples coming
from Physics, we shall emphasize the role of vector fields which are
projectable onto $X$.

Also note that the symmetries of Cartan preserve the horizontal subspaces
for the connection formalism.

\begin{prop}
Assume that $L$ is regular. If $\xi_Z$ is a Cartan symmetry for the De
Donder equations then $\xi_Z$ preserves the horizontal distribution of any
solution $\Gamma$ satisfying \eqref{eq:hdedonder}.
\end{prop}

\textit{Proof}. Since $\xi_Z$ is a Cartan symmetry then
$\pounds_{\xi_Z}\Omega_L=0$. Therefore
\begin{equation*}
\pounds_{\xi_Z}i_{\hbox{\textbf{h}}}\Omega_L=0
\end{equation*}
for any solution $\Gamma$ of \eqref{eq:hdedonder} with horizontal
projector $\hbox{\textbf{h}}$ .

Hence,
\begin{eqnarray*}
0&=&\left( \pounds_{\xi_Z}i_{\hbox{\bf h}}\Omega_L\right) (\xi_0, \xi_1,
\ldots, \xi_n) \\
&=& \xi_Z\left( i_{\hbox{\bf h}}\Omega_L(\xi_0, \xi_1, \ldots, \xi_n)\right)
-\sum_{a=0}^{n}i_{\hbox{\bf h}}\Omega_L(\xi_1, \ldots, [\xi_Z,
\xi_{a}],\ldots, \xi_n) \\
&=&\sum_{b=0}^{n}\xi_Z\left( i_{\hbox{\bf
h}(\xi_{b})}\Omega_L(\xi_0, \ldots, \widehat{\xi_{b}}, \ldots, \xi_n)\right)
\\
&&-\sum_{%
\begin{array}{l}
a,b=0 \\
a\not=b%
\end{array}%
}^{n} (-1)^{b} i_{\hbox{\bf h}(\xi_b)}\Omega_L(\xi_0, \ldots, [\xi_Z,
\xi_a],\ldots, \widehat{\xi_b}, \ldots, \xi_n) \\
&&- \sum_{b=0}^n (-1)^{b+1} i_{\hbox{\bf h}[\xi_Z, \xi_b]}\Omega_L(\xi_1,
\ldots, \widehat{\xi_{b}}, \ldots, \xi_n) \\
&=& \sum_{b=0}^n\left(\pounds_{\xi_Z}i_{\hbox{\bf
h}(\xi_b)}\Omega_L\right) (\xi_0, \ldots, \widehat{\xi_{b}}, \ldots, \xi_n)-
\sum_{b=0}^{n} i_{\hbox{\bf h}[\xi_Z, \xi_b]}\Omega_L(\xi_1, \ldots,
\widehat{\xi_b}, \ldots, \xi_{n})
\end{eqnarray*}

\textsl{First case} $(n>1)$. Since $\Omega_L$ is multisymplectic
and $\pounds_{\xi_Z}\Omega_L=0$ we deduce that
\begin{equation*}
[\xi_Z, \hbox{\bf h}(\xi)]=\hbox{\bf h}[\xi_Z, \xi]\qquad\forall \xi\in
\mathfrak{X}(Z),
\end{equation*}
which implies that the horizontal distribution associated to $\Gamma$ is
\textrm{\textbf{h}}-invariant

\textsl{Second case} $(n=1)$. Taking $\xi=\frac{\partial}{\partial
t}$ then $\hbox{\bf h}(\xi)=\xi_L$ is the Reeb vector field of the
cosymplectic structure $(dt, \Omega_L)$ (being $L$ regular).
Moreover, with the notation $d_t = \frac{d}{dt}$, we have
\begin{equation*}
\hbox{\bf h}[\xi_Z, \frac{\partial}{\partial t}]=-d_t\tau\xi_L,\quad
dt([\xi_Z, \xi_L)]=d_t\tau
\end{equation*}
where $dt(\xi_Z)=\tau$. Therefore,
\begin{equation*}
dt([\xi_Z, \xi_L]-\hbox{\bf h}[\xi_Z, \frac{\partial}{\partial t}])=0
\end{equation*}
Since $(\Omega_L,dt)$ is a cosymplectic structure, we deduce that
\begin{equation}  \label{ass}
[\xi_Z, \xi_L]=\hbox{\bf h}[\xi_Z, \frac{\partial}{\partial t}]=-d_t\tau
\xi_L,
\end{equation}
which implies the invariance of the distribution $\langle \xi_L\rangle$.
Observe that equation (\ref{ass}) is the classical definition of dynamical
symmetry for time-dependent mechanical systems.

Moreover, the boundary conditions are fulfilled since $\xi_Z$ preserves $B$.
\hfill$\ \ \ \vrule height 1.5ex width.8ex depth.3ex \medskip$

Finally, we shall justify that these symmetries are really symmetries, in
the sense that they transform solutions of the De Donder equations into new
solutions of the De Donder equations.

\begin{thm}
\label{thm:DeDonderPreservesSols} The flow of Cartan symmetries maps
solutions of the De Donder equations into solutions of the De Donder
equations.
\end{thm}

\textit{Proof}. If $\sigma$ is a solution of the De Donder
equation, and $\xi\in\mathfrak{X}(Z)$ is a Cartan symmetry having
flow $\phi_t$, and we define for each $t$
\begin{equation*}
\psi_t := \pi_{XZ}\circ\phi_t\circ\sigma
\end{equation*}
then we claim that $\phi_t\circ\sigma\circ\psi_t^{-1}$ is a solution of the
De Donder equations. Being the symmetry tangent to $B$, the boundary
condition will be automatically satisfied.

As $\psi_0 = Id$, $\psi_t$ is a local diffeomorphism for small
$t^{\prime}s$. Therefore, $\phi_t\circ\sigma\circ\psi_t^{-1}$
makes sense for small $t^{\prime}s$. In order to prove
\begin{equation*}
(\phi_t\circ\sigma\circ\psi_t^{-1})^*(\iota_X\Omega_L)=
(\psi_t^{-1})^*\sigma^*\phi_t^*(\iota_X\Omega_L) = 0
\end{equation*}
it suffices to see that
\begin{equation*}
\sigma^*\phi_t^*(\iota_X\Omega_L) = 0
\end{equation*}
for $t$ in a neighbourhood of 0. Now for $t=0$, this equation reduces to the
De Donder equation, therefore, it suffices to see that
\begin{equation*}
\sigma^*(\pounds_\xi\iota_X\Omega_L) = 0
\end{equation*}

Using again the De Donder equation,
\begin{equation*}
0 = \sigma^*(\iota_{[\xi,X]}\Omega_L) =
\sigma^*(\pounds_\xi\iota_X\Omega_L)-\sigma^*(\iota_X \pounds_\xi\Omega_L)
\end{equation*}
But
\begin{equation*}
\pounds_\xi\Omega_L = -d\pounds_\xi\Theta_L = -dd\alpha = 0
\end{equation*}
which completes the proof. \hfill$\ \ \ \vrule height 1.5ex width.8ex
depth.3ex \medskip$

\subsection{Symmetries for singular Lagrangian systems}

For the singular Lagrangian case (described in section
\ref{singularLagrSection}), we consider diffeomorphisms $\Psi:
Z\rightarrow Z$ which preserve the Poincar\'e-Cartan $(n+2)$-form
$\Omega_L$ (i.e. $\phi^*\Omega_L=\Omega_L)$ and are
$\pi_{XZ}$-projectable.0

\begin{prop}
\label{propo1} If the diffeomorphism $\Psi: Z\longrightarrow Z$
verifying $\Psi(B)\subseteq B$ preserves the $(n+2)$-form
$\Omega_L$ and it is $\pi_{XZ} $-projectable, then it restricts to
a diffeomorphism $\Psi_{a}:Z_{a}\longrightarrow Z_a$, where $Z_a$
is the $a$-ry constraint submanifold. Therefore, $\Psi$ restricts
to a diffeomorphism $\Psi_{f}: {Z_f}\longrightarrow {Z_f}$.
\end{prop}

\textit{Proof}. If $z\in Z_1$ then there exists a linear mapping $\hbox{\bf
h}_{z} : T_z Z \longrightarrow T_zZ$ such that $\hbox{\bf h}^2_z = \hbox{\bf
h}_z$, $\ker \hbox{\bf h}_z = (\mathcal{V}\pi_{X Z})_z$ and
\begin{equation*}
i_{\hbox{\bf h}_z} \Omega_{L}(z) = n \Omega_{L}(z)
\end{equation*}
Consider the mapping
\begin{equation*}
\hbox{\bf h}_{\Psi(z)}= T_z\Psi\circ \hbox{\bf h}_z\circ T_{\Psi(z)}
\Psi^{-1}
\end{equation*}
It is clear that $\hbox{\bf h}_{\Psi(z)}$ is linear and $\hbox{\bf
h}_{\Psi_z}^2=\hbox{\bf h}_{\Psi(z)}$ Moreover, since $\Psi$ is
$\pi_{XZ}$ projectable then $\ker \hbox{\bf h}_{\Psi(z)} =
(\mathcal{V}\pi_{X Z})_{\Psi(z)}$. Finally, since
$\Psi^*\Omega_L=\Omega_L$ then
\begin{equation*}
i_{\hbox{\bf h}_{\Psi(z)}} \Omega_{L}(\Psi(z)) = n \Omega_{L}(\Psi(z))
\end{equation*}
Therefore, if $z\in Z_1$ then $\Psi(z)\in Z_1$. Thus, the proposition is
true if $a=1$. Now, suppose that the proposition is true for $a=l$ and we
shall prove that it is also true for $a=l+1$.

Let $z$ be a point in $Z_{l+1}$ then there exists $\hbox{\bf
h}_{z} : T_zZ \longrightarrow T_z Z_l$ linear such that $\hbox{\bf
h}^2_z = \hbox{\bf h}_z$, $\ker \hbox{\bf h}_z =
(\mathcal{V}\pi_{X Z})_z$ and $i_{\hbox{\bf h}_z} \Omega_{L}(z) =
n \Omega_{L}(z)$. Since $\Psi(Z_l)\subseteq Z_l$ and $\Psi$ is a
diffeomorphism, then $T_z\Psi(T_zZ_l)\subseteq T_{\Psi(z)}Z_l$.
Thus, $\hbox{\textbf{h}} _{\Psi(z)}: T_{\Psi(z)} Z\longrightarrow
T_{\Psi(z)} Z_l$ and $\Psi(z)\in Z_{l+1}$. We also have that
$\hbox{\textbf{h}} (TB_f)\subseteq TB_f$. \hfill$\ \ \ \vrule
height 1.5ex width.8ex depth.3ex \medskip$

\begin{cor}
Let $\xi_Z$ be a $\pi_{XZ}$-projectable vector field on $X$ such
that $\pounds_{\xi_Z}\Omega_L=0$, then $\xi_Z$ is tangent to $Z_f$
\end{cor}

\begin{cor}
A Cartan symmetry which is $\pi_{XZ}$-projectable is tangent to $Z_f$
\end{cor}

Proposition \ref{propo1} motivates the introduction of a more
general class of symmetries. If $Z_f$ is the final constraint
submanifold and $i_{f1}: Z_f\longrightarrow Z$ is the canonical
immersion then we may consider the $(n+2)$-form
$\Omega_{Z_f}=i_{f1}^* \Omega_L$, the $(n+1)$-form
$\Theta_{Z_f}=i_{f1}^*\Theta_L$ and now analyze a new kind of
symmetries.

\begin{defn}
A Cartan symmetry for the system $(Z_f, \Omega_{Z_f})$ is a vector
field on $Z_f$ tangent to $Z_f\cap B$ such that
$\pounds_{\xi_{Z_f}}\Theta_{Z_f}=d\alpha_{Z_f}$, for some
$\alpha_{Z_f}\in \Lambda^n Z_f$.
\end{defn}

If it is clear that if $\xi_Z$ is a Cartan symmetry of the De Donder
equations then using Proposition \ref{propo1} we deduce that $X_{|Z_f}$ is a
Cartan symmetry for the system $(Z_f, \Omega_{Z_f})$.

\subsection{Symmetries in the Hamiltonian formalism}

We can define as well symmetries in the Hamiltonian formalism as we did for
the De Donder equation, which are closely related by the equivalence theorem.

\begin{defn}
Given a Hamiltonian $h$, we have the following definitions of symmetries for
the Hamilton equations:

(1) A vector field $\xi_Y$ on $Y$ is said to be a \textbf{Noether symmetry},
or a \textbf{divergence symmetry} if there exists a semibasic $n$-form on $Y$
whose pullback $\alpha$ to $\Lambda^{n+1}_2Y$ (which is exact $\alpha=d\beta$
on $B^*$) and verifies

(a) The $\alpha$-lift of $\xi_Y$ to $\Lambda^{n+1}_2Y$ is projectable to a
vector field $\xi_Y^{(1*)}$

(b) $\pounds_{\xi_Y^{(1*)}}\Theta_h = d\alpha$, $\xi_Y^{(1*)}$ is also
tangent to $B^*$ and verifies $\pounds_{\xi_Y^{(1*)}|_{B^*}}\pi_{XZ^*}=0$.

(2) A vector field $\xi_Z$ on $Z^*$ is a \textbf{Cartan symmetry} if
\begin{equation*}
\pounds_{\xi_Z}\Theta_h = d\alpha
\end{equation*}
where $\alpha$ is an $n$-form on $Z^*$ (which is exact
$\alpha=d\beta$ on $B^*$), $\xi_Z$ is also tangent to $B^*$ and
verifies $\pounds_{\xi_Z|_{B^*}}\pi_{XZ^*}=0$
\end{defn}

As usual, Noether symmetries induce Cartan symmetries on $Z^*$.

Supose that $\xi$ is a vector field on $Y$, and $\alpha$ is the
pull-back to $\Lambda^{n+1}_2 Y$ of a $\pi_{XY}$-semibasic form on
$Y$. If the $\alpha$-lift of $\xi$ to $\Lambda^{n+1}_2 Y$ projects
onto a vector field on $Z^*$ then $\xi_Y$ is a Noether symmetry.

\begin{thm}
\textbf{(Noether's theorem)} If $\xi_{Z^*}$ is a Cartan symmetry,
such that $\pounds_{\xi_{Z^*}}\Theta_h = d\alpha$, then
$\sigma^*d(\iota_{\xi_{Z^*}} \Theta_h-\alpha)=0$ for every
solution $\sigma$ of the Hamilton equations. Furthermore,
$\iota_{\xi_{Z^*}}\Theta_h-\alpha$ is exact on $\partial Z^*$.
\end{thm}

This theorem is entirely analogous to that of the Noether's theorem for De
Donder equations.

Finally, we shall justify that these are real symmetries, in the sense that
they transform solutions of the Hamilton equations into new solutions of the
Hamilton equations.

\begin{thm}
The flow of Cartan symmetries maps solutions of the Hamilton equations into
solutions of the Hamilton equations.
\end{thm}

The proof is identical to that given for the De Donder equations
in theorem \ref{thm:DeDonderPreservesSols}.

\subsection{The Legendre transformation and the symmetries}

In this section we shall finally relate the symmetries of the De Donder
equations to the symmetries of the Hamiltonian formalism, under the
assumption of hyperregularity. Within this section, we shall assume that $L$
is a hyperregular Lagrangian.

\begin{prop}
If $\xi_Z$ is a Cartan symmetry for the De Donder equation, then
$Tleg_L(\xi_Z)$ is a Cartan symmetry for the Hamilton equations.
The converse is also true.
\end{prop}

\textit{Proof}. If we just apply $(leg_L^{-1})^*$ to the Cartan condition
for the De Donder equations we get the Cartan condition for the Hamilton
equations:
\begin{equation*}
0=(leg_L^{-1})^*(\pounds_{\xi_Z}\Theta_L-d\alpha) =
\pounds_{Tleg_L(\xi_Z)}(leg_L^{-1})^*\Theta_L -d\tilde{\alpha}=
\pounds_{Tleg_L(\xi_Z)}\Theta_h-d\tilde{\alpha}.
\end{equation*}
where $leg_L^*\tilde{\alpha}=\alpha$. Boundary preservation is trivial,
because of the way $B^*$ has been defined, and the compatibility with the
Legendre map.\hfill $\ \ \ \vrule height 1.5ex width.8ex depth.3ex \medskip$

In a similar way we prove the following result

\begin{lem}
If $\xi_Y$ is a Noether symmetry for the De Donder equation, such
that $\pounds_{\xi_Y^{(1)}}\Theta_L-d\alpha$, then
$TLeg_L(\xi_Y^{(1)})$ is the $\alpha$-lift of $\xi_Y$.
\end{lem}

From which we can obtain

\begin{prop}
Every Noether symmetry for the De Donder equations is a Noether symmetry for
the Hamilton equations. The converse is also true.
\end{prop}

\textit{Proof}. We have that
\begin{equation*}
Tleg_L(\xi_Y^{(1)}) = (T\mu\circ TLeg_L)(\xi_Y^{(1)}) \;
\end{equation*}
therefore the $\alpha$-lift of $\xi_Y$ projects onto
$Tleg_L(\xi_Y^{(1)})$ on $Z^*$, and as $\xi_Y^{(1)}$ is a Cartan
symmetry, its image $Tleg_L(\xi_Y^{(1)})$ also verifies the Cartan
condition (as $\pounds_{Tleg_L(\xi_Y^{(1)})}\Theta_h
-d\tilde{\alpha} =\pounds_{Tleg_L(\xi_Y^{(1)})}(leg_L^{-1})^*
\Theta_L -d(leg_L^{-1})^*\alpha =
(leg_L^{-1})^*(\pounds_{\xi_Y^{(1)}}\Theta_L-d\alpha)=0$). As
usual, boundary conditions are trivially fulfilled. \hfill$\ \ \
\vrule height 1.5ex width.8ex depth.3ex \medskip$

\subsection{Symmetries in the Hamiltonian formalism for almost regular
Lagrangians}

On the final constraint submanifold $M_f$ we have the following definition.

\begin{defn}
A Cartan symmetry for the system $(M_f, \Omega_{M_f})$ is a vector
field on $M_f$ tangent to $M_f\cap B^*$ such that
$\pounds_{\xi_{M_f}}\Theta_{M_f}=d\alpha_{M_f}$, for some
$\alpha_{M_f}\in \Lambda^n M_f$.
\end{defn}

\begin{prop}
If $\xi_{M_f}$ is a Cartan symmetry of $(M_f, \Omega_{M_f})$ then any vector
field $\xi_{Z_f}$, such that $T leg_f (\xi_{Z_f})=\xi_{M_f}$ is a Cartan
symmetry of $(Z_f, \Omega_{Z_f})$.
\end{prop}

\subsection{Symmetries on the Cauchy data space}

The symmetries of presymplectic systems were exhaustively studied
by two of the authors in \cite{Le-Da1,Le-Da2} (see also
\cite{EMR3,GP}). In \cite{Le-Da1} (Proposition 4.1 and Corollary
4.1) it was proved that for a general presymplectic system given
by $(M, \omega, \Lambda)$, where $M$ is a differentiable manifold,
$\omega$ a closed 2-form and $\Lambda$ a closed 1-form, a vector
field $\xi$ such that
\begin{equation*}
i_{\xi} \omega=dG,
\end{equation*}
where $G: M\rightarrow \mathbb{R}$, is a Cartan symmetry of the
presymplectic system (for $\Lambda=0$). In fact, given a solution $U$ for
the presymplectic system, since $U$ satisfies $\iota_U \, \omega = 0$, then
we have
\begin{equation*}
0 = \iota_U \iota_\xi \omega = U(G).
\end{equation*}

The following proposition explains the relationship between Cartan
symmetries of the De Donder equations and Cartan symmetries for the
presymplectic system $(\tilde{Z}, \widetilde{\Omega})$.

\begin{prop}
Let $\xi_Z$ be a Cartan symetry of the De Donder equations, that
is, $\pounds_{\xi_Z}\Theta_L = d\alpha$. Then the induced vector
field $\xi_{\tilde{Z}}$ in $\tilde{Z}$, defined by
$\xi_{\tilde{Z}}(\gamma)=\xi_Z\circ \gamma$, is a Cartan symmetry
of the presymplectic system $(\tilde{Z}, \widetilde{\Omega_L})$.
\end{prop}

\textbf{Proof:} If $\pounds_{\xi_Z}\Theta_L = d\alpha$, then
\begin{equation*}
i_{\xi_Z}\Omega_L=d(\alpha-i_{\xi_Z}\Theta_L)
\end{equation*}
that is, $\xi_Z$ is a Hamiltonian vector field for the $n$ form
$\beta=\alpha-i_{\xi_Z}\Theta_L$. Then from Proposition 4.8 we
have
\begin{equation*}
i_{\widetilde{\xi_Z}}\widetilde{\Omega_L}=d\tilde{\beta}
\end{equation*}
which shows that $\widetilde{\xi_Z}$ is a Cartan symmetry for the
presymplectic system $(\tilde{Z}, \widetilde{\Omega_L})$. \hfill
$\ \ \ \vrule height 1.5ex width.8ex depth.3ex \medskip$

\subsection{Conservation of preserved quantities along solutions}

\begin{prop}
If $\alpha$ is a preserved quantity, and $c_{\tilde{Z}}$ is a
solution of the De Donder equations \eqref{DeDonderTilde} such
that its projection $c_{\tilde{X}}$ to $\tilde{X}$ splits $X$ and
$\alpha$ is exact on $B \subseteq
\partial Z$ $(\alpha_{_{|B}}=d\beta)$, then $\tilde{\alpha}\circ c_{\tilde{Z}}$
is constant; in other words, the following function
\begin{equation*}
\int_M c_{\tilde{Z}}(t)^*\alpha - \int_{\partial M} c_{\tilde{Z}}(t)^*\beta
\end{equation*}
is constant with respect to $t$.
\end{prop}

\textit{Proof}. Pick $t_1<t_2$ two real numbers in the domain of
the solution curve, and let us denote by $M_1=c_{\tilde{X}}(t_1)$
and $M_2=c_{\tilde{X}}(t_2)$. As $c_{\tilde{X}}$ splits $X$, then
we can consider the piece $U \subseteq X$ identified with
$M\times[t_1,t_2]$, $M_1$ is identified with $M\times t_1$, $M_2$
is identified with $M\times t_2$, and let us denote by $V$ the
boundary piece corresponding to $\partial M \times [t_1,t_2]$. On
view of \eqref{eq:liftcurve}, then

\begin{equation*}
c_{\tilde{Z}}(t)^*d\alpha=0 \quad \hbox{for all}\quad t
\end{equation*}

whence if we integrate and apply Stoke's theorem, we get

\begin{equation*}
0 = \int_{M_2} c_{\tilde{Z}}(t)^*\alpha + \int_V c_{\tilde{Z}}(t)^*\alpha -
\int_{M_1} c_{\tilde{Z}}(t)^*\alpha
\end{equation*}

If we put $\alpha = d\beta$ on $B$, then $0 = \partial \partial U = \partial
M_2 +\partial V - \partial M_1$, whence applying Stoke's theorem again, we
obtain

\begin{equation*}
\int_V c_{\tilde{Z}}(t)^*\alpha = \int_{\partial V}
c_{\tilde{Z}}(t)^*\beta = \int_{\partial M_1}
c_{\tilde{Z}}(t)^*\beta - \int_{\partial M_2}
c_{\tilde{Z}}(t)^*\beta.
\end{equation*}
\hfill $\ \ \ \vrule height 1.5ex width.8ex depth.3ex \medskip$

\begin{cor}
In particular, if $\xi_Y$ is a symmetry of the Lagrangian for the
De Donder equations , then the preceding formula can be applied to
the preserved quantity $\iota_{\xi_Y^{(1)}}\Theta_L$ and we get
that the following integral is preserved along solutions of the De
Donder equations \eqref{DeDonderTilde} such that its projection
$c_{\tilde{X}}$ to $\tilde{X}$ splits $X$
\begin{equation*}
\int_M c_{\tilde{Z}}(t)^*\iota_{\xi_Y^{(1)}}\Theta_L +
\int_{\partial M} c_{\tilde{Z}}(t)^*\iota_{\xi_Y^{(1)}}\Pi
\end{equation*}
\end{cor}

The preceding formula can also be found on \cite{BSF}.

\subsection{Localizable symmetries. Second Noether's theorem}

\begin{defn}
A symmetry of the lagrangian $\xi_Y$ is said to be \textbf{localizable} when
$\xi_Y^{(1)}$ it vanishes on $\partial Z$ and for every pair of open sets $U$
and $U^{\prime}$ in $X$ with disjoint closures, there exists another
symmetry of the lagrangian $\zeta_Y$ such that
\begin{equation*}
\xi_Y^{(1)} = \zeta_Y^{(1)} \qquad\displaystyle{on }\;\pi_{XZ}^{-1}(U)
\end{equation*}
and
\begin{equation*}
\zeta_Y^{(1)}=0 \qquad\displaystyle{on
}\;\pi_{XZ}^{-1}(U^{\prime})\cup\partial Z
\end{equation*}
\end{defn}

\begin{thm}
\textbf{Second Noether Theorem.} If $\xi_Y$ is a localizable
symmetry, and $c_{\tilde{Z}}$ is a solution of De Donder equations
\eqref{DeDonderTilde}, then
\begin{equation*}
\widetilde{(\iota_{\xi_Y}\Theta_L)}(c_{\tilde{Z}}(t))=0
\end{equation*}
for all $t$. Therefore, if $\alpha=\iota_\xi\Theta_L$ is the preserved
quantity, then $\tilde{\alpha}$ is a constant of motion for the De Donder
equations.
\end{thm}

\textit{Proof}. First Noether theorem guarantees that the
preceding application is constant. Pick $t_0$ in the domain of
definition of $c_{\tilde{Z}}$, the space-time decomposition of $X$
guarantees that, for $t\neq t_0$, we can find, using tubular
neighbourhoods, two disjoint open sets $U$ and $U^{\prime}$ with
disjoint closures containing $Im(c_{\tilde{Z}}(t_0))$ and
$Im(c_{\tilde{Z}}(t))$ respectively.

If $\zeta_Y$ is the Cartan symmetry whose existence guarantees the notion of
localizable symmetry, respect to $U$ and $U^{\prime}$, then
\begin{equation*}
\widetilde{(\iota_{\xi_Y}\Theta_L)}(c_{\tilde{Z}}(t_0))=
\widetilde{(\iota_{\zeta_Y}\Theta_L)}(c_{\tilde{Z}}(t_0))=
\widetilde{(\iota_{\zeta_Y}\Theta_L)}(c_{\tilde{Z}}(t))= 0.
\end{equation*}
\hfill $\ \ \ \vrule height 1.5ex width.8ex depth.3ex \medskip$

\section{Momentum map}

In this section we are interested in considering groups of symmetries acting
on the configuration space $Y$, which induce a lifted action into $Z$ which
preserves the Lagrangian form.

\subsection{Action of a group}

If $G$ is a Lie group acting on $Y$, then the action of $G$ on $Y$ can be
lifted to an action of $G$ on $Z$, and the infinitesimal generator of the
lifted action corresponds to the lift of the infinitesimal generator of the
action, in other words,
\begin{equation*}
\xi_Z = \xi_Y^{(1)}
\end{equation*}

\begin{defn}
We shall say that a Lie group $G$ acts as a \textbf{group of symmetries of
the Lagrangian} if it defines an action on $Y$ that projects onto a
compatible action on $X$, which 1-jet prolongation preserves $B$, and if the
flow $\phi_Z$ of $\xi_Z$ verifies
\begin{equation*}
\phi_Z^*\mathcal{L}=\mathcal{L}\newline
\qquad \phi_Z^*\Pi=\Pi\newline
\end{equation*}
\end{defn}

The fact that the action is fibred implies that $\xi_Y$ is a projectable
vector field. Therefore, the condition $\phi_Z^*\mathcal{L}=\mathcal{L}$,
infinitesimally expressed as
\begin{equation*}
\pounds_{\xi_Z}\mathcal{L}=0,
\end{equation*}
jointly with the following two direct consequences of the definition:

(i) $\xi_Z$ is tangent to $B$

(ii) $\pounds_{(\xi_{Z})_{_{|B}}}\Pi=0$,

states the fact that $\xi_Y$ is a symmetry of the Lagrangian.

\subsection{Momentum map}

If we have a group of symmetries of the Lagrangian $G$ acting on $Y$, we can
make use of the Poincar\'e-Cartan $(n+1)$-form on $Z$ to construct the
analogous of the momentum map in Classical Mechanics.

\begin{defn}
The \textbf{momentum map} is a mapping
\begin{equation*}
J : Z \longrightarrow \mathfrak{g}^*\otimes \Lambda^nZ
\end{equation*}
or alternatively,
\begin{equation*}
J : Z \otimes \mathfrak{g} \longrightarrow \Lambda^nZ
\end{equation*}
defined by $J(z,\xi) := (\iota_{\xi_Z}\Theta_L)_z$.

Therefore, $J(\cdot,\xi)$ is a $n$-form, that we shall denote by $J^\xi$.
\end{defn}

\begin{remark}
\textrm{On $B$, since $\pounds_{(\xi_{Z})_{_{|B}}}\Pi=0$ we have
that $\iota_{(\xi_{Z})_{_{|B}}}d\Pi = -
d\iota_{(\xi_{Z})_{_{|B}}}\Pi$, and therefore,
\begin{equation*}
J(z,\xi) = (\iota_{\xi_{Z}} \Theta_L|_B)(z) = (\iota_{\xi_{Z}} d\Pi)(z) = -
(d\iota_{\xi_{Z}} \Pi)(z)
\end{equation*}
}
\end{remark}

Notice that $J^\xi$ is a preserved quantity, and we called $\widetilde{J^\xi}
$ its associated momentum.

\begin{prop}
\label{prop:dJxi}
\begin{equation*}
dJ^\xi = \iota_{\xi_Z}\Omega_L
\end{equation*}
\end{prop}

\textit{Proof}. As $\xi$ is projectable,
$\pounds_{\xi_Z}\Theta_L=0$ (by \ref{prop:propertiesF}), whence
\begin{equation*}
0 = \pounds_{\xi_Z}\Theta_L = \iota_{\xi_Z}d\Theta_L +
d\iota_{\xi_Z}\Theta_L = -\iota_{\xi_Z}\Omega_L+dJ^\xi.
\end{equation*}
\hfill $\ \ \ \vrule height 1.5ex width.8ex depth.3ex \medskip$

\subsection{Momentum map in Cauchy data spaces}

If $G$ is a Lie group acting on $Y$ as symmetries of the Lagrangian, it
induces an action on $\tilde{Z}$ defined pointwise on the image of every
curve in $\tilde{Z}$.

For $\xi\in\mathfrak{g}$, the vector field $\xi_{\tilde{Z}}$ is precisely
the vector on $\tilde{Z}$ induced by the vector field $\xi_Z$ on $Z$. And
since $\xi_Z$ is a Cartan symmetry, so is $\xi_{\tilde{Z}}$.

In a similar manner, the presymplectic form $\widetilde{\Theta_L}$ induces a
momentum map
\begin{equation*}
\tilde{J}: \tilde{Z} \longrightarrow \mathfrak{g}^*
\end{equation*}
defined using its pairing (for $\xi\in\mathfrak{g})$
\begin{equation*}
\tilde{J}^\xi = \langle \tilde{J}, \xi \rangle : \tilde{Z} \longrightarrow
\mathbb{R}
\end{equation*}
by
\begin{equation*}
\tilde{J}^\xi := \iota_{\xi_{\tilde{Z}}}\widetilde{\Theta_L}
\end{equation*}
One immediately has that $\widetilde{J^\xi} = \tilde{J}^\xi$. As
we know that a Cartan symmetry for the De Donder equations in $Z$,
then $\tilde{\xi}$ is a Cartan symmetry for the De Donder
equations in $\tilde{Z}$, thus $\tilde{J}^\xi$ is a preserved
quantity for the presymplectic setting.

By repeating the arguments in \eqref{prop:dJxi}, we have:

\begin{prop}
\begin{equation*}
d\tilde{J}^\xi = \iota_{\xi_{\tilde{Z}}}\widetilde{\Omega_L}
\end{equation*}
\end{prop}

\section{Examples}

\subsection{The Bosonic string}

Let $X$ be a 2-dimensional manifold, and $(B, g)$ a $(d+1)$-dimensional
spacetime manifold endowed with a Lorentz metric $g$ of signature $(-,+,
\dots ,+)$. A \textsl{bosonic string} is a map $\phi : X \longrightarrow B$
(see \cite{BGP,gimmsy1}).

In the folllowing, we shall follow the Polyakov approach to
clasical bosonic string theory. Let $S^{1,1}_2(X)$ be the bundle
over $X$ of symmetric covariant rank two tensors of Lorentz
signature $(-,+)$ or $(1,1)$. We take the vector bundle
$\pi:Y=X\times B\times S^{1,1}_2(X) \longrightarrow X$. Therefore,
in this formulation, a field $\psi$ is a section $(\phi, s)$ of
the vector bundle $Y=X\times B\times S^{1,1}_2(X)\longrightarrow
X$, where $\phi : X\longrightarrow X\times B$ is the bosonic
string and $s$ is a Lorentz metric on $X$.

\subsubsection{Lagrangian description}

We have that $Z= J^1(X\times B)\times_{X} J^1(S^{1,1}_2(X))$. Taking
coordinates $(x^{\mu})$, $(y^i)$ and $(x^{\mu}, s_{\mu\zeta})$ on $X$, $B$
and $S^{1,1}_2(X)$ then the canonical local coordinates on $Z$ are $%
(x^{\mu}, y^i ,s_{\zeta\xi}, y^i_{\mu}, s_{\zeta\xi\mu})$. In this system of
local coordinates, the Lagrangian density is given by
\begin{equation*}
\mathcal{L} = -\frac{1}{2}\sqrt{-\det (s)}s^{\zeta\xi}g_{ij}y^i_{\zeta}
y_{\xi}^jd^2 x\; .
\end{equation*}
The Cartan 2-form is
\begin{equation*}
\Theta_L = \sqrt{-\det(s)}\left(-s^{\mu\nu}g_{ij}y^j_\nu dy^i\wedge d^1x_\mu
+ \frac{1}{2}s^{\mu\nu}g_{ij}y^i_\mu y^j_\nu d^2x\right)
\end{equation*}
and the Cartan 3-form is
\begin{eqnarray*}
\Omega_L&=& d y^i\wedge d\left( -\sqrt{-\det (s)}s^{\zeta\xi}g_{ij}y^j_{\xi}
\right)\wedge d^1x_{\zeta} \\
&&- d\left( \frac{1}{2}\sqrt{-\det (s)}s^{\zeta\xi}g_{ij}y^i_{\zeta}
y_{\xi}^j \right)\wedge d^2x \\
&=& -\frac{1}{2}\left(\frac{\partial \sqrt{-\det(s)}}{\partial
s_{\rho\sigma}}s^{\zeta\xi}g_{ij}y^i_{\zeta}y^j_{\xi}-\sqrt{-\det
(s)}s^{\zeta \rho}s^{\xi
\sigma}g_{ij}y^{i}_{\eta}y^j_{\xi}\right) ds_{\rho\sigma}\wedge d^2x \\
&&-\frac{1}{2}\sqrt{-\det(s)} s^{\zeta\xi} \frac{\partial g_{ij}}{\partial
y^k} y^i_{\zeta}y^j_{\xi}\,dy^k\wedge d^2x-\sqrt{-\det (s)} s^{\zeta\xi}
g_{ij}y^i_{\zeta}\, dy^j_{\xi}\wedge d^2x \\
&&+\left( \frac{\partial \sqrt{-\det(s)}}{\partial h_{\rho\sigma}}
s^{\zeta\xi} g_{ij}y^j_{\xi}-\sqrt{-\det(s)}s^{\zeta \rho}s^{\xi
\sigma}g_{ij}y^j_{\xi} \right)d s_{\rho\sigma}\wedge dy^i\wedge
d^1x_{\zeta}
\\
&&+\sqrt{-\det(s)} s^{\zeta\xi} \frac{\partial g_{ij}}{\partial y^k}
y^j_{\xi}\, dy^k\wedge dy^i\wedge d^1x_{\zeta} \\
&&+\sqrt{-\det(s)} s^{\zeta\xi} g_{ij}\, d y^j_{\xi}\wedge dy^i\wedge
d^1x_{\zeta}.
\end{eqnarray*}

If we solve the equation $i_{\mathbf{h}}\Omega_L=\Omega_L$, where
\begin{equation*}
\mathbf{h} = dx^{\mu}\otimes\left(\frac{\partial}{\partial x^\mu}
+ {\Gamma}^i_{\mu}\frac{\partial}{\partial y^i} +
{\gamma}_{\zeta\xi\mu}\frac{\partial}{\partial s_{\zeta\xi}} +
{\Gamma}^i_{\zeta\mu}\frac{\partial}{\partial y^i_{\zeta}}
+{\gamma}_{\zeta\xi\rho\mu}\frac{\partial}{\partial
s_{\zeta\xi\rho}}\right) \;,
\end{equation*}
we obtain that:
\begin{eqnarray*}
\Gamma^i_{\mu}&=& y^i_{\mu} \\
0&=&\frac{1}{2}\sqrt{-\det(s)} s^{\zeta\xi}\frac{\partial
g_{ij}}{\partial y^k} y^i_{\zeta}y^j_{\xi} -\sqrt{-\det(s)}
s^{\zeta\xi}\frac{\partial g_{kj}}{\partial y^i}
y^i_{\zeta}y^j_{\xi} -\sqrt{-\det(s)} s^{\zeta\xi} g_{kj}
\Gamma^j_{\xi\zeta} \\
&&- \left( \frac{\partial \sqrt{-\det(s)}}{\partial
s_{\rho\sigma}}s^{\zeta\xi}
g_{kj}y^j_{\xi}-\sqrt{-\det(s)}s^{\zeta \rho}s^{\xi
\sigma}g_{kj}y^j_{\xi} \right)\gamma_{\rho\sigma\zeta}\;,
\end{eqnarray*}
and the constraints given by the equations
\begin{equation*}
\frac{\partial }{\partial s_{\rho\theta}}\left(\sqrt{-\det(s)}
s^{\zeta\xi}\right)g_{ij} y^i_{\zeta}y^j_{\xi}=0 \; .
\end{equation*}
The previous equation corresponds to the three following constraints
\begin{eqnarray*}
\left[ s^{\zeta 0}s^{\xi 0} (s_{01}^2-s_{00}s_{11}) + \frac{1}{2}
s^{\zeta\xi}s_{11}\right] g_{ij} y^i_{\zeta} y^j_{\xi} & = & 0 \\
\left[ s^{\zeta 1}s^{\xi 1} (s_{01}^2-s_{00}s_{11}) + \frac{1}{2}
s^{\zeta\xi}s_{00}\right] g_{ij} y^i_{\zeta} y^j_{\xi} & = & 0 \\
\left[ s^{\zeta 0}s^{\xi 1} (s_{01}^2-s_{00}s_{11}) -
s^{\zeta\xi}s_{01}\right] g_{ij} y^i_{\zeta} y^j_{\xi} & = & 0 \\
\end{eqnarray*}
which determine $Z_2$.

\subsubsection{Hamiltonian description}

The Legendre transformation is given by
\begin{equation*}
Leg_L(x^{\mu}, y^i ,s_{\zeta\xi}, y^i_{\mu}, s_{\zeta\xi\mu})=(x^{\mu}, y^i
,s_{\zeta\xi}, -\sqrt{-\det(s)}\, s^{\mu\zeta}g_{ij}y^j_{\zeta}, 0)
\end{equation*}
Therefore, the Lagrangian ${L}$ is almost-regular and, moreover,
$\tilde{M}_1=\hbox{Im } Leg_L\cong M_1 =leg_L(Z)\cong J^1(X\times
B)\times_{X} S^{1,1}_2(X)$. Take now coordinates $(x^{\mu}, y^i,
s_{\zeta\xi}, p_i^{\mu})$ on $M_1$ and consider the mapping $s_1:
M_1 \rightarrow \tilde{M_1}$ given by
\begin{equation*}
s_1 (x^{\mu}, y^i ,s_{\zeta\xi}, p_i^{\mu})=(x^{\mu}, y^i
,s_{\zeta\xi},
p=\frac{1}{2\sqrt{-\det(s)}}s_{\zeta\xi}g^{ij}p^i_{\zeta}p^j_{\xi},
p_i^{\mu})
\end{equation*}

Then, we have
\begin{equation*}
\Omega_{M_1}
=-d\left(\frac{1}{2\sqrt{-\det(s)}}s_{\zeta\xi}g^{ij}p_i^{\zeta}p_j^{\xi}\right)\wedge
d^2x+dy^i\wedge dp^{\mu}_i\wedge d^1x_{\mu}
\end{equation*}
and the Hamilton equations are given by $i_{\tilde{\mathbf{h}}
}\Omega_{M_1}=\Omega_{M_1}$. Putting
\begin{equation*}
\tilde{\mathbf{h}}= dx^{\mu}\otimes\left(\frac{\partial}{\partial
x^\mu} + \tilde{\Gamma}^i_{\mu}\frac{\partial}{\partial y^i} +
\tilde{\gamma}_{\zeta\xi\mu}\frac{\partial}{\partial s_{\zeta\xi}}
+ \tilde{\Gamma}^{\zeta}_{i\mu}\frac{\partial}{\partial
p_i^{\zeta}} \right)
\end{equation*}
we obtain
\begin{eqnarray*}
\tilde{\Gamma}^i_{\mu}&=&-\frac{1}{\sqrt{-\det(s)}}s_{\zeta\mu}g^{ij}p^{\zeta}_j \\
\tilde{\Gamma}^{\mu}_{i\mu}&=&\frac{1}{2\sqrt{-\det(s)}}s_{\zeta\xi}\frac{\partial
g^{ij}}{\partial y^k} p^i_{\zeta} p^j_{\xi} \;,
\end{eqnarray*}
and the secondary constraints
\begin{equation*}
\frac{g^{ij}}{\sqrt{-\det(s)}}\left(\frac{1}{2\det(s)}\frac{\partial
\det(s)}{\partial
s_{\rho\sigma}}s_{\zeta\xi}p^{\zeta}_ip^{\xi}_j-p^{\rho}_ip^{\sigma}_j\right)=0
\end{equation*}
determining $M_2$.

\subsubsection{Symmetries}

Let $\lambda$ be an arbitrary function on $X$, and we denote also by $\lambda
$ its pullback to $Y$ and $Z$.

Consider the following $\pi_{XY}-$projectable vector field on $Y$
\begin{equation*}
\xi_Y := \lambda s_{\sigma\rho} \frac{\partial }{\partial s_{\sigma\rho}}
\end{equation*}
Its 1-jet prolongation is given by
\begin{equation*}
\xi_Z := \xi_Y^{(1)} = \lambda s_{\sigma\rho} \frac{\partial
}{\partial s_{\sigma\rho}} + \left(\frac{\partial
\lambda}{\partial x^\mu}s_{\sigma\rho}+\lambda s_{\sigma\rho,\mu}
\right)\frac{\partial }{\partial s_{\sigma\rho,\mu}}
\end{equation*}

\medskip

We shall prove that $\xi_Y$ is a symmetry of the Lagrangian. Note that

\begin{align*}
\pounds_{\xi_Z} \Theta_L = \pounds_{\xi_Y}
&(\sqrt{-\det(s)})\left(-s^{\mu\nu}g_{ij}y^j_\nu dy^i\wedge
d^1x_\mu + \frac{1}{2}s^{\mu\nu}g_{ij}y^i_\mu y^j_\nu d^2x\right) \\
&+ \sqrt{-\det(s)}\left(-\pounds_{\xi_Y}(s^{\mu\nu})g_{ij}y^j_\nu dy^i\wedge
d^1x_\mu + \frac{1}{2}\pounds_{\xi_Y}(s^{\mu\nu})g_{ij}y^i_\mu y^j_\nu
d^2x\right)
\end{align*}
And a little computation shows that
\begin{equation*}
\xi_Y(\sqrt{-\det(s)}) = \lambda \sqrt{-\det(s)}
\end{equation*}
and
\begin{equation*}
\pounds_{\xi_Y}(s^{\mu\nu}) = - \lambda s^{\mu\nu}
\end{equation*}
Therefore, $\xi_Y$ is a symmetry of the Lagrangian, and as the corresponding
Cartan symmetry $\xi_Z$ is $\pi_{XZ}$ projectable, then the symmetry
projects onto the final constraint manifold.

The preserved quantity given by Noether's theorem is given by
\begin{equation*}
J^{\xi_Y} = \sum_{\sigma,\rho,\mu} \lambda s_{\sigma\rho,\mu} s_{\sigma\rho}
d^1x_\mu
\end{equation*}

Note that the vector field
\begin{equation*}
\xi_Y = 2\lambda s_{\sigma\rho} \frac{\partial }{\partial s_{\sigma\rho}}
\end{equation*}
is the infinitesimal generator of the action of the group
$N=\mathcal{C}S_2^{1,1}(X)\equiv\mathcal{F}(X,\mathbb{R}^+)$ of
the conformal transformations of a metric of signature $(1,1)$
given by
\begin{equation*}
\lambda (\phi,s) := (\phi, \lambda^2s)
\end{equation*}

We have that
\begin{equation*}
det(\lambda^2 s) = \lambda^4 det(s)
\end{equation*}
and
\begin{equation*}
(\lambda^2 s)^{\mu\nu} = \lambda^{-2} s^{\mu\nu};
\end{equation*}
therefore, the action preserves the constraint equations.

In a similar manner, we can consider the action of $H=Diff(X)$ by
\begin{equation*}
\eta (\phi,s) := (\phi\circ\eta^{-1},(\eta^{-1})^* s)
\end{equation*}
or more generally, consider the semidirect product $G=H[N]$, where the
action of elements $\eta\in H$ on elements $\lambda\in N$ is given by
\begin{equation*}
\eta\cdot\lambda := \lambda\circ\eta^{-1}
\end{equation*}
The group $G$ is a group of symmetries for $Y$, and the action is given by
\begin{equation*}
(\eta,\lambda)\cdot (\phi,s) := (\phi\circ\eta^{-1},\lambda^2 (\eta^{-1})^*
s)
\end{equation*}

\subsubsection{Symmetries on the Hamiltonian side}

Not being $L$ regular, we cannot guarantee that $\xi_Y$ is a symmetry of the
Lagrangian for the Hamiltonian side. However, an easy computation gives us
that
\begin{equation*}
\xi_Y^{(1)} = \lambda s_{\sigma\rho} \frac{\partial }{\partial
s_{\sigma\rho}} - \lambda p^\mu_{\sigma\rho}\frac{\partial
}{\partial p^\mu_{\sigma\rho}}
\end{equation*}

Thus,
\begin{equation*}
\pounds_{\xi_Y^{(1)}} \Theta_L =
\pounds_{\xi_Y^{(1)}}(p^\mu_{\sigma\rho}ds_{\sigma\rho}d^nx_\mu) =
p^\mu_{\sigma\rho} s_{\sigma\rho}\frac{\partial \lambda}{\partial x^\mu}d^2
x
\end{equation*}

However, note that in $M_1$ we have that $p^\mu_{\sigma\rho}=0$,
therefore $\xi_Y$ restricts to a symmetry there of the form
\begin{equation*}
\lambda s_{\sigma\rho} \frac{\partial }{\partial s_{\sigma\rho}}
\end{equation*}
Furthermore, this is the infinitesimal generator of the restriction of the
lifted action on $Z^*$, and one easily deduces, on view of the form of the
secondary constrain equation, that the action restricts as well to the
secondary constraint submanifold.

\subsubsection{More symmetries}

In general, one can consider the invariance of the equations and the
Lagrangian respect to diffeomorphisms of $X$. If $\eta$ is one of such
diffeomorphisms, then $\eta(\phi,s) = (\phi\circ\eta^{-1},(\eta^{-1})^*s)$,
having infinitesimal generator
\begin{equation*}
-(s_{\sigma\mu}\frac{\partial \xi^\mu}{\partial
x^\rho}+s_{\rho\mu}\frac{\partial \xi^\mu}{\partial
x^\sigma})\frac{\partial }{\partial s_{\sigma\rho}
}+\xi^\mu\frac{\partial }{\partial x^\mu}
\end{equation*}
where $\xi^{\mu}\frac{\partial }{\partial x^\mu}$ is the infinitesimal
generator of $\eta$.

The most general situation arises when considering the semidirect
product $H[N]$ of the group $H=Diff(X)$ and the group $N$ of the
positive real functions on $X$ defined above, given by
\begin{equation*}
\eta\cdot\lambda := \lambda\circ\eta^{-1}
\end{equation*}
The action is defined as follows
\begin{equation*}
(\eta,\lambda)(\phi,s)=(\phi\circ\eta^{-1},\lambda^2(\eta^{-1})^*s),
\end{equation*}
and the infinitesimal generator is
\begin{equation*}
2\lambda s_{\sigma\rho}\frac{\partial }{\partial s_{\sigma\rho}}
-(s_{\sigma\mu}\frac{\partial \xi^\mu}{\partial
x^\rho}+s_{\rho\mu}\frac{\partial \xi^\mu}{\partial
x^\sigma})\frac{\partial }{\partial s_{\sigma\rho}
}+\xi^\mu\frac{\partial }{\partial x^\mu}
\end{equation*}
This is proved to be a symmetry of the Lagrangian (see \cite{gimmsy1}), and
the corresponding preserved quantity is
\begin{equation*}
\frac{\partial L}{\partial y^i}(y^i_\mu \xi^\nu) + \frac{\partial
L}{\partial s_{\sigma\rho}}(s_{\sigma\rho,\nu}\xi^\nu-2\lambda
s_{\sigma\rho}+s_{\sigma\nu}\frac{\partial \xi^\nu}{\partial
x^\rho}+s_{\rho\nu}\frac{\partial \xi^\nu}{\partial x^\sigma})=0
\end{equation*}
for $\lambda, \xi^\nu and \frac{\partial \xi^\nu}{\partial x^\rho}$
arbitrary, which gives in particular the equation $\partial L/\partial
s_{\sigma\rho} = 0$, which is expanded into
\begin{equation*}
\frac{1}{2}s^{\mu\nu}g_{ij}y^i_\mu y^j_\nu s_{\sigma\rho}=g_{ij}y^i_\sigma
y^j_\rho
\end{equation*}
which amounts to say that $h$ is a metric conformally equivalent to $\phi^* g
$ and that the conformal factor is precisely $\frac{1}{2}s^{\mu%
\nu}g_{ij}y^i_\mu y^j_\nu$.

\subsection{Klein-Gordon equations}

\subsubsection{Lagrangian setting}

For the Klein-Gordon equation, we set $(X,g)$ be a Minkovski
space, and $Y:=X\times\mathbb{R}$, where $\pi:Y\longrightarrow X$
is the first canonical projection. A section $\phi$ of $\pi$ can
be identified with a smooth function on $X$, say $\varphi\in
\mathcal{C}^\infty(X)$, where $y(j^1\phi(x))=\varphi(x)$ and
$\displaystyle{z_\mu(j^1\phi(x))=\frac{\partial \varphi}{\partial
x^\mu}(x)}$.

The chosen volume form will be $\eta := \sqrt{-det \, g}$.

\subsubsection{Lagrangian setting}

The Lagrangian function will be
\begin{equation*}
L(x^\mu,y,z_\mu) := \frac{1}{2}\left(g^{\mu\nu}z_\mu z_\nu + m^2y^2\right)
\end{equation*}

which is regular, as
\begin{equation*}
\hat{p}^\mu = \frac{\partial L}{\partial z_\mu} = g^{\mu\nu}z_\nu
\end{equation*}
and thus the Hessian matrix is precisely $(g^{\mu\nu})$.

The Poincar\'e-Cartan 4-form is
\begin{equation*}
\Theta_L = \sqrt{-det \, g}\left(g^{\mu\nu}z_\mu dy\wedge d^3x_\mu
- \frac{1}{2} (g^{\mu\nu}z_\mu z_\nu - m^2y^2)d^4x\right)
\end{equation*}

The boundary condition will be $B=0$, that is, $\sigma(\partial X)=0$, and
this restriction is required as an asymptotic condition to replace the
restrictions of compactness that we have placed on $X$.

And the Euler-Lagrange equations in terms of $\varphi$ become
\begin{equation*}
m^2\varphi = g^{\mu\nu}\frac{\partial ^2\varphi}{\partial x^\mu \partial
x^\nu}
\end{equation*}
that is, the Klein-Gordon equation.

\subsubsection{Legendre transformation and Hamiltonian setting}

We compute
\begin{equation*}
\hat{p} = \frac{1}{2}(-g^{\mu\nu}z_\mu z_\nu + m^2y^2)\sqrt{- det \, g}
\end{equation*}

Thus we can write the Hamiltonian
\begin{equation*}
H(x^\mu,y,p^\mu) = \frac{1}{2}(g_{\mu\nu}p^\mu p^\nu + m^2y^2),
\end{equation*}
and the Hamilton equation for $\varphi$ corresponding to a section
$\phi(x^\mu)=(x^\mu,\varphi(x^\mu),\varphi^\mu(x^\mu))$ become
\begin{align*}
\frac{\partial \varphi}{\partial x^\mu} &= g_{\mu\nu}p^\nu \\
\sum_\mu\frac{\partial \varphi^\mu}{\partial x^\mu} &= (\sqrt{- det g\, }
)m^2\varphi
\end{align*}

\subsubsection{Symmetries}

Let $\xi_X$ be a Killing vector field on $X$, with coordinates
\begin{equation*}
\xi_X = \xi_\mu\frac{\partial }{\partial x^\mu}
\end{equation*}

Let us call $\xi_Y$ the vector field $\xi_X$ as seen in $Y$, that is,
locally,
\begin{equation*}
\xi_Y(x,t) := \xi_\mu\frac{\partial }{\partial x^\mu}
\end{equation*}

Its 1-jet prolongation $\xi_Z$ is given by
\begin{equation*}
\xi_Z = \xi_\mu\frac{\partial }{\partial x^\mu} -
z_\nu\frac{d\xi^\nu}{dx^\mu }\frac{\partial }{\partial z_\mu}
\end{equation*}

These vector fields are symmetries of the Lagrangian, and the associated
preserved quantity is written as
\begin{equation*}
\left[-g^{\mu\nu}z_\mu\xi^\gamma dy\wedge d^2x_{\nu\gamma}-\frac{\xi^\nu}{2}
\left(g^{\mu\nu}z_\mu z_\nu-m^2y^2\right)d^3x_\gamma\right]\sqrt{-det \, g}
\end{equation*}

\subsubsection{Cauchy surfaces}

The general integral expression for the preserved quantity for an
arbitrary Cauchy surface $M$ and for sections
$\phi(x^\mu)=(x^\mu,\varphi(x^\mu),\frac{\partial
\varphi}{\partial x^\mu}(x^\mu))$ solutions of the Euler-Lagrange
equations, and verifying the boundary condition, is given by
\begin{equation*}
\int_M \sqrt{-det \, g} \left[ g^{\mu\gamma}\frac{\partial
\varphi}{\partial x^\mu}\xi^\nu\frac{\partial \varphi}{\partial
x^\nu}+ g^{\mu\nu}\frac{\partial \varphi}{\partial
x^\mu}\xi^\gamma\frac{\partial \varphi}{\partial x^\nu}-
\frac{\xi^\gamma}{2}\left(g^{\mu\nu}\frac{\partial
\varphi}{\partial x^\mu}\frac{\partial \varphi}{\partial
x^\nu}-m^2\varphi^2\right) \right] d^3x_\gamma
\end{equation*}

In the particular case in which we have $M$ to be a space-like Cauchy
surface, $g$ induces a positive definite metric $g_M$ on $M$, and we have
that the preserved quantity is expressed as
\begin{equation*}
\int_M \sqrt{-det \, g} \left[ \frac{\partial \varphi}{\partial
x^0}\xi^\nu \frac{\partial \varphi}{\partial x^\nu}+
g^{\mu\nu}\frac{\partial \varphi}{\partial
x^\mu}\xi^0\frac{\partial \varphi}{\partial x^\nu}-
\frac{\xi^0}{2} \left(g^{\mu\nu}\frac{\partial \varphi}{\partial
x^\mu}\frac{\partial \varphi }{\partial x^\nu}-m^2\varphi^2\right)
\right]d^3x_0
\end{equation*}

Whenever $\xi_X$ is space-like (that is, parallel to $M$), we obtain that
the preserved quantity gets
\begin{equation*}
\int_M \left[ \frac{\partial \varphi}{\partial x^0}\frac{\partial
\varphi}{\partial x^\nu}\xi^\nu \right]d^3x_0
\end{equation*}
which is the angular momentum whenever $\xi_X$ is an infinitesimal rotation,
and linear momentum whenever it is an infinitesimal translation.

For the contrary, if $\xi_X = \frac{\partial }{\partial x^0}$ we get
\begin{equation*}
\frac{1}{2}\int_M \left[ \frac{\partial \varphi}{\partial
x^0}\frac{\partial \varphi}{\partial x^0}+ g^{AB}\frac{\partial
\varphi}{\partial x^A}\frac{\partial \varphi}{\partial x^B}+
m^2\varphi^2 \right]d^3x_0
\end{equation*}
which is the energy of the field $\varphi$ on the Cauchy surface $M$.

\section*{Acknowledgments}

This work has been supported by grant BFM2001-2272 from the Ministry of
Science and Technology. A. Santamar{\'\i}a--Merino wishes to thank the
Programa de formaci\'on de Investigadores of the Departamento de
Educaci\'on, Universidades e Investigaci\'on of the Basque Government
(Spain) for financial support.

\bibliographystyle{amsplain}
\bibliography{bibliograf'ia}

\end{document}